\documentclass[fleqn]{2017SCGE}
\usepackage{multirow}
\usepackage{makecell}

\usepackage{slantsc}
\usepackage{aas_macros}
\usepackage{amsmath}
\usepackage[square,sort,comma,numbers]{natbib}
\usepackage{xcolor,colortbl}
\usepackage[normalem]{ulem}
\graphicspath{{figures/}}

\usepackage{bm}
\usepackage{multirow}

\newcommand{\ud}{\mathrm{d}}

\newcommand{\rc}{\rowcolor{lightgray}}

\newcommand{\msun}{\mathrm{M}_\odot}

\begin{document}

\ensubject{subject}

\ArticleType{Article}
\Year{2019}
\Month{December}
\Vol{1}
\No{1}
\DOI{}
\ArtNo{000000}
\ReceiveDate{000, 2019}
\AcceptDate{000, 2019}

\title{The mass of our Milky Way}{The mass of our Milky Way}

\author[1,2]{Wenting Wang}{{wenting.wang@ipmu.jp}}%
\author[1,2]{Jiaxin Han}{jiaxin.han@sjtu.edu.cn}%
\author[3]{Marius Cautun}{}%
\author[1]{Zhaozhou Li}{}%
\author[4]{Miho N. Ishigaki}{}%

\AuthorMark{Wang W.}

\AuthorCitation{Wang W., Han J., et al}

\address[1]{Department of Astronomy, Shanghai Jiao Tong University, Shanghai 200240, China}
\address[2]{Kavli IPMU (WPI), UTIAS, The University of Tokyo, Kashiwa, Chiba 277-8583, Japan}
\address[3]{Leiden Observatory, Leiden University, PO Box 9513, NL-2300 RA Leiden, the Netherlands}
\address[4]{Astronomical Institute, Tohoku University, Aoba-ku, Sendai, Miyagi 980-8578, Japan}

\abstract{We perform an extensive review of the numerous studies and methods used to determine the total mass of the Milky Way. 
We group the various studies into seven broad classes according to their modeling approaches. The classes include: 
i) estimating Galactic escape velocity using high velocity objects; ii) measuring the rotation curve through terminal and circular
velocities; iii) modeling halo stars, globular clusters and satellite galaxies with the Spherical Jeans equation and iv) with 
phase-space distribution functions; v) simulating and modeling the dynamics of stellar streams and their progenitors; vi) modeling 
the motion of the Milky Way, M31 and other distant satellites under the framework of Local Group timing argument; and vii) measurements 
made by linking the brightest Galactic satellites to their counterparts in simulations. For each class of methods, we introduce their 
theoretical and observational background, the method itself, the sample of available tracer objects, model assumptions, uncertainties, 
limits and the corresponding measurements that have been achieved in the past. Both the measured total masses within the radial range 
probed by tracer objects and the extrapolated virial masses are discussed and quoted. We also discuss the role of modern numerical 
simulations in terms of helping to validate model assumptions, understanding systematic uncertainties and calibrating the measurements. 
While measurements in the last two decades show a factor of two scatters, recent measurements using \textit{Gaia} DR2 data are 
approaching a higher precision. We end with a detailed discussion of future developments in the field, especially as the size and 
quality of the observational data will increase tremendously with current and future surveys. In such cases, the systematic 
uncertainties will be dominant and thus will necessitate a much more rigorous testing and characterization of the various mass 
determination methods.}



\maketitle

\begin{multicols}{2}


\section{Introduction}\label{sec:intro}

In the current structure formation paradigm of $\Lambda$ cold dark matter ($\Lambda$CDM), 
gas cools in the center of an evolving population of dark matter halos \citep{1978MNRAS.183..341W}, 
which forms galaxies. Dark matter halos grow in mass and size through both smooth accretion 
of diffuse matter and from mergers with other halos \citep[e.g.][]{2011MNRAS.413.1373W}. 
Smaller halos together with their own central galaxies fall into larger halos and become 
``subhalos'' and ``satellites'' of the galaxy in the center of the dominant host halo. 
Orbiting around the central galaxy of the host halo, these satellites and subhalos lose 
mass due to tidal effects. Stars are stripped from them to form stellar streams, 
which then gradually mix in phase space. These stars form the stellar halo around the 
central galaxy \citep[e.g.][]{2005ApJ...635..931B, 2006MNRAS.365..747A,2010MNRAS.406..744C}. 
In the end, satellite galaxies and stripped material from these satellites merge with the central 
galaxy and contribute to its growth. 

Compared with other distant galaxies, the distances and velocities of individual stars that form 
the diffuse stellar halo of our Milky Way (hereafter MW) can be directly observed, because we are 
embedded within our MW. The observed dynamics of luminous objects in the MW stellar halo, such 
as bright stars, satellite galaxies, globular clusters, maser sources, HI gas clouds and tidal streams, 
which serve as dynamical tracers, contain valuable information about the underlying potential. Given 
a reasonable model which describes their dynamics or phase-space distributions within a realistic 
potential profile with free parameters, one can constrain the total mass distribution and infer the 
total or virial mass of our MW. 

\begin{figure*}
    \centering
    \includegraphics[width=\textwidth]{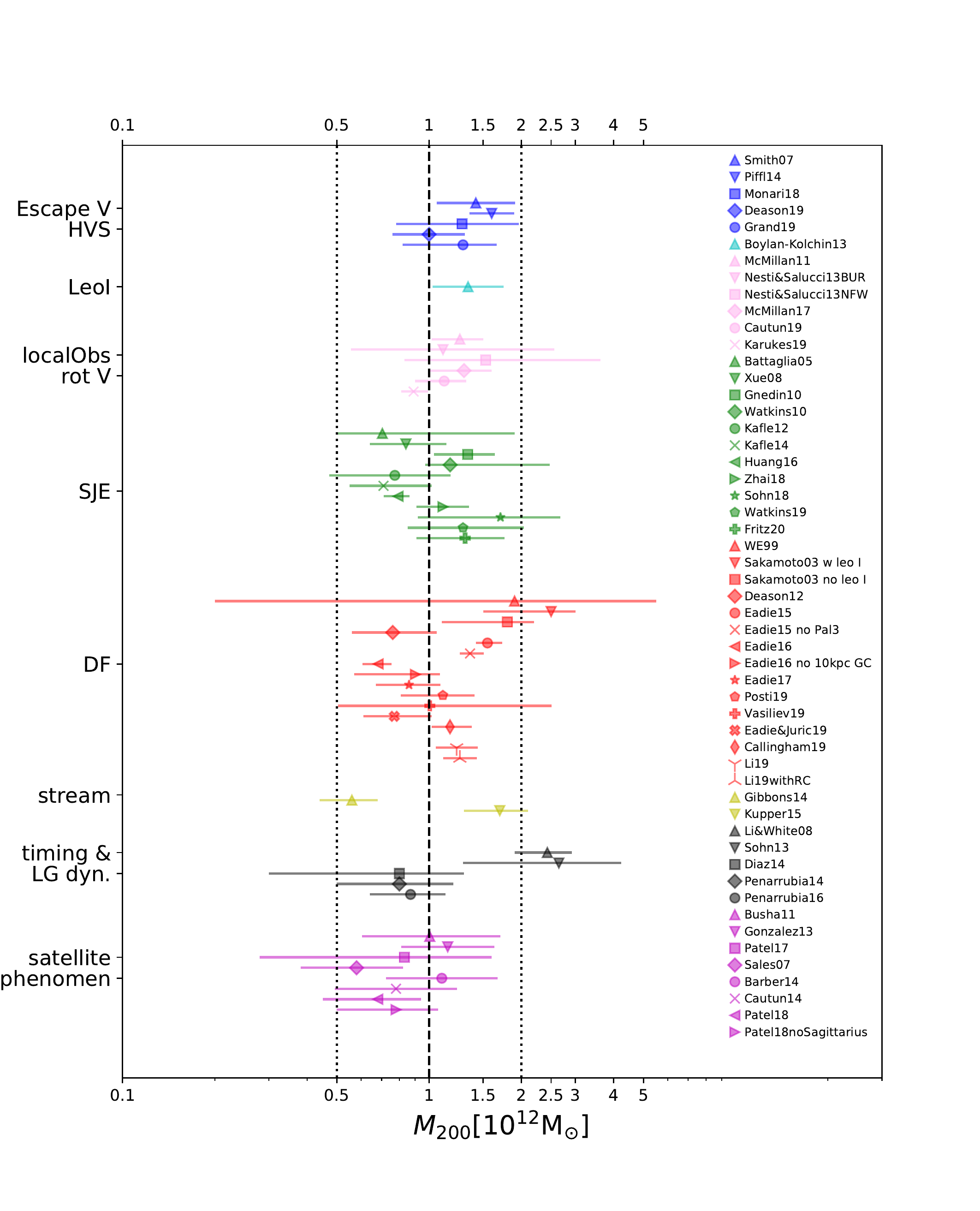}
    \vskip -.5cm
    \caption{Literature compilation of inferred virial masses for the MW. Classes of methods 
    are marked in different colors. Measurements have been converted to $M_{200}$, assuming NFW profiles. 
    95\% or 90\% confidence regions have been converted to 1-$\sigma$ (68\%) errors, assuming the errors 
    are either Gaussian in linear space if the reported upper and lower errors have comparable size, or 
    Gaussian in log space if the upper and lower errors have very different size in linear scale but are 
    more comparable in log space. However, the assumption of Gaussian errors does not always hold. We 
    just keep the original confidence regions \citep{2015ApJ...806...54E,2016ApJ...829..108E,2017ApJ...835..167E,2019ApJ...875..159E} 
    or decrease the errors by about 10\% for a few studies based on Bayesian analysis \citep{2013JCAP...07..016N}. 
    A few measurements have considered systematic uncertainties in their errors, for which we 
    also keep the original errors \citep{1999MNRAS.310..645W,2003A&A...397..899S,2010MNRAS.406..264W}.
    The vertical dashed line at $1\times10^{12}~\msun$, and two vertical dotted lines at 0.5 and 2 $\times10^{12}~\msun$ 
    are plotted to guide the eye. The readers can see \ref{app:measurements} for a table summarizing these measurements, 
    as well as the enclosed masses within fixed radii covered by tracer objects. \emph{A figure showing a subset of measurements 
    using \textit{Gaia} DR2 data are presented and discussed in Sec.~\ref{sec:disc} (Fig.~\ref{fig:massplot_gaiadr2}).}}
    \label{fig:massplot}
\end{figure*}

We provide in Fig.~\ref{fig:massplot} a literature summary of measured virial masses for the MW. It 
is an updated version of Figure 1 in Wang et al.\ \cite{2015MNRAS.453..377W} and Figure 7 in 
Callingham et al. \cite{2019MNRAS.484.5453C}. The figure provides a general impression of the multitude 
of studies and the variety of methods used to constrain the virial mass of our Galaxy. For clarity, we 
grouped the various approaches into several categories, with each category shown in a different color. 
The figure shows only measurements with quoted statistical errors or confidence intervals, and does not 
include measurements without associated uncertainties. The exact $M_{200}$ values shown in Fig.~\ref{fig:massplot} 
and their corresponding errors are provided in the second column of the table in \ref{app:measurements}.

The measurements in Fig.~\ref{fig:massplot} show a very large scatter. Part of the scatter is due 
to model extrapolations. For many of the studies, there were no or limited number of luminous 
tracers out to large enough distances, and thus to estimate the mass outside the radius of the most 
distant object, extrapolations of the model potential profile were made in these studies. For example, 
Taylor et al. in 2016 \cite{2016MNRAS.461.3483T} reported that an accurate measurement of the mass 
within 50~kpc can result in a 20\% uncertainty on the virial mass of the Galaxy. Moreover, the virial 
mass plotted as the $x$-axis in Fig.~\ref{fig:massplot} is defined as the total mass enclosed 
within a radius $R_{200}$, inside which the density is 200 times the critical density of the universe. 
The virial mass defined in this way is denoted as $M_{200}$. In fact, studies in Fig.~\ref{fig:massplot} 
used varying definitions of virial masses. We have made conversions to change these different definitions 
to $M_{200}$, assuming that the underlying mass profiles follow the NFW halo mass profile 
\citep{1996ApJ...462..563N,1997ApJ...490..493N}. If the original studies have provided constraints 
for the halo concentration or relations to calculate the concentration, we take their concentration 
when making the conversion to $M_{200}$. Otherwise, we use a mean virial mass versus concentration relation 
provided by Duffy et al. in 2008 \cite{2008MNRAS.390L..64D} to obtain the concentration and make the conversion.
Additional uncertainties can be introduced through such conversions. 

The remaining scatter in Fig.~\ref{fig:massplot} is very likely caused by systematics in the models 
or peculiar assumptions when coping with incomplete data. For example, the velocity anisotropies for 
the observed tracer objects have to be known in order to break the mass-anisotropy degeneracy and 
properly constrain the rotation velocity or the underlying potential. However, tangential velocities 
in reality are often not available, and thus the velocity anisotropy has to be assumed as a constant, 
as a radius-dependent function with free parameters, inferred from numerical simulations or marginalized 
over, which unavoidably introduced additional uncertainties to the measured virial mass. Furthermore, 
many dynamical models rely on steady state and spherical assumptions, which might 
not be valid for our MW. Dynamically hot streams and coherent movements of satellite galaxies can 
violate the steady state assumption, and dark matter halos are triaxial \citep{2002ApJ...574..538J}.

\begin{figure*}
    \centering
    \includegraphics[width=\textwidth]{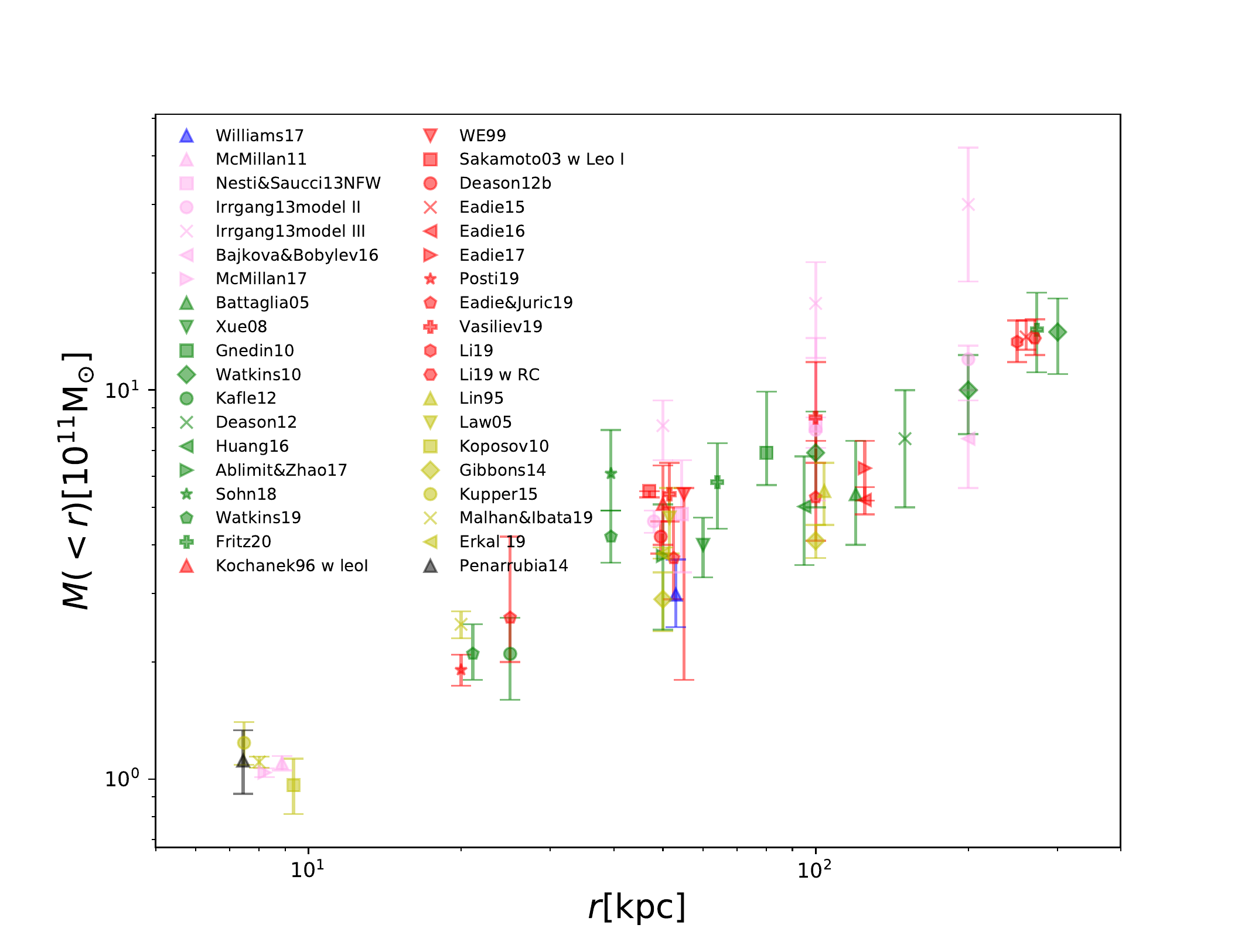}
    \vskip -.5cm
    \caption{Literature compilation of enclosed masses within the radii which can be covered by observed dynamical tracers.
    The same color scheme as Fig.~\ref{fig:massplot} is maintained for measurements grouped in the same category of method.
    Small offsets have been added to the groups of data points at $r=8$, 50, 100 and 260~kpc. 
    }
    \label{fig:massplot2}
\end{figure*}

In fact, many measurements in the past provided constraints on the circular velocities or the enclosed masses within the 
radii which can be covered by observed dynamical tracers, and we summarize these measurements in Fig.~\ref{fig:massplot2}. 
In \ref{app:measurements}, we provide in the third column of the table the enclosed masses within fixed radii, together 
with available circular velocities and local escape velocities at the solar radius, if these were provided. The readers 
can also find a similar figure from, for example, Eadie \& Juri{\'c} in 2019 \cite{2019ApJ...875..159E}, and a table from, 
for example, the review paper by Bland-Hawthorn et al. in 2016 \cite{2016ARA&A..54..529B}. The mass within the maximum 
radii of tracers should in principle be less model dependent and more reliable compared with the extrapolated virial 
mass in many cases. In fact, a general feature of dynamical modeling is that the best constrained mass for a given 
tracer is located around the median tracer radius~\citep[e.g.]{Walker,2016MNRAS.456.1003H}.

Although the enclosed mass within a fixed radius, which is covered by the radial distribution of employed tracers, has 
less uncertainty than the extrapolated virial mass, the latter is still a very important and useful quantity in many 
applications. The virial mass is critical for comparing observed properties of the Milky Way with cosmological predictions. 
The so-called missing satellite problem \cite{1999ApJ...522...82K,1999ApJ...524L..19M} is one of the examples. Very 
early on it was pointed out that the observed number of satellite galaxies is significantly lower than the predicted 
number of dark matter subhalos by numerical simulations. Although this problem can be alleviated by newly discovered 
faint MW satellites \citep[e.g.][]{2018PASJ...70S..18H,2019PASJ...71...94H}, explained by galaxy formation physics 
\cite[e.g.][]{2000ApJ...539..517B,2002MNRAS.333..156B} which predicts that a significant number of small subhalos 
do not host a galaxy, or explained by warm dark matter which predicts significantly less number of small subhalos 
\citep[e.g.][]{2010MNRAS.404L..16M,2014MNRAS.442.2487K,2011PhRvD..83d3506P}, the total mass is closely related to 
the number of predicted subhalos \citep[e.g.][]{2018MNRAS.479.2853N,2019arXiv190702463F}. A ``light'' MW contains 
fewer subhalos of a given mass and thus can help to alleviate the problem. 

More recently, another problem, so-called ``too big 
to fail'',  was raised by Boylan-Kolchin et al. in 2011 \cite{2011MNRAS.415L..40B}. It concerns an 
apparent lack of MW satellite galaxies with central densities as high as those of the most massive dark 
matter subhalos predicted by $\Lambda$CDM simulations of MW-like hosts. Proper mechanisms are required 
to explain how the central density of the most massive subhalos in simulations can be 
reduced in order to match observations. However, the number of massive halos in simulations 
strongly depends on the total mass of the MW, and the problem disappears if the mass of our MW is 
smaller than $1\times10^{12}\msun$ \cite[e.g.][]{2012MNRAS.424.2715W,2014MNRAS.445.1820C}. 

The total mass and the underlying potential of our MW are also crucial for studies that focus on 
reconstructing the orbital evolution of individual objects. For example, it was discovered that MW 
satellite galaxies tend to be distributed in a highly inclined plane, and Pawlowski et al. \cite{2012MNRAS.423.1109P} 
reported the discovery of a vast polar structure (VPOS) of satellite galaxies, globular clusters and streams 
around the MW, indicating anisotropic spatial distribution and infall of these objects. If planes of satellite 
galaxies are ubiquitous across the Universe, it poses great challenges to the standard cosmological model 
\cite{2015MNRAS.449.2576C,2015MNRAS.452.3838C}. With more available proper motion data from \textit{Gaia} DR2, 
it has become possible to look into details of the reconstructed orbits for these objects and examine whether 
they indeed move in the same plane \citep[e.g.][]{2018A&A...616A..12G,2018A&A...619A.103F,2019arXiv190402719S}. 
Such a study inevitably requires a fiducial potential model for the orbital integration. Any uncertainty in the 
potential model can affect the orbit integration and hence bias our understandings. 


Knowing the MW mass is critical for predicting the future fate of our Galaxy, since having a more massive 
MW leads to a rapid merger of our Galaxy with its brightest satellite, the Large Magellanic Cloud (LMC) 
\cite{2019MNRAS.483.2185C}, and with its nearest giant neighbor, M31 \cite{2012ApJ...753....9V}.

Nowadays, we are in the large astronomical data survey era. Not only photometric and astrometric 
quantities, such as the magnitude, color, parallax (hence the Heliocentric distance) of each 
observed object can be measured, but also more and more objects with line-of-sight velocities, 
which approximately equal to the radial velocities for distant sources, have been collected through 
deep spectroscopic surveys, including the Radial Velocity Experiment \citep[RAVE;][]{2006AJ....132.1645S,
2017AJ....153...75K}, the LAMOST\footnote{Large Sky Area Multi-Object Fiber Spectroscopic Telescope} 
Experiment for Galactic Understanding and Exploration \citep[LEGUE;][]{2012RAA....12..735D,
2012RAA....12.1197C,2012RAA....12..723Z}, the Sloan Extension for Galactic Understanding and Exploration
\citep[SEGUE;][]{2009AJ....137.4377Y}, the Apache Point Observatory Galactic Evolution Experiment 
\citep[APOGEE;][]{2011AJ....142...72E}, the Galactic Archeology with HERMES survey \citep[GALAH][]
{2017MNRAS.465.3203M} and the \textit{Gaia}-ESO Survey \citep[GES;][]{2012Msngr.147...25G}. Moreover, 
with high precision astrometric instruments, proper motions of stars can be measured 
\citep[e.g.][]{2014ASPC..480...43V} by comparing imaging data taken at different epochs, after 
correcting for our own motion and controlling systematics \citep[e.g.][]{2017ApJS..232....4T}. More 
recently, with the launch of \textit{Gaia} \citep{2001A&A...369..339P}, a considerable amount of proper 
motion data are being collected. The mean proper motions of satellites and globular clusters based on 
their member stars have been refined and expanded \citep{2018A&A...616A..12G,2018ApJ...863...89S,
2018A&A...619A.103F,2018ApJ...867...19K,2018A&A...620A.155M,2019ApJ...875...77P,2019MNRAS.484.2832V,
2019MNRAS.482.5138B}.




It is thus a good time to revisit the existing methodologies of measuring the total mass of our MW, 
and think about how to improve the modeling by better controlling systematic uncertainties and 
observational errors. Thus in this review, we provide detailed descriptions of existing methods
measuring the total mass of our MW, the type of luminous objects which can be used as dynamical 
tracers of the underlying potential and modeling uncertainties. We hope to provide the reader better 
understandings towards these methods and broader views about how to make improvements in future studies. 
In addition, we hope our paper can help to summarize existing measurements for the mass of our MW 
in a clear and self-consistent way, and hence be useful for people who want to compare with these 
compiled measurements. 

Note, however, although the baryonic mass makes an important contribution to the total mass of 
the inner MW, in this review we focus on methods of modeling and measuring the total mass. Details 
such as how to measure the mass in the nuclear region of the MW, stellar mass of the bulge and
surface density in the local disk region through observations are beyond the scope of this 
review. The readers can check this information from the review paper of \cite{2016ARA&A..54..529B}. 



We start by introducing the method of measuring Galactic escape velocities using high velocity objects, 
in particular halo stars in the solar neighborhood (Sec.~\ref{sec:HVS}), and move on to 
introduce other local observables including terminal and circular velocities which can be used to measure 
the rotation curve for the inner MW (Sec.~\ref{sec:localobs}). Going further beyond the local observables, 
we introduce other methods including the spherical Jeans equation (Sec.~\ref{sec:jeans}) and the 
phase-space distribution function (Sec.~\ref{sec:DF}), which model more distant dynamical tracer objects 
including halo stars, globular clusters and satellite galaxies. We describe the dynamical modeling of tidal 
streams in Sec.~\ref{sec:stream}. Sec.~\ref{sec:timing} introduces the Local Group timing argument and the 
local Hubble flow approach by modeling mainly the radial motion of MW versus M31, and the motion of more 
distant satellite galaxies in the Local Group. The group of methods linking classical satellite galaxies in 
our MW to simulated subhalos is described in Sec.~\ref{sec:SatPheno}. Finally, we briefly mention a non-dynamical 
measurement in Sec.~\ref{sec:other}. We summarize these methods and discuss the role of modern numerical 
simulations in Sec.~\ref{sec:disc}.

The readers will see that almost all methods have to assume a realistic potential model at the first place. 
Methods from Sec.~\ref{sec:HVS} to Sec.~\ref{sec:stream} mainly stem on the framework of modeling the observed 
positions and velocities (or phase-space distribution) of tracers. Many of the measurements described in 
Sec.~\ref{sec:timing} and Sec.~\ref{sec:SatPheno} rely on calibrations made through modern numerical simulations 
of MW-like galaxies in a cosmological context. 

Throughout the paper, unless otherwise stated, we quote 
the enclosed mass within a given radius and the $M_{200}$ virial mass, with 1-$\sigma$ errors. Different 
virial mass definitions are converted to $M_{200}$ assuming the NFW model profile, and different percentage 
of confidence regions are converted to 1-$\sigma$ errors assuming the errors are Gaussian in linear space 
if the upper and lower errors have comparable sizes, or are Gaussian in log space if the upper and lower 
errors are more comparable in log space. 


\section{Galactic escape velocities: high and hypervelocity objects}
\label{sec:HVS}

The Galactic escape velocity is a fundamental quantity reflecting the depth of the underlying potential 
for our MW. It can be constrained using a variety of tracers, such as the high velocity stars in the 
tail of the velocity distribution for the population of halo stars, hypervelocity stars which are believed 
to be ejected from the Galactic center, and a few satellite galaxies moving with high velocities. In the 
following we introduce those fast moving objects and the approaches to model them. Measurements in 
this section fall in the categories of ``Escape V HVS'' and ``Leo I'' in Fig.~\ref{fig:massplot}.

\subsection{The high velocity tail distribution of halo stars}
Early attempts of measuring the Galactic escape velocity can be traced back to the 1920s and 1960s
\citep[e.g.][]{1928BAN.....4..269O,1926Obs....49..302O}. The measurements were based on modeling 
the observed high velocity stars with an analytical functional form describing the velocity distribution 
of these stars near the high velocity tail. The readers can find more details about the full phase-space 
distribution function of dynamical tracer objects within a given potential in Sec.~\ref{sec:DF}. Here 
we only briefly introduce the idea. The Jeans theorem states that the distribution of tracers in a dynamical 
system can be described by integrals of motion. The asymptotic form of the distribution function near the 
high velocity tail can be approximated as a power law
\begin{equation}
 f(E)\propto E^k,
\end{equation}
where the energy $E$ is defined through $E=-\Phi-v^2/2$, with $\Phi$ and $v^2/2$ being the potential and kinematic 
energy. $k$ describes the shape of the distribution at the high velocity end. The potential energy is defined such 
that $\Phi(r_{\rm max})=0$ at some maximum radius, $r_{\rm max}$, of the Galaxy, beyond which the star is considered 
to have ``escaped". Under such a definition, the escape velocity is simply given by
\begin{equation}
    \Phi(r)=-\frac{1}{2}v_\mathrm{esc}^2(r). \label{eqn:escapev}
\end{equation} 
Thus
\begin{equation}
 f(v|v_\mathrm{esc},k) \propto (v_\mathrm{esc}^2-v^2)^k\ \ \ \ \ \ \ \ \ \ (\mathrm{for} \ v<v_\mathrm{esc}), 
 \label{eqn:fhighvfull}
\end{equation}
where $v=|\textbf{v}|$.

In 1990, Leonard and Tremaine \cite{1990ApJ...353..486L} suggested that the term $(v_\mathrm{esc}+v)^k$ can be 
dropped and the velocity distribution of stars at the high velocity end can be modeled by the following formula
\begin{equation}
 f(v|v_\mathrm{esc},k) \propto (v_\mathrm{esc}-v)^k\ \ \ \ \ \ \ \ \ \ (v<v_\mathrm{esc}).
 \label{eqn:fhighv}
\end{equation}

Integrating Eqn.~\ref{eqn:fhighvfull} or Eqn.~\ref{eqn:fhighv} over tangential velocities, the radial velocity 
distribution is
\begin{equation}
 f(v_r|v_\mathrm{esc},k)=\int f(v|v_\mathrm{esc},k) \delta(v_r-\textbf{v}\cdot \textbf{n}) \mathrm{d}^3{v},
 \label{eqn:vtint}
\end{equation}
where $\textbf{n}$ is a unit vector along the line of sight. 

Basically, spectroscopic observations can be used to measure line-of-sight velocities with respect to the 
Sun. If we know the solar motion\footnote{Details about how to measure the solar motion are provided in 
Sec.~\ref{sec:localobs}}, Heliocentric distances and velocities can be used to obtain radial velocities 
with respect to the Galactic center. When proper motions are not available and hence tangential velocities 
are difficult to be robustly inferred\footnote{By fitting analytical velocity-space or phase-space distributions 
to observed line-of-sight velocities, it is still possible to model and infer the tangential velocity components 
(see more details in Sec.~\ref{sec:jeansbetadirect})}, Eqn.~\ref{eqn:fhighv} can be used to compare with the 
measured radial velocities of high velocity stars, and constrain the escape velocity, $v_\mathrm{esc}$, at the 
Galactocentric radius, $r$, of the star. Besides, the measurement errors of line-of-sight velocities were 
typically much smaller than the uncertainties of tangential velocities inferred from proper motions. 
Based on simulated data, Leonard and Tremaine in 1990  \cite{1990ApJ...353..486L} showed that estimates made 
using only radial velocities were as accurate as those made when employing proper motion data with large 
uncertainties.


Using Eqn.~\ref{eqn:fhighv}, the local escape velocity at the solar neighborhood was estimated to be in the range of 
450 to 650~km/s (90\% confidence level) by Leonard and Tremaine in 1990 \cite{1990ApJ...353..486L}. A subsequent work 
by Kochanek in 1996 \cite{1996ApJ...457..228K} adopted Eqn.~\ref{eqn:fhighvfull} and refined the escape velocity to be 
in the range of 489~km/s to 730~km/s (90\% confidence level). These early studies were limited by the small sample size 
of available high velocity stars. More recently, with continuously growing spectroscopic observations, the sample of 
high velocity stars with available radial velocity measurements had significantly increased. 

Based on high velocity stars selected from an early internal data release of the RAVE survey, Smith et al. in 2007 
\cite{2007MNRAS.379..755S} modeled their radial velocity distributions following Eqn.~\ref{eqn:fhighvfull}. 
Cosmological simulations of disk galaxies were used in their study to determine the limit for the parameter $k$. The 
local escape velocity was estimated to be $544_{-46}^{+64}$km/s (90\% confidence), and the mass within 50~kpc was 
found to be in the range of 3.6 to $4.0\times10^{11}\msun$. Adopting an adiabatically contracted 
NFW halo model, the virial mass of our MW was estimated to be $M_{200}=1.42_{-0.36}^{+0.49}\times 10^{12}\msun$. 

With increased sample of stars from the fourth data release (DR4) of RAVE, Piffl et al. in 2014 \cite{2014A&A...562A..91P} 
repeated the modeling using Eqn.~\ref{eqn:fhighv}, and used hydrodynamical simulations to validate the functional 
form, set a prior for the parameter $k$, and choose a minimum velocity cut for stars and the $r_\mathrm{max}$ 
escape radius. Because the increased sample of stars covered a broader distance range than those in previous studies, 
the position information of stars was also incorporated into their modeling. This was achieved by either scaling the 
escape velocity at different distances to the solar position through Eqn.~\ref{eqn:escapev}, or by analyzing stars at 
different distances separately. The local escape velocity was updated to be $533_{-41}^{+54}$km/s (90\% confidence). 
Assuming an NFW profile for the dark matter halo, the virial mass was estimated to be $M_{200}=1.60_{-0.25}^{+0.29}\times 10^{12}\msun$, 
which is in a good statistical agreement with the earlier study of Smith et al. in 2007 \cite{2007MNRAS.379..755S}.


The sample of stars used by Smith et al. in 2007 and Piffl et al. in 2014 was rather small. In 2017, Williams et al. 
\cite{2017MNRAS.468.2359W} selected intrinsically bright main sequence turn off (hereafter MSTO) stars, blue horizontal branch
(hereafter BHB) stars and K-giants with measured distances and line-of-sight velocities from SDSS/DR9, among which $\sim$2000 
stars are above their minimum velocity cut as high velocity halo stars. Their sample of high velocity stars spans $\sim$~40~kpc
in distance, from the solar vicinity to $\sim$50~kpc. \cite{2017MNRAS.468.2359W} considered in their Bayesian modeling the 
radial dependence of the escape velocity, the distance errors and possible contamination by outliers. The local escape 
velocity was constrained to be $521_{-30}^{+46}$~km/s, and the escape velocity drops to $379_{-28}^{+34}$~km/s at 50~kpc (94\% 
confidence region). The prior for the parameter $k$ was allowed to be flat over a much broader region given their larger sample 
of stars, which served to directly constrain the values of $k$ from data. $k$ does not seem to be a strong function 
of positions. For MSTO and K-giants, $k$ was approximately constrained to be $4\pm1$, while the value for BHBs was slightly 
favored to be higher. Given Eqn.~\ref{eqn:escapev} and $M(<r)=\frac{r^2}{G}\frac{\ud \Phi}{\ud r}$. The escape velocity measured 
by \cite{2017MNRAS.468.2359W} over 6 and 50~kpc can be converted to the enclosed mass or rotation velocity as a function of 
distance. The mass within 50~kpc was best constrained to be $2.98_{-0.52}^{+0.69}\times 10^{11}\msun$. 

The launch of \textit{Gaia} had led to a significant increase in the sample of high velocity stars within a few kpc from the Sun. 
Based on \textit{Gaia} DR2, Monari et al. in 2018 \cite{2018A&A...616L...9M} selected a sample of 2,850 counter-rotating halo stars, 
to be distinguished from stars in the MW disk. They measured the escape velocity curve between 5~kpc and 10.5~kpc, and the local 
escape velocity was updated to be $580\pm 63$km/s. Adopting an NFW profile plus a disk and a bulge component given by Irrgang 
et al. \cite{2013A&A...549A.137I}, the virial mass of our MW was estimated to be $M_{200}=1.28_{-0.50}^{+0.68}\times 10^{12}\msun$.

Very recently in 2019, Deason et al. \cite{2019MNRAS.485.3514D} selected $\sim$2,300 counter-rotating halo stars, out of which 
$\sim$240 have total velocities larger than 300~km/s, and are between Galactocentric distances of 4 and 12~kpc. Deason et al. 
\cite{2019MNRAS.485.3514D} adopted both analytical distributions and the \textsc{auriga} simulations \citep{2017MNRAS.467..179G} 
to investigate the dependence of the parameter $k$ on various properties, including the velocity anisotropies (see Sec.~\ref{sec:jeans} 
for more details about the definition of velocity anisotropy) and number density profiles of stars. The recent discovery of the 
``Gaia Sausage'' structure \citep[e.g.][]{2018MNRAS.478..611B} in our MW, which was due to the merger of a dwarf galaxy and shows that  
halo stars in the solar vicinity have strong radially biased velocity anisotropy, helps to set a prior of $1<k<2.5$. This is smaller than 
those adopted by Monari et al. \cite{2018A&A...616L...9M} and Piffl et al. \cite{2014A&A...562A..91P}. The escape velocity at solar 
radius was estimated by Deason et al. \cite{2019MNRAS.485.3514D} to be $528_{-25}^{+24}$~km/s. Assuming NFW profiles, the virial 
mass was best constrained to be $M_{200}=1.00_{-0.24}^{+0.31}\times 10^{12}\msun$. 

In a follow-up study by Grand et al. in 2019 \cite{2019MNRAS.487L..72G}, the effects of substructures were visited by applying 
the approach of Deason et al. \cite{2019MNRAS.485.3514D}, with slight modifications, to the set of \textsc{auriga} simulations
\citep{2017MNRAS.467..179G}. The recovered virial masses had a median falling $\sim$20\% below the true values, with a scatter 
of roughly a factor of 2. After correcting for the bias, the MW virial mass was revised as $M_{200}=1.29_{-0.47}^{+0.37}\times 
10^{12}\msun$, with extra systematic uncertainties to be kept in mind.

\subsection{Bound and unbound hypervelocity stars}

Unbound hypervelocity stars exceed the Galactic escape velocity and are usually believed to form through exotic mechanisms 
such as ejections by the super massive black hole in the Galactic center. Such hypervelocity stars have been detected in 
the outer stellar halo (see the review paper by Brown et al. in 2015 \cite{2015ARA&A..53...15B}). Due to the strategy 
and instruments used for detection, many previously detected hypervelocity stars are early-type stars. 

\begin{figure}[H]
    \centering
    \includegraphics[width=0.49\textwidth]{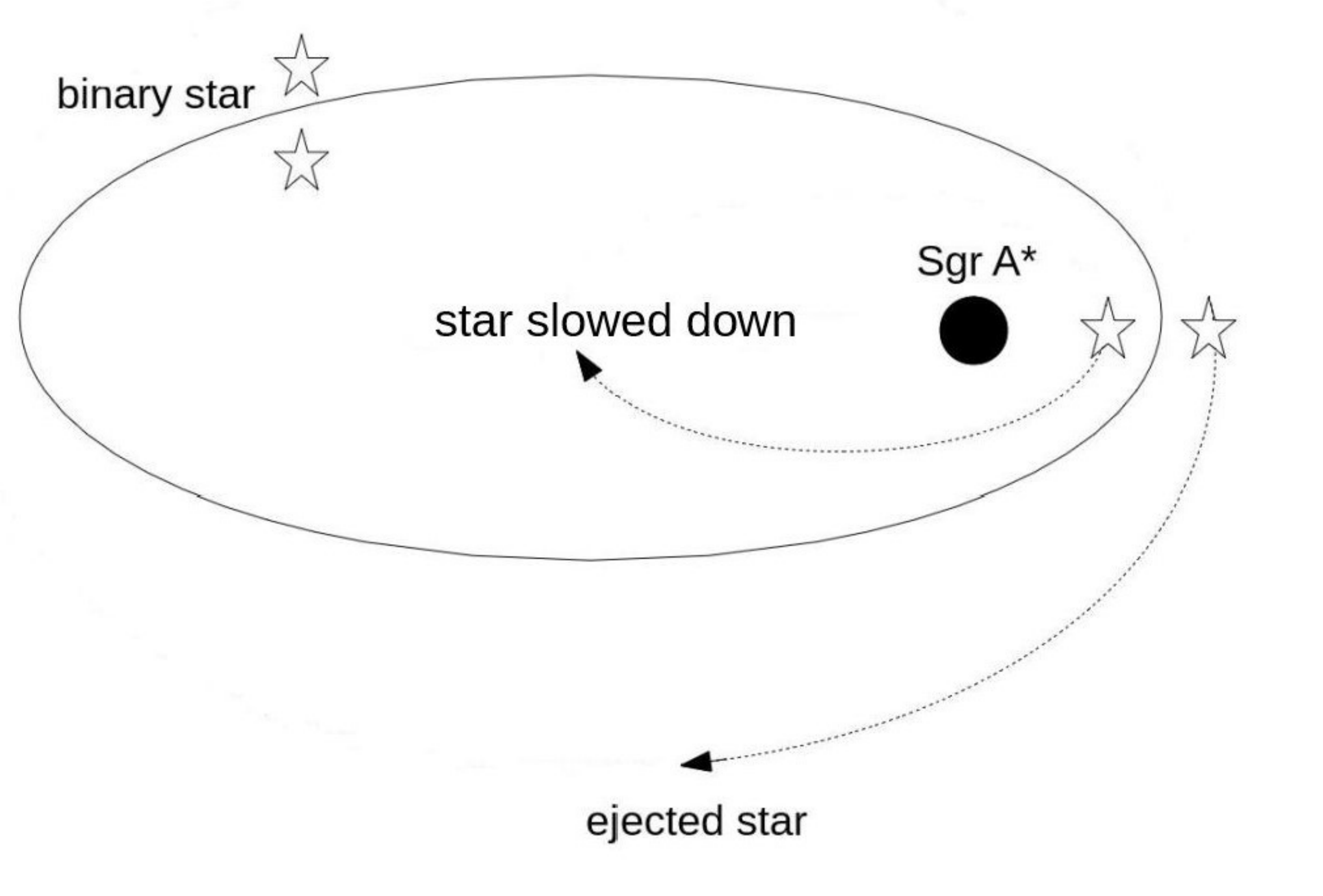}
    \vskip -.5cm
    \caption{Demonstration of the Hills mechanism. The binary star system is dissolved by the super massive 
    black in the Galactic center, Sgr A*. One member star of the binary system gets slowed down and stays close to 
    the Galactic center of our MW. According to energy conservation, the other star gains very high velocity and 
    gets ejected. This plot referenced Fig.~2 of \cite{2015ARA&A..53...15B}.}
    \label{fig:hills}
\end{figure}

It was found that the trajectories of these early-type hypervelocity stars are consistent with coming from the Galactic 
center. Thus they are very likely formed through the gravitational interaction with the super massive black hole (Sgr A*) 
in the Galactic center and as a result gained such high velocities \citep[e.g.][]{2003ApJ...599.1129Y,2006MNRAS.368..221G,
2007MNRAS.376..492G,2008MNRAS.383...86O,2015MNRAS.454.2677C,2016MNRAS.458.2596F}. More specifically, observations are 
consistent with a scenario that each of the hypervelocity stars originally belonged to a binary star system, and the system 
was tidally dissociated by Sgr A*. The process is called the ``Hills'' mechanism \citep{1988Natur.331..687H}. This 
mechanism is demonstrated in Fig.~\ref{fig:hills}. One member of the binary system got ejected, while the other stayed in 
the Galactic center. In particular, this picture is consistent with the observational fact that within $r\sim0.5$~kpc to 
the Galactic center, young stars have been observed \citep[e.g.][]{2006ApJ...643.1011P,2014MNRAS.444.1205J,2018ASSL..424...69L,
2018MNRAS.478L.127N}, which otherwise might challenge our knowledge of star formation because molecular clouds are difficult 
to survive strong tidal forces in the Galactic center.

A similar sample of bound hypervelocity stars, which are likely stars ejected from the Galactic center through the same 
mechanism \citep{2006ApJ...653.1194B}, but whose velocities are still below the escape velocity, have been observed as 
well \citep{2007ApJ...660..311B,2007ApJ...671.1708B,2014ApJ...787...89B}. 

There are alternative models for the formation of hypervelocity stars. 
For example, they could be ejected by the Large Magellanic Cloud or be runaways from the MW disk \citep{2016ApJ...825L...6B, 
2017MNRAS.469.2151B,2019MNRAS.483.2007E}. Despite these debates, if the early-type hypervelocity stars indeed form through 
the ejection by Sgr A*, they not only contain information about processes happened in the Galactic center, but can also be 
used to constrain the shape \citep[e.g.][]{2005ApJ...634..344G,2007MNRAS.379.1293Y} and depth \citep[e.g.][]{2009ApJ...697.2096P,
2017NewA...55...32F,2017MNRAS.467.1844R,2007MNRAS.379L..45S} of the potential. 

Assuming the Sgr A* ejection scenario is true, Rossi et al. \cite{2017MNRAS.467.1844R} modeled the velocity distribution of 
observed hypervelocity stars in their 2017 study, following the model in a series of earlier papers \citep{2012ApJ...748..105K,
2014ApJ...795..125R}. 

If there is only the black hole potential, the velocity for the ejected star at infinity is given by 
\begin{equation}
    v_\mathrm{eject}=\sqrt{\frac{2Gm_2}{a}}\left(\frac{M}{m_t}\right)^{1/6}, 
    \label{eqn:veject}
\end{equation}
where $M$ is the central black hole mass, $m_t$ is the total mass of the binary, $a$ is the binary separation and $m_2$ 
is the mass of the companion star in the binary system \citep{2010ApJ...708..605S}. 

Rossi et al. \cite{2017MNRAS.467.1844R} modeled the distribution of binary separations and mass ratios as power-law forms, 
which then predicted the distribution of ejecting velocities through Eqn.~\ref{eqn:veject}. After ejection, the change in velocity 
of each star can be either calculated by assuming some escape velocity out to 50~kpc or be calculated by adopting some
model potential including components of the Galactic disk, bulge and dark matter halo. Assuming the ejection rate and 
life time of stars, the predicted velocity distribution of these stars can be compared with the true velocities of observed 
hypervelocity stars, and thus constrain the adopted escape velocity or potential models. Their analysis favored halos with 
escape velocity from the Galactic center to 50~kpc smaller than 850~km/s, which then favored $\Lambda$CDM halos with 
$M_{200}$ in the range of 0.5 and $1.5\times10^{12}\msun$. 

Perets et al. \cite{2009ApJ...697.2096P} proposed an independent method of using the asymmetric distribution for ingoing and 
outgoing hypervelocity stars to constrain the MW potential in 2009. Ejected stars exceeding the escape velocity are unbound 
and will leave our MW, whereas bound ejected stars will reach the apocenter and turn back. Thus bound stars can contribute to 
both ingoing stars with negative velocities \citep{2009ApJ...706..925B,2008ApJ...680..312K,2015ARA&A..53...15B} and outgoing 
stars with positive velocities. Unbound stars only contribute to the outgoing population, which introduces an asymmetry in 
the high velocity tail of the velocity distribution. Indeed, such asymmetry has been observed, with a significant excess of stars 
traveling with radial velocities larger than 275~km/s \citep{2007ApJ...660..311B}. 

The asymmetry is also related to the lifetime of such ejected stars. Some stars might have evolved to a different stage before 
reaching to a large enough Galactocentric distance and turning back. Note they do not totally disappear, but may have evolved 
out of the detection range of corresponding instruments. For example, the MMT (Multiple Mirror Telescope) hypervelocity star 
survey \citep{2009ApJ...690.1639B,2014ApJ...787...89B} mostly target the main sequence B stars. Fragione and Loeb in 2017
\cite{2017NewA...55...32F} modeled the observed asymmetry by varying both the potential model and the life or travel time 
of hypervelocity stars. If fixing the travel time of hypervelocity stars to 330~Myr for typical B stars, the MW virial mass 
$M_{200}$ was found to be in the range of 1.2 to $1.9\times 10^{12}\msun$. 

More recently, with \textit{Gaia} DR2, previously identified hypervelocity stars with only line-of-sight velocities were revisited 
and extended to have full 6-dimensional phase-space information based on \textit{Gaia} proper motions \citep[e.g.][]{2018A&A...620A..48I,
2018MNRAS.479.2789B,2018ApJ...866...39B}. More hypervelocity star candidates, especially late-type stars \footnote{Stars whose 
spectral types are F, G, K or M, including both dwarf and giant stars.}, have been reported and predicted \citep[e.g.][]{2018MNRAS.tmp.2466M,
2018MNRAS.476.4697M,2018ApJ...866..121H}. \textit{Gaia} proper motions enabled further and more robust investigations on the origin 
for hypervelocity star candidates. 

While those previously discovered early-type hypervelocity stars are very likely from the Galactic center, the origin of late-type 
hypervelocity stars is not clear. Based on proper motions and radial velocities, Boubert et al. in 2018 \cite{2018MNRAS.479.2789B} 
concluded that in fact almost all previously-known late-type hypervelocity stars are very likely bound to our Milky Way. A similar 
conclusion was　reached by Hawkins and Wyse in 2018 \cite{2018MNRAS.481.1028H} based on chemical abundance patterns, that a few 
candidate hypervelocity stars are most likely bound high velocity halo stars, which are close to the high velocity tail of the 
distribution, but are unlikely hypervelocity stars ejected from the Galactic center or from the Large Magellanic Cloud. 

Hattori et al. in 2018 \cite{2018ApJ...866..121H} discovered 30 stars with extreme velocities ($\geq $480~km/s) from 
\textit{Gaia} DR2. Tracing their orbits, they reported that at least one of the stars is consistent with having been ejected 
from the Galactic center. Unlike previous observations of early-type hypervelocity stars, these stars are quite old, with 
chemical properties similar to the Galactic halo. Assuming these stars are bound, the virial mass of our MW should be 
higher than $1.4\times 10^{12} \msun$.

\subsection{Bound and unbound satellite galaxies}
\label{sec:boundunboundsat}

Other types of fast moving objects such as dwarf satellite galaxies can be used to constrain the mass of our MW 
as well. Among the MW satellite galaxies, Leo I plays an important role since it has a large Galactocentric distance 
and a high velocity \citep[e.g.][]{2013ApJ...768..139S}, which could suggest that Leo I is only weakly bound (if at 
all) to the MW. Incorporating Leo I into the analysis has to rely on the assumption that Leo I is bound to our MW. 
As a result, a heavy MW is often required to keep Leo I bound given its large distance and high velocity (see more 
details in Sec.~\ref{sec:jeanestimator}, Sec.~\ref{sec:DFEL} and Sec.~\ref{sec:timing}). Based on subhalos in MW-like 
halos from the Aquarius simulation \citep{2008MNRAS.391.1685S}, Boylan-Kolchin et al. \cite{2013ApJ...768..140B} 
demonstrated in 2013 that Leo I is very unlikely to be unbound, because 99.9\% subhalos in their simulations are 
bound to their host halos. To keep Leo I bound, Boylan-Kolchin et al. \cite{2013ApJ...768..140B} estimated the virial 
mass of our MW to be $M_{200}=1.34_{-0.31}^{+0.41}\times 10^{12}\msun$. 

  \begin{figure}[H]
\centering
\includegraphics[width=0.4\textwidth]{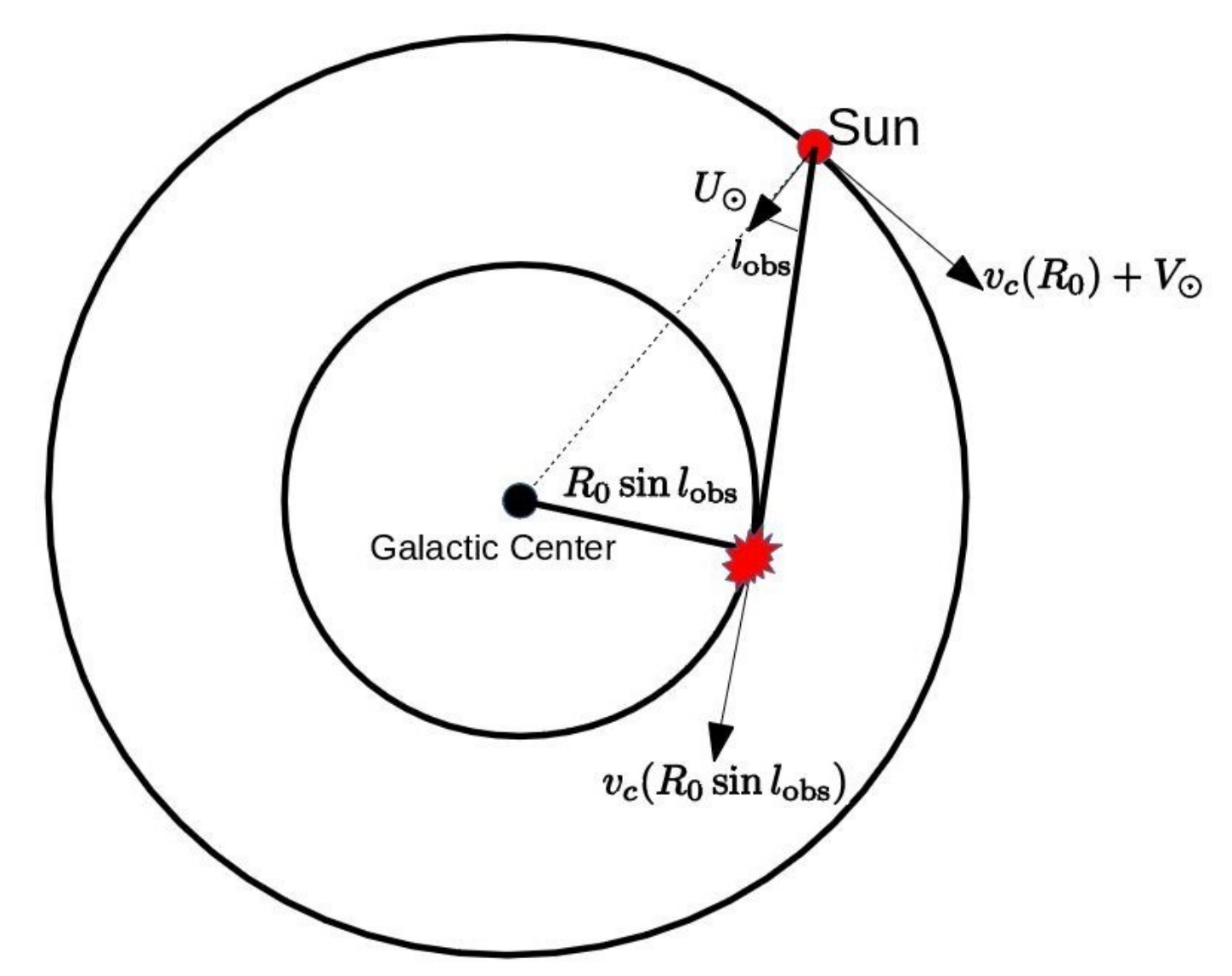}\\
\includegraphics[width=0.4\textwidth]{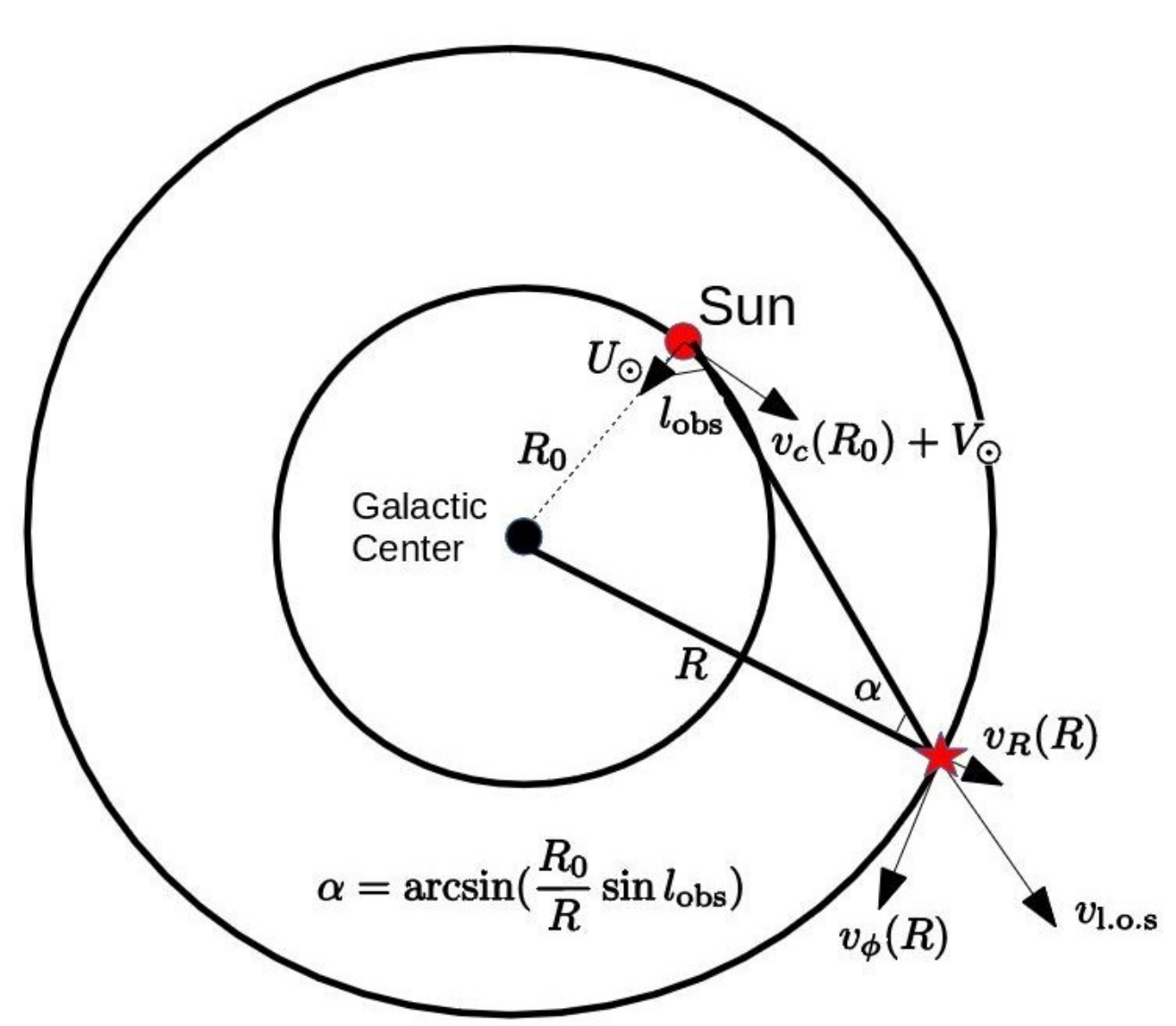}
    \caption{{\bf Top:} Plot showing the concept of terminal velocity, for a gas cloud inside the Galactic disk and 
    within the solar radius. The gas is moving along a circular orbit, and the maximum velocity which can be observed 
    along that circular orbit happens at the tangent point with Galactic longitude of $l_\mathrm{obs}$. $R_0$ is the 
    Galactocentric distance of our Sun, and $R_0\sin l_\mathrm{obs}$ is the Galactocentric distance for the gas cloud. 
    $v_c(R_0\sin l_\mathrm{obs})$ is the circular velocity at the radius of the gas cloud. The terminal velocity is 
    $v_c(R_0\sin l_\mathrm{obs})$ minus the corresponding velocity components of the rotation velocity for the Local 
    Standard of Rest ($v_c(R_0)$) and the peculiar solar motion ($U_\odot$ and $V_\odot$) with respect to the Local 
    Standard of Rest, projected along the line of sight. Both $U_\odot$ and $V_\odot$ are in fact much smaller than $v_c(R_0)$. 
    {\bf Bottom:} Plot showing the concept of the line-of-sight velocity for a star within the Galactic disk and outside 
    the solar radius. The star is assumed to be observed at Galactic longitude of $l_\mathrm{obs}$. $R$ and $R_0$ are 
    the Galactocentric distance of the star and our Sun. $v_R(R)$ and $v_\phi(R)$ are the radial and tangential velocities 
    of the star with respect to the Galactic center. $U_\odot$ and $V_\odot$ reflect the peculiar motion of our Sun, and 
    $v_c(R_0)$ is the circular velocity of the Local Standard of Rest. They are the same as those defined in the top plot. 
    Both $U_\odot$ and $V_\odot$ are in fact much smaller than $v_c(R_0)$. $v_R(R)$ is much smaller than $v_\phi(R)$.
    The line-of-sight velocity of the star with respect to us is the velocity difference between the star and our Sun 
    projected along the line-of-sight direction.}
\label{fig:vtermivcirc}
\end{figure}

Using \textit{Gaia} DR2 proper motion data, member stars of a few MW satellite galaxies can be more robustly identified, 
which then provide the averaged proper motions of these satellite galaxies. Fritz et al. in 2018 \cite{2018A&A...619A.103F} 
derived proper motions for 39 companion galaxies of our MW out to 420 kpc. Based on arguments of keeping acceptable 
distributions of orbital apocenters and having a reasonable fraction of bound satellites, they reported that a heavy 
MW (~$1.6\times10^{12}\msun$) is more preferable than a light MW (~$0.8\times10^{12}\msun$).

\section{Rotation velocities of the inner MW: local observables}
\label{sec:localobs}

The Galactic rotation curve (circular or rotation velocity as a function of radius) can directly reflect the mass 
enclosed within different radii. In this section, we introduce local observables and corresponding methods 
of measuring the rotation velocities of the inner MW. Measurements in this section fall in the category of 
``LocalObs rot V'' in Fig.~\ref{fig:massplot}.

To measure the shape of the rotation curve within the solar orbit, previous studies have typically employed the 
terminal velocities of the interstellar medium (ISM) or HI gas clouds \citep[e.g.][]{1954BAN....12..117V,
1979AJ.....84.1181G,2008ApJ...679.1288L,2008ASSP....4..178S}. The basic idea relies on the fact that for circular 
orbits in an axis-symmetric potential and within the solar orbital radius, the observed peak velocity of ISM along 
any line of sight in the Galactic disk plane corresponds to the gas at the tangent point. The approximation of 
circular orbits is reasonable given the fact that the inner region of our MW is dominated by the disk component. 
In other words, the terminal velocity tells that there is a particular direction, along which the rotation velocity 
of circular orbits entirely contributes to the line-of-sight velocity. This is demonstrated in the top plot of 
Fig.~\ref{fig:vtermivcirc}.

We use $R_0$ to represent the Galactocentric distance of our Sun, and we assume the peak velocity is observed 
at Galactic latitude of $b=0$ (in the Galactic disk plane) and Galactic longitude of $l=l_\mathrm{obs}$.
The Galactocentric distance of the observed IGM is $R=R_0 \sin l_\mathrm{obs}$, and the terminal velocity 
at $R$ \citep{2017MNRAS.465...76M} is
\begin{align}
 v_\mathrm{terminal}(R_0 \sin l_\mathrm{obs}) & = v_c(R_0 \sin l_\mathrm{obs})\nonumber \\
 & - \left(v_c(R_0)+V_\odot\right) \sin l_\mathrm{obs} - U_\odot \mathrm{cos}l_\mathrm{obs}. 
\label{eqn:termiv}
 \end{align}

$v_c(R_0 \sin l_\mathrm{obs})$ and $v_c(R_0)$ are the rotation velocities at $R=R_0 \sin l_\mathrm{obs}$ for 
the observed IGM and at $R_0$ for our Sun, respectively. The second term refers to the rotation of Local 
Standard of Rest (hereafter LSR), $v_c(R_0)$, and the motion of our Sun with respect to the LSR in the 
direction of Galactic rotation, $V_\odot$. Note the LSR follows the mean motion of material in the neighborhood 
of the Sun, which is often assumed to be circular, and the Sun has a small peculiar motion relative to the LSR. 
$U_\odot$ is the velocity towards Galactic center. The solar motion with respect to the Galactic center 
is a combination of the velocity of the LSR and the peculiar motion of the Sun with respect to the LSR in the 
same direction 
\begin{equation}
 \textbf{V}_\odot= \left(U_\odot,v_c(R_0)+V_\odot,W_\odot \right),
\end{equation}
where $W_\odot$ is the velocity component of the solar peculiar motion perpendicular to the Galactic disk.
Note the velocity components for solar peculiar motion, $U_\odot$, $V_\odot$ and $W_\odot$ are all much smaller 
than the rotation velocity of the LSR, $v_c(R_0)$. 

Assuming the peculiar motions of the Sun, $U_\odot$, $V_\odot$ and $W_\odot$, are well determined, which we will 
discuss later, terms of $U_\odot$ and $V_\odot$ can be moved to the left side as known quantities. The right hand 
side of Eqn.~\ref{eqn:termiv} becomes $v_c(R_0 \sin l_\mathrm{obs}) - v_c(R_0) \sin l_\mathrm{obs}$, which can be 
reduced to $\frac{v_c(R_0 \sin l_\mathrm{obs})}{R_0 \sin l_\mathrm{obs}}-\frac{v_c(R_0)} {R_0}$ after divided by 
$R_0 \sin l_\mathrm{obs}$ \citep[e.g.][]{1998MNRAS.294..429D}. Hence given the observed terminal velocities, plus 
the Galactocentric distance of our Sun and the solar motion, the shape of the rotation curve can be determined. 

The normalization of the rotation curve can be determined, by measuring the absolute rotation velocity for 
the LSR, $v_c(R_0)$, for example. We will discuss later in this section about how to measure $v_c(R_0)$. 

Terminal velocities are usually adopted to measure the shape of the rotation curve within the orbit of our Sun. For 
regions slightly outside the Sun's Galactocentric radius but still on the Galactic disk, measurements of the rotation 
velocities can be made by modeling observed distances and line-of-sight velocities of maser sources and disk stars 
\citep[e.g.][]{2008AJ....136..118F}. In particular, astrophysical maser sources are associated with high-mass star 
forming regions, which are expected to be on nearly circular orbits in the Galactic disk. Because the emission of 
masers is a narrow spectral line, the Heliocentric parallaxes, proper motions and line-of-sight velocities of maser 
sources can be very well measured based on radio interferometry.

This is demonstrated in the bottom plot of Fig.~\ref{fig:vtermivcirc}.
The observed line-of-sight velocity, $v_\mathrm{l.o.s}$, of a maser source or disk star outside the solar radius at 
Galactic latitude $b=0$, Galactic longitude $l=l_\mathrm{obs}$ and Galactocentric distance $R$ is given by

\begin{align}
 v_\mathrm{l.o.s}=v_\phi(R)\sin (\arcsin(\frac{R_0}{R}\sin l_\mathrm{obs}) )\nonumber \\
 -(v_c(R_0)+V_\odot)\sin (l_\mathrm{obs})\nonumber \\
 +v_R(R) \cos(\arcsin(\frac{R_0}{R}\sin l_\mathrm{obs}) )\nonumber \\
 -U_\odot \cos (l_\mathrm{obs}),
 \label{eqn:vcirc}
\end{align}
where $U_\odot$, $V_\odot$ and $R_0$ are still the peculiar solar motions towards Galactic center and in the direction 
of Galactic rotation, and the Galactocentric distance of our Sun. $v_c(R_0)$ is the rotation velocity of the LSR. 
$v_\phi(R)=v_c(R)-v_a(R)$ is the tangential velocity component of the maser source or disk star at $R$. $v_c(R)$ is 
the rotation velocity at $R$, and $v_a(R)$ is introduced to describe the asymmetric drift by Binney and Tremaine (2008). 
$v_R(R)$ is the radial motion of the observed maser source or star, which is much smaller than $v_c(R)$ or $v_\phi(R)$.

Similarly, once we know the solar motion and its Galactocentric distance, the rotation velocity of the LSR and the 
asymmetric drift term, we can constrain the rotation velocity $v_c(R)$ for the source observed at $R$ through 
Eqn.~\ref{eqn:vcirc}. The radial motion, $v_R(R)$, can be treated as a free parameter or averaged to zero over a 
sample of sources. 

We now briefly introduce how to measure the rotation velocity of the LSR and the peculiar motion of our Sun. These 
can be inferred through the apparent motion of Sgr A* in the Galactic Center \citep[e.g.][]{2004ApJ...616..872R,
2009ApJ...692.1075G,2012MNRAS.427..274S}. If Sgr A* is at rest in the Galactic frame, the apparent motion of Sgr A* 
reflects the absolute motion of our Sun with respect to the Galactic center. The peculiar motion of our Sun can then 
be decoupled from the rotation velocity of the LSR through the observation of kinematics from nearby stars. In addition, 
the accurate Heliocentric distances and line-of-sight velocities of maser sources can be used to jointly model and 
constrain the rotation velocities of masers themselves with respect to the Galactic center, the rotation velocity of 
the LSR, the Galactocentric distance and the peculiar motion of our Sun \citep[e.g.][]{2009ApJ...700..137R,2009ApJ...704.1704B,
2010MNRAS.402..934M,2011AN....332..461B,2012PASJ...64..136H}.

While the terminal velocities within the solar radius and the line-of-sight velocities of sources slightly outside 
the solar radius but within the Galactic disk are traditional observables to constrain the rotation velocities for the 
inner MW, it is necessary to mention that in 2012, Bovy et al. \cite{2012ApJ...759..131B} was probably the first to use 
hot stellar tracers out of the MW disk to measure the MW rotation curve between 4~kpc and 14~kpc, based on the spherical 
Jeans equation and phase-space distribution functions. More recently in 2018, with spectroscopic data from APOGEE and 
photometric data from WISE, 2MASS and \textit{Gaia} to get precise parallaxes and hence full six-dimensional phase-space 
coordinates, Eilers et al. \cite{2018arXiv181009466E} measured the rotation velocity curve based on the Jeans equation 
with an axisymmetric potential from 5~kpc to 25~kpc to the Galactic center. We briefly mention their efforts here, and 
postpone discussions about the (spherical) Jeans equation and the distribution function to Sec.~\ref{sec:jeans} and 
Sec.~\ref{sec:DF}. The readers can also check Sec.~\ref{sec:stream} about constraining the local rotation velocity using 
the GD-1 stream \citep{2010ApJ...712..260K,2019MNRAS.486.2995M}.

Started from the last century, numerous efforts had been spent to constrain potential models for the Galactic disk, 
bulge and the dark matter halo using the measured circular velocities of the inner MW. These studies were often combined 
with a few other observables for the inner MW (typically based on stellar dynamics or star counting) in the following.

\begin{description}
 \item[$\bullet$] The local vertical force some distance above the Galactic disk or the total surface density within a 
 cylinder crossing the disk \citep[e.g.][]{1991ApJ...367L...9K,2013ApJ...779..115B,2013ApJ...772..108Z,2004MNRAS.352..440H},
 measured with the observed distances and radial velocities of stars in the Galactic pole and the vertical Jeans equation.
 \item[$\bullet$] Total local volume density \citep[e.g.][]{2000MNRAS.313..209H}, measured with stars in the solar vicinity
 \item[$\bullet$] Local surface density of visible matter in the disk \citep[e.g.][]{1989MNRAS.239..605K,2004MNRAS.352..440H}
 \item[$\bullet$] The velocity dispersion in Baade's window\footnote{A sky area with relatively low amounts of interstellar dust 
 along the line of sight, which is an observational window to the obscured Galactic Center of the MW.} to the Galactic center 
 \citep[e.g.][]{1988AJ.....95..828R}.
 \item[$\bullet$] The mass in the very central parsec regions.
\end{description}

In addition, as the readers will see, in order to constrain the mass of our MW out to large distances, the above local observables 
are not enough, and measurements made by other alternative methods based on more distant tracer objects should be adopted.

Early attempts of this kind can be traced back to 1998 \citep[e.g.][]{1998MNRAS.294..429D}, when Dehnen and Binney jointly 
modeled the observed terminal velocities, distances and line-of-sight velocities of maser sources, local vertical forces,
surface densities and the observed velocity dispersion of the bulge in Baade’s window. Combined with other contemporary 
measurements of the enclosed mass within 100~kpc to the Galactic center by Kochanek \citep{1996ApJ...457..228K}, which was 
mainly based on phase-space distribution functions (see Sec.~\ref{sec:DF} for more details), the rotation curves out to 
100~kpc were obtained for different models. The total mass within 100~kpc was constrained to be in the range of $3.41\times
10^{11}\msun$ and $6.95\times 10^{11}\msun$. 

In 2002, Klypin et al. \cite{2002ApJ...573..597K} presented a set of gravitational potential models for our MW, based on standard 
disk formation theory and adiabatic compression of baryons within cuspy dark matter halos. Models with and without the exchange 
of angular momentum between baryons and dark matter were both considered. The models with a range of different parameters were 
tested against the terminal velocities, the circular velocities slightly outside the solar radius, the local surface density of 
gas and stars, the vertical force at 1.1~kpc above the Galactic disk and the mass in the very central parsec regions of our MW. 
Klypin et al. \cite{2002ApJ...573..597K} also included the enclosed mass within 100~kpc to the Galactic center from other studies 
\citep{1996ApJ...457..228K,1998MNRAS.294..429D}, measured with distribution functions, and found that their modeling preferred our 
MW to have virial mass of about $M_{200}=0.86\times 10^{12}\msun$, though their analysis was not based on strict statistical 
inferences.


Weber and Boer in 2010 \cite{2010A&A...509A..25W} constrained the local dark matter density. They made best fit to observed
data including the total mass within the solar orbital radius, the total density and the surface density of visible matter at the 
solar position, the local vertical force, the shape of the rotation curve within the Galactic disk \citep{2008ASSP....4..178S}, and the 
constrained mass of our MW within given radii from other studies \citep{1999MNRAS.310..645W,2005MNRAS.364..433B,2008ApJ...684.1143X}.
Given the correlations among local dark matter density, the scale length of the dark matter halo and the Galactic disk, and since 
the scale lengths were poorly constrained, their best-fit local dark matter density varied from 0.005 to 0.01 $\msun \mathrm{pc}^{-3}$, 
which allowed the total mass of our MW up to $2\times 10^{12}\msun$. 

More recently in 2011, McMillan \cite{2011MNRAS.414.2446M} jointly modeled the observed terminal velocities, 
maser sources and the local vertical force using their disk, bulge plus dark halo models. However, for the mass enclosed 
within even larger distances, McMillan still had to refer to other contemporary measurements based on the distribution 
function \citep{1999MNRAS.310..645W}. Their best-fit rotation curve extended to $\sim$100~kpc. The MW virial mass was 
measured to be $M_{200}=1.26\pm 0.24 \times 10^{12} \msun$. Later on, with more available maser observations, the 
measured virial mass was updated to be $M_{200}=1.3\pm 0.3 \times 10^{12} \msun$ in a follow-up paper 
\citep{2017MNRAS.465...76M}.


In 2013, Irrgang et al. \cite{2013A&A...549A.137I} adopted three different model potentials with disk, 
bulge and dark matter halo to constrain the mass of our MW within 50, 100 and 200~kpc. They jointly modeled 
the solar motion, the terminal velocities, the observations of maser sources, the local total mass density and the local 
surface mass density of the Galactic disk. The velocity dispersion in the bulge was used to constrain the inner most 
region, and an hypervelocity halo BHB star was assumed to be bound to our MW and hence put further constraints on the 
potential out to 200~kpc. The mass enclosed within 200~kpc was constrained to be $1.9_{-0.8}^{+2.4}\times 10^{12}\msun$, 
$1.2_{-0.2}^{+0.1}\times 10^{12}\msun$ and $3.0_{-1.1}^{+1.2}\times 10^{12}\msun$ (90\% confidence) for the three potential 
models adopted in their analysis, respectively. 

Nesti and Salucci in 2013 \cite{2013JCAP...07..016N} included in their analysis the observed velocity dispersion of halo 
stars out to 80~kpc, and used the spherical Jeans equation (see Sec.~\ref{sec:jeans} for details) to obtain the rotation 
velocities out to such distances. They adopted both the Burkert (core) and NFW (cusp) profiles for the modeling, and the best 
constrained masses within 50~kpc were $4.5^{+3.5}_{-2.0}\times 10^{11}\msun$ and $4.8^{+2.0}_{-1.5}\times 10^{11}\msun$ 
for the Burkert and NFW model profiles, respectively. The masses within 100~kpc were $6.7^{+6.7}_{-3.3}\times 10^{11}\msun$ 
and $8.1^{+6.0}_{-3.2} \times 10^{11}\msun$, respectively. The virial masses of the best-fit Burkert and NFW profiles were 
extrapolated to be $1.11^{+1.6}_{-0.61}\times 10^{12}\msun$ and $1.53^{+2.3}_{-0.77}\times 10^{12}\msun$\footnote{The 
errors are 95.45\% confidence level (2-$\sigma$). Through private communications with the authors, the main driving uncertainty 
in the errors was the velocity anisotropy $\beta$, for which no prior can be known. So the errors cannot be simply converted 
to 1-$\sigma$ uncertainties assuming Gaussian errors. The associated error in Fig.~\ref{fig:massplot} is simply the 2-$\sigma$ 
errors decreased by 10\%.}.

The \textit{galpy} software, which is a python package for galactic-dynamics calculations, was developed by Bovy in 2015 
\citep{2015ApJS..216...29B}. It incorporated an example potential model with disk, bulge and halo components. The model 
potential was based on fits to local observables including the terminal velocities, the velocity dispersion through the Baade's 
window, the local vertical force, the local visible surface density and the local total density, in combination with the rotation 
curve measured by Bovy et al. in 2012 \cite{2012ApJ...759..131B} at the solar neighborhood (see above) and the total mass within 
60~kpc to the Galactic center of \cite{2008ApJ...684.1143X} through the spherical Jeans equation (see 
Sec.~\ref{sec:jeans} for details). Their model potential preferred a virial mass of about $M_{200}=0.7\times 10^{12}\msun$.


In two subsequent papers, Bajkova and Bobylev in 2016 \cite{2016AstL...42..567B,2017AstL...43..241B} used the spherical Jeans 
equation to fit a bulge and a disk together with a few different halo models to the line-of-sight velocities of hydrogen clouds 
at the tangent points, kinematic and parallax data of 130 maser sources within 25~kpc, as well as more distant rotation velocity 
measurements by \cite{2014ApJ...785...63B}. If adopting the NFW model profile for the halo, the mass within 200~kpc was constrained 
to be $7.5\pm1.9\times 10^{11}\msun$.

Recently, Cautun et al. \citep{Cautun2019_MW_model} in 2019 have combined the stellar rotation curve measured by {\it Gaia} \cite{2018arXiv181009466E} 
with the outer mass measurements from satellite dynamics \cite{2019MNRAS.484.5453C} to constrain both the stellar and the dark matter 
distribution of the MW. They have used a contracted dark matter halo model with free mass and concentration, and stellar bulge and disk 
components with several free parameters. Their best-fit model corresponds to a total MW mass, $M_{200}=1.12_{-0.22}^{+0.20}\times
10^{12}~\msun$, and a dark matter halo concentration (before baryonic contraction), $c=8.2_{-1.5}^{+1.7}$, which is typical of a 
$10^{12}~\msun$ halo. Furthermore, Cautun et al. \citep{Cautun2019_MW_model} have shown that the same data is equally well fit by an 
NFW halo profile, but with a 20 percent lower halo mass, much higher concentration and a 20 percent higher stellar mass. It illustrates 
that the rotation curve for distances below $20~\mathrm{kpc}$ cannot break the degeneracy between the halo and the stellar mass profiles, 
and thus, because the MW baryonic profile is still poorly understood, the inferred halo mass depends on the baryonic model employed in a 
given study. 

Combining observations of rotation velocities for the inner MW compiled by \cite{2017SoftX...6...54P,2016MNRAS.463.2623H} and the 
rotation velocities up to 100~kpc obtained through the Spherical Jeans Equation by \cite{2016MNRAS.463.2623H}, Karukes et al. in 2019 
constrained the virial mass of our MW to be $M_{200}=0.89_{-0.08}^{+0.10}\times 10^{12}\msun$. 

Almost all the above studies had to rely on observations of more distant luminous objects and other alternative methods 
to infer the mass distribution out to larger distances, such as the spherical Jeans equation and the distribution function. We 
now move on to introduce how the mass of our MW can be constrained through the spherical Jeans equation, through the phase-space 
distribution function of dynamical tracer objects and through dynamical modeling of tidal streams in the following three sections 
(Sec.~\ref{sec:jeans}, Sec.~\ref{sec:DF} and Sec.~\ref{sec:stream}).

\section{The rotation velocity out to large distances from the Jeans Equation: halo stars, globular clusters and satellite galaxies}
\label{sec:jeans}

In the previous section, we introduced how the rotation curve of the inner MW can be inferred through the terminal 
and circular velocities. To obtain the rotation curve out to large distances, the spherical Jeans Equation 
has been frequently used. In the following, we start by introducing the spherical Jeans Equation and then move on 
to describe relevant studies in literature. Measurements in this section fall in the category of ``SJE'' in 
Fig.~\ref{fig:massplot}.

\subsection{The spherical Jeans Equation}

The dynamical structure of a system can be fully specified by its phase-space distribution function, or the number density of objects in 
phase space, $f(\textbf{x}, \textbf{v}, t)\equiv {\ud^3 N}/{\ud^3x\ud^3v}$. In absence of collision, the phase-space density is conserved 
along the orbits of the particles, i.e., ${\ud f}/{\ud t}=0$, leading to the so-called collisionless Boltzmann equation \begin{equation}
    \frac{\partial f}{\partial t}+\frac{\ud \textbf{v}}{\ud t}\cdot\nabla_{\textbf{v}} f+ \textbf{v}\cdot \nabla_\textbf{x} f=0,\label{eqn:cbe}
\end{equation} a manifestation of the Liouville theorem in classical mechanics. For a smooth distribution of particles, the particle acceleration 
is determined by the smooth potential field, ${\ud \textbf{v}}/{\ud t}=-\nabla_\textbf{x}\Phi$.
Taking the first moment of the collisionless Boltzmann equation over velocity  one can derive the more frequently used Jeans equation, which 
is a 6-dimensional analogy to the 3-dimensional Euler equations for fluid flow. 

When the system is in a steady-state, both the underlying potential and the distribution function are independent of time, i.e., 
$\Phi(\textbf{x},t)=\Phi(\textbf{x})$ and ${\partial f}/{\partial t}=0$. The Jeans equation then relates the potential gradients to 
observable quantities including the number density distribution, the mean velocity and velocity dispersions of different velocity 
components for observed objects. 

Adopting the Jeans equation to constrain the potential gradient requires the knowledge of spatial derivatives of the 
velocity dispersions for different velocity components (e.g. the vertical, radial and azimuthal velocity dispersions 
and cross terms), which is not easy. Studies using the Jeans equation to constrain the underlying distribution of luminous and 
dark matter were traditionally limited to the solar neighborhood\citep[e.g.][]{1989MNRAS.239..605K,1998A&A...329..920C,2000MNRAS.313..209H}, 
within a few kilo-parsecs from the Galactic plane \citep[e.g.][]{1991ApJ...367L...9K,2003A&A...399..531S,2004MNRAS.352..440H,
2012ApJ...746..181S,2012ApJ...751..131B,2012ApJ...756...89B,2012MNRAS.425.1445G,2013ApJ...772..108Z} and out to about $\sim$10~kpc 
with photometric distances \citep[e.g.][]{2014ApJ...794..151L}. 

If further assuming the Galactic halo is spherical, we can derive the simplified and so far more frequently used spherical 
Jeans equation (hereafter SJE; Binney and Tremaine 1987):
\begin{equation}
    \frac{1}{\rho_\ast}\frac{\ud(\rho_\ast \sigma^2_r)}{\ud r}+\frac{2\beta \sigma^2_r}{r}=-\frac{\ud \Phi}{\ud r}=-\frac{V_c^2}{r},
 \label{eqn:pot}
\end{equation}
where quantities on the left side are the radial velocity dispersion of tracers in the system, $\sigma_r$, 
their radial density profile, $\rho_\ast$, and their velocity anisotropy, $\beta$. The velocity anisotropy is 
defined as
\begin{equation}
\beta=1-\frac{\sigma^2_\theta+\sigma^2_\phi}{2\sigma^2_r}
=1-\frac{\langle {v^2_\theta} \rangle- \langle v_\theta \rangle^2+\langle {v^2_\phi}\rangle-\langle v_\phi \rangle^2}{2(\langle {v^2_r} \rangle-\langle v_r \rangle^2)}, 
\label{eqn:beta}
\end{equation}
where $\sigma_\theta$ and $\sigma_\phi$ are velocity dispersions of the two tangential components. $v_r$ is 
the radial velocity. $v_\theta$ and $v_\phi$ are the two components of the tangential velocity. 
When the quantities on the left-hand side of Eqn.~\ref{eqn:pot} can be measured for observed luminous dynamical tracers, 
such as halo stars, globular clusters and satellite galaxies, the rotation velocity (or the potential gradient) on the right-hand 
side of the same equation can be directly inferred. 

In reality, the observed quantity is the radial velocity dispersion of dynamical tracers, $\sigma_r$, converted from 
the Heliocentric line-of-sight velocities, and the tracer density profile, $\rho_\ast$. However, the velocity anisotropy, $\beta$, 
is more difficult to be properly measured if proper motions are not available, especially for tracer objects at large distances. It 
is obvious from Equation~\ref{eqn:pot} that the velocity anisotropy term is degenerate with the gravity term, so that an overestimate 
of $\beta$ leads to an underestimate in mass. This is known as the mass-anisotropy degeneracy. 

Assuming $\beta$ is constant, the solution to Equation~\ref{eqn:pot} is
\begin{equation}
 \sigma^2_r(r)=\frac{1}{r^{2\beta}\rho_\ast(r)}\int_r^{\infty}\ud r' r'^{2\beta}\rho_\ast(r')\ud \phi/ \ud r ,
 \label{eqn:sigmar}
\end{equation}
subject to the boundary condition that $\lim_{r \to \infty}r^{2\beta}\rho_\ast\sigma^2_{r,\ast}=0$ 
\citep[e.g.][]{2005MNRAS.364..433B,2012ApJ...761...98K}.

\subsection{Measurements with assumed or externally calibrated anisotropy}

Based on distances and radial velocities of 240 tracer objects in the stellar halo of our MW, including BHB stars , 
red giant stars, globular clusters and satellite galaxies, from the spectroscopic \textsc{Spaghetti} survey, Battaglia et al. 
\cite{2005MNRAS.364..433B} used Eqn.~\ref{eqn:sigmar} to constrain the mass of our MW in 2005. Constant $\beta$ was 
adopted in their analysis. The density profile of tracers was measured to have a power-law form of $\rho_\ast(r)\propto 
r^{-\alpha}$, with $\alpha\sim$3.5 out to $\sim$50~kpc from the Galactic center \citep{2000AJ....119.2254M,2000ApJ...540..825Y}. 
Assuming the power-law density profile of tracers is valid out to large distances, the radial velocity dispersion was measured 
to be almost a constant of 120~km/s out to 30~kpc and continuously drops to $\sim$50~km/s at 120 kpc. The best-fit NFW model 
led to the virial mass of our MW of $M_{200}=0.7^{+1.2}_{-0.2}\times 10^{12}\msun$, and the best-fit mass within 120~kpc to 
the Galactic center was constrained to be $5.4^{+2.0}_{-1.4}\times 10^{11}\msun$. 


In addition to a constant $\beta$, Battaglia et al. \cite{2005MNRAS.364..433B} investigated alternative functional forms 
of $\beta$ profiles as a function of the Galactocentric distance. For a given mass model, although not all functional forms 
of $\beta$ can produce reasonable fits to the data, the best-fit virial mass is strongly dependent of the chosen functional 
form for $\beta$ (see also, e.g., \cite{2014ApJ...785...63B} and \cite{2018RAA....18..113Z}). In a follow-up study, Dehnen 
et al.~in 2006 \cite{2006MNRAS.369.1688D} revisited the results of Battaglia et al. \cite{2005MNRAS.364..433B}, and found a 
virial mass of about $1.5\times 10^{12}\msun$. In contrast to Battaglia et al. \cite{2005MNRAS.364..433B}, Dehnen et al.~\cite{2006MNRAS.369.1688D} 
claimed that the observed radial velocity dispersions are consistent with a constant velocity anisotropy of tracers, if the 
density profile of tracers is truncated beyond 160~kpc. These studies demonstrate that \emph{the mass to be constrained is 
very sensitive to assumptions behind both velocity anisotropies and tracer density profiles}. 

Given the strong $\beta$-dependence, some other studies attempted to rely on numerical simulations to estimate the anisotropy 
when applying the Jeans equation. Xue et al. in 2008 \cite{2008ApJ...684.1143X} also adopted the SJE as part of their analysis, 
but instead of directly fitting the observed radial velocity dispersions with assumptions of $\beta$, Xue et al. \cite{2008ApJ...684.1143X} 
constrained the rotation curve of our MW out to 60~kpc, which relies on the distribution of radial versus circular 
velocities of star particles in two simulated halos of hydrodynamical simulations. The circular velocity as a function 
of radius within 60~kpc was determined by matching the observed distribution of radial versus circular velocities to 
the corresponding distribution in simulated halos. In their analysis, the SJE was used to scale their simulated halos, 
which have slightly different radial profiles of star particles compared with the best estimated power-law slope of our 
MW. The mass within 60~kpc to the Galactic center was estimated to be $4.0\pm 0.7\times 10^{11}\msun$. Adopting the 
NFW model profile, the virial mass was constrained to be $M_{200}=0.84^{+0.3}_{-0.2}\msun$.

Based on halo stars with radial velocity measurements from the Hypervelocity Star Survey, Gnedin et al. in 2010 
\cite{2010ApJ...720L.108G} measured the radial velocity dispersion between 25 and 80~kpc from the Galactic center. 
The velocity anisotropy was inferred from numerical simulations, with a plausible range of $0 \leq \beta \leq 0.5$ 
and a most likely value of 0.4. Over the probed radial range, the power-law index of the tracer density profile 
was between 3.5 and 4.5. The plausible range of circular velocity at 80~kpc inferred from the SJE was between 175 
and 231~km/s. Gnedin et al. \cite{2010ApJ...720L.108G} constrained the mass within 80~kpc to the Galactic center 
as $6.9_{-1.2}^{+3.0}\times 10^{11}\msun$. The virial mass within 300~kpc was extrapolated to be 
$M_{200}=1.3\pm0.3\times 10^{12}\msun$.

Very recently, Zhai et al. in 2018 \cite{2018RAA....18..113Z} used the SJE to model the differentiation of line-of-sight velocity 
dispersions based on 9627 K giant stars from LAMOST DR5, with distances between 5 and 120~kpc from the Galactic center. If 
$\beta$ was assumed as 0.3, the MW virial mass was constrained to be $1.11_{-0.20}^{+0.24}\times 10^{12}\msun$. 

\subsection{Inferring mass and anisotropy from data}
\label{sec:jeansbetadirect}


To overcome the mass-anisotropy degeneracy, many studies devoted efforts to either directly infer or indirectly model $\beta$ 
from observational data. In the solar vicinity, $\beta$ was measured to be $\sim$0.6 based on proper motions of stars 
\citep[e.g.][]{2009MNRAS.399.1223S,2010ApJ...716....1B}. Without proper motions, tangential velocities with respect to the 
Galactic center can still be inferred from the line-of-sight velocities for objects in the inner MW, given the fact that our 
Sun is about 8~kpc from the Galactic center \citep[e.g.][]{2004AJ....127..914S,2011MNRAS.411.1480D,2012MNRAS.424L..44D,
2012ApJ...761...98K, 2015ApJ...813...89K}. The observed line-of-sight velocities are contributed by both radial and tangential 
velocities with respect to the Galactic center. The fraction of tangential velocities contained in the line-of-sight velocities 
depends on both Galactocentric distances and Galactic coordinates of the observed object \citep{2015ApJ...813...89K}. For tracer 
objects at large distances, the line-of-sight velocities are dominated by the radial components and contain very little 
information about the tangential velocity components. 

Early in 1997, Sommer-Larsen et al. \cite{1997ApJ...481..775S} analyzed 679 BHB stars between 7 and 65~kpc from the Galactic center. 
Assuming dynamical equilibrium in a logarithmic Galactic potential, they found indications that the tangential velocity dispersion 
further beyond the solar radius should be larger than the value in our solar neighborhood. In 2005, Sirko et al. \cite{2004AJ....127..914S} 
fitted an ellipsoidal velocity distribution to 1170 BHB stars from SDSS, and reported that the halo beyond our solar vicinity is 
close to isotropic. Adopting a power-law distribution function with a constant $\beta$ (see Sec.~\ref{sec:DF} for more details of 
the distribution function), Deason et al. in 2011 \cite{2011MNRAS.411.1480D} fitted 3549 BHB stars from SDSS/DR7 and reported 
a tangential halo between 10 and 25~kpc and a radial halo between 25 and 50~kpc. Then in a later study, Deason et al. in 2012 
\cite{2012MNRAS.424L..44D} allowed both the potential parameter and velocity anisotropy in their model distribution function to 
vary and reported $\beta=0.5$ between 16 and 48~kpc. It was discussed by Kafle et al. in 2012 \cite{2012ApJ...761...98K} that the 
tangential halo claimed by Deason et al. in 2011 \cite{2011MNRAS.411.1480D} within 25~kpc is very likely due to the broad radial 
binning.

In fact, the study by Kafle et al. in 2012 \cite{2012ApJ...761...98K} involved modeling of the anisotropy profile between 9 and 
25~kpc based on maximum likelihood analysis with analytical distribution functions. The best constrained $\beta$ is close to 0.5 
in the very inner part of our MW and falls sharply beyond 13~kpc, reaching a minimum of -1.2 at 17~kpc and rises again on larger 
scales. Beyond 25~kpc, radial velocities can still be measured, but it was impossible to properly measure the tangential components 
from their line-of-sight velocities. Kafle et al. \cite{2012ApJ...761...98K} fitted three-component potential models of Galactic 
disk, bulge and halo to the estimated circular velocity profile out to 25~kpc based on the SJE, and the mass enclosed within 25~kpc 
was measured to be $2.1\times10^{11}\msun$. The virial mass was extrapolated to be $M_{200}=0.77^{+0.40}_{-0.30}\times 10^{12}\msun$. 
With the extrapolated potential profile, tracer density profile and the measured radial velocity dispersion on distances larger than 
what can be probed by their sample of tracers, they used the SJE to predict $\beta$ to be roughly 0.5 over the radial range of 25 to 
56~kpc. 

Note, however, the very inner region of our MW, which is close to the Galactic disk, is not spherically symmetric, but the SJE assumes 
spherical symmetry. This can bias the estimated mass. In addition, the underlying potential of the outer halo is not ideally 
spherical because dark matter halos are triaxial \citep{2002ApJ...574..538J}. There are efforts of applying the SJE to simulated galaxies 
and halos \cite[e.g.][]{2018MNRAS.476.5669W,2018MNRAS.475.4434K}. More importantly, the assumption of a steady-state is also non-trivial 
and can lead to significant systematics. For example, Wang et al. in 2018 \cite{2018MNRAS.476.5669W} have shown evidences of how violations 
of the spherical and the steady state assumptions behind the SJE can potentially affect the constrained halo mass of MW-like halos. We 
provide more detailed discussions in Sec.~\ref{sec:sims}. 

Using proper motions of 13 main sequence halo stars from the multi-epoch \textit{HST}/ACS photometry, Deason et al. in 2013 \cite{2013ApJ...766...24D} 
reported that $\beta$ is consistent with zero (isotropic) at 24~kpc. In addition, King III et al. in 2015 \cite{2015ApJ...813...89K} found 
a minimum in their measured anisotropy profiles at $\sim$20~kpc, based on 6174 faint F-type stars from the radial velocity sample of the MMT 
telescope, and 13480 F-type stars from SDSS. However, compared with Kafle et al. \cite{2012ApJ...761...98K}, the minimum in their anisotropy 
profile is more negative, and they claimed that the less negative measurements in other studies is likely due to their broader binning.

Direct measurements of $\beta$ and the mass distribution up to and beyond the Galactocentric distance of 100~kpc are even more challenging, 
where the line-of-sight velocities are almost entirely dominated by the radial velocities with respect to the Galactic center. Using a sample of 
halo stars out to $\sim$150~kpc, Deason et al. in 2012 \cite{2012MNRAS.425.2840D} found that the radial velocity dispersion of these stellar 
tracers falls rapidly on such large distances. Assuming the tracer density falls off between 50 and 150~kpc with a power-law index smaller 
than 5 and assuming radial orbits, the mass within 150~kpc to the Galactic center was estimated to lie in the range between $5\times 10^{11}$ 
and $10^{12}\msun$.

In a later study by Kafle et al. in 2014 \cite{2014ApJ...794...59K}, the radial velocity dispersion profile out to $\sim$160~kpc was measured 
with K giants from SDSS/DR9, and the SJE was used to constrain the mass distribution and the velocity anisotropy out to such large distances. 
Kafle et al. \cite{2014ApJ...794...59K} modeled the inner tracer density profile as double power law with a break radius. Beyond 100~kpc, the 
profile was assumed to be truncated beyond a characteristic radius plus an exponential softening quantified by some scale length. Within 25~kpc, 
$\beta$ can be known from previous studies. Beyond 50~kpc, they assumed $\beta$ to be a constant, and the change of $\beta$ was assumed to be 
linear between 25 and 50~kpc. The break and truncation radii, softening scale length and the constant $\beta$ beyond 50~kpc were all treated 
as free parameters, in combination with other free parameters in their three-component potential model. The virial mass was best fit to be
$M_{200}=0.71^{+0.31}_{-0.16}\times 10^{12} \msun$, and $\beta$ of the outer halo was estimated to be $0.4\pm 0.2$.

Using multiple species of halo stars and combining the terminal velocity measurements with the SJE analysis, Bhattacharjee et al. in 2014 
\cite{2014ApJ...785...63B} measured the rotation curve of our MW up to $\sim$200~kpc. Since the circular velocity decreases with the increase 
of $\beta$ at a given radius, the maximum value of $\beta=1$ corresponds to the lower limit of mass enclosed within 200~kpc, which was 
constrained by Bhattacharjee et al. \cite{2014ApJ...785...63B} to be $6.8\pm4.1 \times 10^{11}\msun$. 

Huang et al. in 2016 \cite{2016MNRAS.463.2623H} used about 16,000 primary red clump giants in the outer disk from the LSS-GAC (LAMOST 
Spectroscopic Survey of the Galactic Anticancer ) of the on-going LAMOST experiment and the SDSS-III/APOGEE survey, plus 5,700 K giants 
from the SDSS/SEGUE survey to derive the rotation curve of our MW out to 100~kpc. In the inner MW region, the rotation velocity was 
deduced from line-of-sight velocities following the approaches in Sec.~\ref{sec:localobs}, whereas the rotation curve in the outer halo 
was obtained from the SJE, with the values of $\beta$ taken from all the previous studies mentioned above and interpolated. Their 
best-fit potential model led to the virial mass of $M_{200}=0.85^{+0.07}_{-0.08}\times 10^{12} \msun$.

In 2017, Ablimit and Zhao \cite{2017ApJ...846...10A} adopted 860 ab-type RR Lyrae stars in the Galactic halo to look at the rotation 
velocities out to $\sim$50~kpc using the SJE. Their sample of stars were identified from the Catalina Surveys DR1, combined with spectroscopic 
data from SDSS DR8 and LAMOST DR4 to obtain radial velocities. They adopted two different choices of $\beta$, a constant value of $\beta=0$ 
within 50~kpc and the radius-dependent $\beta$ profile by \cite{2015MNRAS.454..698W} (see Sec.~\ref{sec:DFaction} for more details), in 
which $\beta$ changes from $\sim$0.32 to $\sim$0.67 between 10 and 50~kpc. For $\beta=0$, the circular velocity at 50~kpc was constrained  
to be 180.00$\pm$31.92~km/s, and the enclosed mass within 50~kpc was estimated as $3.75\pm 1.33 \times 10^{11}\msun$. 
 

More recently, since much more proper motion data has been measured by \textit{Gaia}, Bird et al. in 2019 \cite{2019AJ....157..104B} 
directly measured $\beta$ for more than 8,600 metal poor K giants, based on distances and line-of-sight velocities from LAMOST and cross 
matched to \textit{Gaia} DR2 to obtain proper motions. Between the solar radius and 25~kpc to the Galactic center, their sample is highly 
radial ($\beta\sim0.8$). $\beta$ gradually becomes less radial beyond 25~kpc, reaching $\sim$0.3 at $\sim$100~kpc. In contrast to previous 
measurements made by Kafle et al. in 2012 \cite{2012ApJ...761...98K}, Deason et al. in 2013 \cite{2013ApJ...766...24D} and King III et al. 
in 2015 \cite{2015ApJ...813...89K}, Bird et al. \cite{2019AJ....157..104B} did not report any minimum for the $\beta$ profile within 25~kpc. 
In addition, they claimed the sensitivity of their measured $\beta$ profile to substructures.

To conclude, the measurement of the velocity anisotropy or $\beta$ profile, from the Galactic center to the outer stellar halo, still 
suffers from inconsistencies and debates. The robustness of the measurements depends on a variety of factors. First of all, whether proper 
motions are available is very important, and deriving $\beta$ from line-of-sight velocities might suffer from systematics due to the assumed 
tracer velocity models. As we have already mentioned, many previous studies constrained the tangential velocity components from the observed 
line-of-sight velocities have to rely on fitting ellipsoidal models in velocity space or distribution functions in phase space. Moreover, 
the particular type of tracer objects and the sample selection may both result in different measurements, because different tracer populations 
are not expected to have the same velocity distributions. Also, the influence of substructures is still uncertain. This has been discussed by 
Loebman et al. in 2018 \cite{2018ApJ...853..196L}, in which the possible origin and persistence of the dip feature in $\beta$ profiles were 
investigated using both N-body and hydrodynamical simulations. 

\subsection{Variants to the SJE: Mass estimators}
\label{sec:jeanestimator}
The Jeans equation itself can be regarded as a mass estimator in which the enclosed mass is related to the velocity dispersion profile. 
Starting from the collisionless Boltzmann equation or the SJE, alternative forms of mass estimators may be derived, usually under some 
more specific forms of the potential and tracer profiles. These estimators may be more compact and convenient to apply to data than the 
SJE. However, one should bear in mind that the limitations and caveats in the Jeans modeling as discussed above are generally also 
relevant to these alternative mass estimators, in addition to model-specific systematics in case extra assumptions are made to derive 
the estimator.

Early attempts of deriving mass estimators which relate observed positions and velocities of tracer objects to the enclosed mass can be 
traced back to the 1960s, which is called the virial theorem \citep{1960ApJ...132..286L}. The virial theorem estimator, however, results in 
a biased estimate of the true mass as pointed out by Bahcall and Tremaine in 1981 \cite{1981ApJ...244..805B}.

In 2010, Watkins et al. \cite{2010MNRAS.406..264W} derived mass estimators for scale-free (single power law) potentials 
and tracer density profiles, which relate the enclosed mass within the maximum radius of the tracer sample to the observed 
velocities and distances of these tracers. When tangential velocities are unknown due to missing proper motions, the 
estimator relates the average radial velocities and distances of tracers to the enclosed mass, while the velocity 
anisotropy of tracers, the power-law indexes of the potentials and tracer density profiles have to be assumed in advance. 

Applying the derived mass estimator to satellite galaxies in our MW and assuming isotropic velocity anisotropy ($\beta=0$), 
the mass within 300~kpc was estimated to be $1.17\pm 0.3 \times 10^{12}\msun$ by Watkins et al. \cite{2010MNRAS.406..264W}. 
In their analysis, six satellites have proper motions. Two estimators with or without tangential velocities when proper 
motions are available or not have been applied separately. The final result was a weighted average between results from 
the two estimators, and Monte Carlos simulations were adopted to estimate the measurement errors. The 
estimated mass within 300~kpc dropped by $\sim$60\% if only radial velocities were used. If considering the plausible range 
of anisotropies based on both numerical simulations and observations, the uncertainty was in fact very large, ranging from 
$\sim 1.0$ to $\sim 2.7\times 10^{12}\msun$. This corresponds to the large errorbar in Fig.~\ref{fig:massplot}. In their 
analysis Leo I played 27\% of role compared with the other satellites. 

Watkins et al. \cite{2010MNRAS.406..264W} drew the plausible power-law index value for their potential model by 
looking at the best-fit slopes of NFW profiles over the radial range of their satellites. For MW-analogous concentrations 
and virial radii, the power-law index of the potential changes slowly and is close to 0.5. However, the NFW profile, which 
can be used to well approximate the dark matter halo profiles in modern cosmological simulations, does not produce a strictly 
single power law potential outside 10~kpc. In a few follow-up papers, mass estimators with more generalized potentials have 
been further developed \citep[e.g.][]{2011MNRAS.413.1744A,2012MNRAS.420.2562A,2011ApJ...730L..26E}.

In fact, not only the potential profile, but also the tracer density profile is not a single power law. Many late-time 
studies found breaks in the number density profiles of halo stars at $\sim$16 to $\sim$30~kpc \citep[e.g.][]{
2008ApJ...680..295B,2009MNRAS.398.1757W,2011ApJ...731....4S,2011MNRAS.416.2903D,2013AJ....146...21S,2014ApJ...794...59K,
2015A&A...579A..38P,2015ApJ...809..144X,2016MNRAS.463.3169D,2019PASJ..tmp...68F}. The power-law indexes reported by many 
of these studies are shallower within the breaking radii, and become steeper outside. For example, \cite{2008ApJ...680..295B} 
found values of $\alpha=2$ and $\alpha=4$ for inner and outer profiles. \cite{2009MNRAS.398.1757W} investigated RR Lyrae 
stars out to 100~kpc and found values of 2.4 and 4.5. \cite{2011MNRAS.416.2903D} reached similar conclusions (power-law 
indexes of 2.3 and 4.6) using BHB stars out to $\sim$40~kpc. Using main-sequence turnoff stars from the Canada-France-Hawaii 
Telescope Legacy Survey, \cite{2011ApJ...731....4S} reported a slightly shallower outer slope of 3.8 beyond 28~kpc, and 
in a follow-up paper, \cite{2013AJ....146...21S} found a smaller breaking radius of 16~kpc. The tension among different 
measurements might be due to significant variations of the power-law index values over the sky, in that substructures can 
potentially affect the measured values, especially if one relies on narrow pencil-beam surveys \citep[e.g.][]{2015MNRAS.446.2274L}. 
It was pointed out by Lowing et al. in 2013 \cite{2013ApJ...763..113D} that the break in tracer density profiles is 
likely associated with an early and massive accretion event. 

More recently, Sohn et al. in 2018 \cite{2018ApJ...862...52S} adopted proper motion measurements from \textit{HST} for 
globular clusters between 10 and 40~kpc from the Galactic center. Still based on the mass estimator developed by Watkins et al. 
in 2010 \cite{2010MNRAS.406..264W} (the one with proper motion and based on scale-free potential and tracer density profiles), 
Sohn et al. \cite{2018ApJ...862...52S} constrained the mass within 39.5~kpc to the Galactic center as $6.1^{+1.8}_{-1.2}
\times10^{11}\msun$. $\beta$ was measured from their sample of globular clusters as $0.609_{-0.229}^{+0.130}$. The virial 
mass was extrapolated to be $M_{200}=1.71_{-0.79}^{+0.97}\times 10^{12} \msun$.

Using 34 globular clusters with proper motions from the second data release (DR2) of \textit{Gaia}, together within the sample 
of globular clusters from \textit{HST} \citep{2018ApJ...862...52S}, Watkins et al. in 2019 \cite{2019ApJ...873..118W} further 
estimated the mass within 21.1~kpc and 39.5~kpc to the Galactic center as $2.1^{+0.4}_{-0.3}\times 10^{11}\msun$ and 
$4.2^{+0.7}_{-0.6}\times 10^{11}\msun$, respectively. The virial mass was extrapolated to be $M_{200}=1.29^{+0.75}_{-0.44}
\times 10^{12}\msun$. $\beta$ was estimated from the data as $0.52^{+0.11}_{-0.14}$.

A very recent study by Fritz et al. \cite{2020arXiv200102651F} applied the Watkins et al. mass estimator to 45 satellites with 
{\it Gaia} proper motions. Subhalos from dark matter only simulations are used to test and calibrate their measurements. A 
significant bias has been reported after applying the mass estimator to simulations, which was mainly attributed to the deviation 
of satellite density profiles from a single power law. Systematic uncertainties arising from LMC mass and LMC satellites were also 
discussed. The mass enclosed within 64 and 273~kpc was estimated to be $5.8_{-1.4}^{+1.5}\times 10^{11}\msun$ and $14.3_{-3.2}^{+3.5}\times 10^{11}\msun$, 
taking into account potential influences of the LMC. The mass out to $\sim$308~kpc was extrapolated to be $1.51_{-0.40}^{+0.45}\times 10^{12}\msun$.
 
\section{Distribution functions: halo stars, globular clusters and satellite galaxies}\label{sec:DF}
As we have introduced in Section~\ref{sec:jeans}, the phase-space distribution function fully specifies the dynamical structure 
of the system, and in principle contains the most complete information for dynamical modeling. As solutions to the collisionless 
Boltzmann equation (Equation~\ref{eqn:cbe}), these functions connect the phase-space coordinates of tracers to the underlying 
potential, and thus can be used to fit to the observed positions and velocities of tracer objects and constrain the model potential.
In the following, we discuss how available functional forms can be chosen, in terms of classical integrals of motion or actions, 
and we will introduce the efforts of using the distribution functions to constrain the mass of our MW. In addition, we also review 
the class of simulation-based distribution functions by linking observed MW satellite galaxies to simulated subhalos and the set 
of orbital probability distributions. Measurements in this section fall in the category of ``DF'' in Fig.~\ref{fig:massplot}.

\subsection{Distribution function in terms of classical integrals of motion}
\label{sec:DFEL}

The phase-space distribution function, $f(\textbf{x},\textbf{v},t)$, describes a dynamical system in terms of positions, $\textbf{x}$, 
velocities, $\textbf{v}$, and time, $t$. As we have mentioned in Sec.~\ref{sec:jeans}, the system is independent of time when it 
is in steady state. The Jeans theorem states that the steady state solution of the distribution function is connected to positions 
and velocities only through the integrals of motions. The strong Jeans theorem further states that there are only three independent 
integrals of motions. For systems with a spherical symmetry, the distribution function of steady state systems can be written down 
in terms of energy and the magnitude of angular momentum, i.e., $f(E,L)$ \citep{1962MNRAS.124....1L}.

The phase-space distribution function of tracer objects bound to the underlying potential (binding energy $E>0$) can be described 
by the Eddington formula \citep{1916MNRAS..76..572E}, which is basically the Abell inversion of the tracer density profile, $\rho_\ast(r)$, 
under the spherical assumption. The simplest isotropic and spherically symmetric case is 

\begin{equation}
F(E)=\frac{1}{\sqrt{8}\pi^2}\frac{\ud}{\ud E}\int_{\Phi(r_\mathrm{max,t})}^E \frac{\ud \rho_\ast(r)}{\ud\Phi(r)}\frac{\ud\Phi(r)}{\sqrt{E-\Phi(r)}},
\label{eq:isoform} 
\end{equation} 
where the distribution function only depends on the binding energy per unit mass, $E=\Phi(r)-\frac{v^2}{2}$. $\Phi(r)$ and $\frac{v^2}{2}$ 
are the underlying dark matter halo potential and kinetic energy per unit mass of tracers. The integral goes from the potential at the tracer 
boundary to the binding energy of interest\footnote{To define the binding energy, we adopt the convention that $\Phi(r_\mathrm{max,t})=0$, and 
$\Phi(r)>0$, which differs from the potential used in previous sections by a sign.}.  Usually both the zero point of potentials and the tracer 
boundary, $r_\mathrm{max,t}$, are chosen at infinity, and thus $\Phi(r_\mathrm{max,t})=0$. 

In reality the velocity distribution of tracers is anisotropic. One possible anisotropic form introduces an 
angular momentum ($L$) dependence as

\begin{equation}
F(E,L)=L^{-2\beta}f(E),  
\label{eq:anisoform}
\end{equation}
where the energy part, $f(E)$, is expressed as \citep{1991MNRAS.253..414C}

\begin{equation}
\begin{split}
f(E)=\frac{2^{\beta-3/2}}{\pi^{3/2}\Gamma(m-1/2+\beta)\Gamma(1-\beta)}\times\\
\frac{\ud}{\ud E}\int_{\Phi(r_\mathrm{max,t})}^E (E-\Phi)^{\beta-3/2+m} \frac{\ud^m [r^{2\beta}\rho_\ast(r)]}{\ud \Phi^m} \ud \Phi\\
=\frac{2^{\beta-3/2}}{\pi^{3/2}\Gamma(m-1/2+\beta)\Gamma(1-\beta)}\times\\
\int_{\Phi(r_\mathrm{max,t})}^E (E-\Phi)^{\beta-3/2+m} \frac{\ud^{m+1} [r^{2\beta}\rho_\ast(r)]}{\ud \Phi^{m+1}} \ud \Phi, 
\end{split}
\end{equation} which assumes the energy, $E$, and the magnitude of angular momentum, $L$, are separable. Here $\beta$ is the velocity anisotropy 
parameter. $m$ is an integer chosen to make the integral converge and depends on the value of $\beta$. If the allowed region of $\beta$ is 
$-0.5<\beta<1$, then $m=1$. 

When the tangential velocities of tracers are not available, the phase-space distribution in terms of radius, $r$, and radial velocity, $v_r$, is 
given by the integral over tangential velocities, $v_t$,

\begin{equation}
P(r,v_r|C)=\int L^{-2\beta}f(E) 2\pi v_t \ud v_t,
\end{equation}
where $C$ denotes a set of model parameters. Via the Laplace transform, this can be written as
\begin{equation}
P(r,v_r|C)=\frac{1}{\sqrt{2}\pi r^{2\beta}} \int_{\Phi(r_\mathrm{max,t})}^{E_r} \frac{\ud\Phi}{(E_r-\Phi)^{1/2}}\frac{\ud r^{2\beta}\rho_\ast}{\ud\Phi},
\label{eq:radialonly}
\end{equation}
where $E_r=\Phi(r)-v_r^2/2$. All factors of $m$ cancel in the Laplace transform and hence Eqn.~\ref{eq:radialonly} does not depend on $m$. 

The more explicit form of the above phase-space distribution function depends on the chosen model for the underlying potential ($\Phi$) 
and tracer density ($\rho_\ast$) profiles. Free parameters in the potential and tracer models are often constrained through the Bayesian 
framework. 

Early attempts of using such distribution functions to constrain the mass of our MW can be traced back to a few decades ago. In 1987, 
Little and Tremaine \cite{1987ApJ...320..493L} devised the method of fitting the model distribution function to observed positions and 
velocities of satellite galaxies under the Bayesian framework, to constrain the mass within 50~kpc to the Galactic center. In 1989, Zaritsky 
et al. \cite{1989ApJ...345..759Z} repeated the analysis with improved velocity estimates of satellites. Assuming a point mass, the median 
mass of the Galaxy was estimated to be $9.3_{-1.2}^{+4.1}\times10^{11}\msun$ for radial satellite orbits, and $12.5_{-3.2}^{+8.4}\times
10^{11}\msun$ for isotropic satellite orbits. The estimated mass can be significantly smaller if excluding Leo I from their sample of 
satellite galaxies. 

Slightly later in 1991, similar estimates were reached by Norris and Hawkins \cite{1991ApJ...380..104N} and by Kulessa and 
Lynden-Bell in 1992 \cite{1992MNRAS.255..105K}. Then in 1996, Kochanek \cite{1996ApJ...457..228K} applied such phase-space 
distribution functions to constrain the mass of our MW using the Jaffe model, in combination with other approaches including the local 
escape velocity of stars (see Sec.~\ref{sec:HVS}), which suppressed the low-velocity solutions, and the rotation curve of the disk (see 
Sec.~\ref{sec:localobs}), which eliminated solutions predicting too high rotation velocities at the solar neighborhood. The Local Group 
timing argument (see Sec.~\ref{sec:timing} for details) was also adopted to suppress high-mass solutions. Kochanek \cite{1996ApJ...457..228K} 
discussed cases when only radial velocities were available and when both radial and tangential velocities were available. The median 
mass within 50~kpc was estimated to be $5.1_{-1.1}^{+1.3} \times 10^{11}\msun$ (90\% confidence level and with Leo I). 

Wilkinson and Evans in 1999 \cite{1999MNRAS.310..645W} adopted the truncated flat-rotation model potential in their distribution 
function, and used the distances and velocities of satellite galaxies and globular clusters to constrain the model (27 objects in total and 
6 with proper motions). Assuming the density profile of satellites and globular clusters was a power law with index of 3.4 and by including 
Leo I, the mass enclosed within 50~kpc was estimated to be $5.4_{-3.6}^{+0.2}\times 10^{11}\msun$, and the total mass of the halo was 
estimated as $1.9^{+3.6}_{-1.7}\times10^{12}\msun$ \footnote{The total mass is a free parameter in their truncated flat-rotation potential 
model. The errors are determined by both the small sample size and the measurement errors, which are thus very large. }.

The estimates were updated by Sakamoto et al. in 2003 \cite{2003A&A...397..899S} using a larger sample of dynamical tracers including 11 
satellites, 127 globular clusters and 413 field horizontal branch stars, among which half of the objects had proper motions. The total mass 
was constrained to be $2.5_{-1.0}^{+0.5}\times10^{12}\msun$ if including Leo I and $1.8_{-0.7}^{+0.4}\times10^{12}\msun$ if excluding Leo I. 
The mass within 50~kpc was estimated to be $5.5_{-0.2}^{+0.0}\times10^{11}\msun$ if including Leo I and $5.4_{-0.4}^{+0.1}\times10^{11}\msun$ 
without Leo I.

Later in 2012, Deason et al. \cite{2012MNRAS.424L..44D} adopted the single power-law potential and tracer density profiles in the model distribution 
function proposed by \citep{1997MNRAS.286..315E}, and considered a constant flattening of the halo, by using 1933 blue horizontal branch stars within 
18 and 48~kpc from the Galactic center as dynamical tracers. The mass within 50~kpc was constrained to be $4.2\pm0.4 \times 10^{11}\msun$. 
Combined with the measured rotation velocities and enclosed masses within fixed radii smaller than the virial radius from other studies, 
\cite{2012MNRAS.424L..44D} claimed that their results favored an NFW halo with concentration of about 20. We extrapolated the NFW profile 
to get the virial mass of $M_{200}=0.94^{+0.22}_{-0.20}\times 10^{12}\msun$. 

In a series of papers since 2015 \citep{2015ApJ...806...54E,2016ApJ...829..108E,2017ApJ...835..167E,2018ApJ...865...72E,
2019ApJ...875..159E}, the distribution functions were revisited. In their first paper, Eadie et al. in 2015 \cite{2015ApJ...806...54E} 
used the Hernquist and the isotropic Jaffe models in their analysis to model the incomplete data. Their sample of dynamical tracers 
involved 59 globular clusters and 29 dwarf galaxies, out of which 26 globular clusters and 18 dwarf galaxies did not have proper motions. 
For tracers with large enough Galactocentric distances, their observed line-of-sight velocities can be approximated as radial velocities 
with respect to the Galactic center, while the unknown tangential velocities were treated as nuisance parameters, with the parameter 
space sampled through the hybrid-Gibbs sampler. The total mass of our MW, which is a parameter in the Hernquist model, was estimated as 
$1.55^{+0.18}_{-0.13}\times 10^{12}\msun$ (95\% confidence region, isotropic anisotropy). The mass within 260~kpc was constrained as 
$1.37_{-0.10}^{+0.14}\times 10^{12}\msun$ (95\% confidence region). It was found that since the proper motion of Pal 3 has very large 
uncertainties, Pal 3 played a very important weight in determining the mass of our MW. If excluding Pal 3 in their analysis, the MW 
mass through the Hernquist parameter was best constrained as $1.36^{+0.15}_{-0.10} \times 10^{12}\msun$ (95\% confidence region). 

The first paper \cite{2015ApJ...806...54E} was based on simple potential models and it was assumed that the dynamical tracers and the 
underlying dark matter have the same spatial distributions. In a follow-up study of 2016, Eadie et al. \cite{2016ApJ...829..108E} 
adopted power-law potential and tracer density profiles of \citep{1997MNRAS.286..315E} and \cite{2012MNRAS.424L..44D}, and adopted 
different radial distributions for tracers and dark matter. They used globular clusters only as tracers, because a single power-law 
density profile better models a single population of objects. The catalog of \cite{1996AJ....112.1487H,2010arXiv1012.3224H} were used 
as the starting point of their sample globular clusters, which contains 157 objects. After exclusions, in total 89 globular clusters 
were selected, out of which 18 did not have complete velocity measurements. A series of different scenarios were tried, by choosing to 
either fix or free the parameters for the underlying potential profile, the tracer density profile and the velocity anisotropy. When 
all the parameters were free, the mass within 125~kpc was constrained to be $5.22_{-0.43}^{+0.41}\times 10^{11}\msun$ (50\% confidence). 
Extrapolated to the virial radius, the virial mass was found to be $M_{200}=0.682_{-0.076}^{+0.071}\times 10^{12}\msun$ (50\% confidence 
region). If only using globular clusters outside 10~kpc where a single-power law potential model can better approximate the outer slope 
represented by the NFW model, the virial mass was constrained as $M_{200}=0.902^{+0.184}_{-0.333}\times 10^{12}\msun$ (50\% confidence ). 

Eadie et al. in 2017 \cite{2017ApJ...835..167E} further applied the hierarchical Bayesian approach to model the phase-space distribution 
of globular clusters, which includes more meaningful treatment of measurement errors. With the sample of globular clusters from 
their previous study in 2016 \cite{2016ApJ...829..108E}, the mass within 125~kpc to the Galactic center was measured as $6.3\pm 1.1
\times 10^{11}\msun$ (95\% confidence). The virial mass was extrapolated to be $M_{200}=0.86_{-0.19}^{+0.23}\times 10^{12}\msun$ (95\% 
confidence). Further extrapolated to 300~kpc, the enclosed mass within 300~kpc was found to be $1.14\pm 0.22 \times 10^{12}\msun$. 

More recently, the {\it Gaia} team measured and released the mean proper motions through member stars for 75 globular clusters \citep{2018A&A...616A..12G}, 
which cover about half of the objects previously provided by \cite{1996AJ....112.1487H} and \cite{2010arXiv1012.3224H}. In their 2019 
paper, Eadie et al. \cite{2019ApJ...875..159E} replaced the data of \cite{1996AJ....112.1487H} and \cite{2010arXiv1012.3224H} 
by the \textit{Gaia} proper motions with some exceptions. When proper motions were not available from \textit{Gaia}, the HST proper motions 
\citep{2018ApJ...862...52S}, if available, were adopted. If the proper motions were still missed, the measurements from other studies 
(see \cite{2016ApJ...829..108E} for details) were adopted. In total, the sample contained 154 objects, out of which 52 had incomplete 
measurements. In the mean time, both Vasiliev \cite{2019MNRAS.484.2832V} and Baumgardt et al. \cite{2019MNRAS.482.5138B} in 2019 identified 
member stars for the $\sim$150 globular clusters from \cite{1996AJ....112.1487H} and \cite{2010arXiv1012.3224H}, and provided independent 
mean proper motion estimates for all of them. Eadie \& Juri{\'c} in 2019 \cite{2019ApJ...875..159E} afterwards made their measurements using
both their own extended globular cluster sample based on \textit{Gaia} DR2 plus \textit{HST} and the catalog of Vasiliev \cite{2019MNRAS.484.2832V}. 
The two catalogs gave very similar constraints. Based on globular clusters with Galactocentric distances larger than 15~kpc, the median 
estimate of the MW virial mass was $M_{200}=0.70_{-0.12}^{+0.17}\times 10^{12}\msun$ (68\% confidence reading from their Fig.~7). If further 
excluding 4 globular clusters with Galactocentric distances between 15 and 20~kpc from \citep{2018MNRAS.475.1537M}, the measurement became 
slightly larger $M_{200}=0.77_{-0.16}^{+0.25}\times 10^{12}\msun$. They provided the cumulative mass profile out to slightly beyond $\sim$~100~kpc.

As the readers may have found, most of the measurements adopted simplified potential models such as the truncated flat rotation  
and power-law potentials. However, the NFW profile, which well approximates the dark matter halo profiles in cosmological simulations, 
lead to more complicated analytical forms of the phase-space distribution function \citep{2015MNRAS.453..377W}, and have been evaluated 
numerically \citep[e.g.][]{2000ApJS..131...39W}, but have never been applied to real tracer objects.

The performance of single power-law potential models and the functional form of Eqn.~\ref{eq:anisoform} both await further tests. 
Moreover, the incompleteness of available data and measurement errors can both affect the results. There have been studies which 
applied such phase-space distribution functions to simulated galaxies and dark matter halos to test their performances \citep[e.g.][]{
2015MNRAS.453..377W,2018ApJ...865...72E,2009MNRAS.399..812W}, for which we make more detailed discussions in Sec.~\ref{sec:sims}. 
Briefly, Eadie et al. in 2018 \cite{2018ApJ...865...72E} showed that the model has difficulties predicting both the inner and outer 
regions of the true underlying cumulative mass profiles of simulated galaxies, after using tracers at all Galactocentric distances. 
By extending the distribution function to be based on the NFW profile, Wang et al. in 2015 \cite{2015MNRAS.453..377W} demonstrated 
different levels of systematic biases from halo to halo. Deviations from the NFW potential, violations of the steady state assumption and 
invalid functional forms of the distribution function can all be responsible for such biases. On the other hand, a careful examination 
of the goodness of fit may serve as a way to discriminate between different models. As shown by the second paper of Li et al. in 2019 
\cite{li2019constraining}, it is possible to verify or select proper distribution function models for a given observation.

\subsection{distribution function in terms of actions}
\label{sec:DFaction}

Actions are defined as integrals of the generalized momenta along a path of the generalized coordinate $J_i=\int p_i \ud q_i$. 
The components, $J_i$, are linked to separable potentials. Each momentum, $p_i$, is a function of one coordinate, $q_i$, plus 
three integrals of motion, which are introduced through the procedure of separation upon solving the Hamilton-Jacobian equation. 
In a spherical potential, for example, the radial action is defined as

\begin{equation}
 J_r=\frac{1}{\pi}\int_{r_p}^{r_a} \ud r \sqrt{2E-2\Phi-\frac{L^2}{r^2}}, 
 \label{eqn:Jrspherical}
\end{equation}

where $r_p$ and $r_a$ are the radii at pericenter and apocenter. The other two actions in a spherical potential can be 
chosen as $J_\phi=L_z$ and $L-L_z$, i.e., the component of the angular momentum along the $z$-axis and the difference 
between the magnitude of angular momentum and its $z$-component. 

The spherical isochrone potential \citep[e.g.][]{1959AnAp...22..126H,1990MNRAS.244..111E} is a special case that have 
analytical solutions to Eqn.~\ref{eqn:Jrspherical}. The triaxial St{\"a}ckel potentials \citep{1985MNRAS.216..273D} are 
the most general separable potentials, allowing exact evaluations of actions using a single quadrature. For all the other 
potentials, actions have to be evaluated numerically. 

Due to the difficulties in computing actions, these quantities were not very commonly used in stellar dynamics to constrain 
the MW potential model in the past. In spite of the difficulties, the usage of actions is still very appealing, mainly because 
of two reasons. On one hand, they are adiabatically invariant, meaning that they are conserved quantities in a slowly varying 
potential. On the other hand, when combined with canonically conjugate angles, they can form a complete system of canonical 
coordinates. 

The actions and their canonically conjugate angles have very useful properties. For example, finite triple of ($J_r,J_\phi,J_z$) 
with $J_r>0$ and $J_z>0$ defines a bound orbit. Along any orbit, the canonically conjugate angle, $\theta_i$, increases linearly 
with time at a fixed rate of $\Omega_i(\textbf{J})$, i.e., $\theta_i(t)=\theta_i(0)+\Omega_i(\textbf{J})t$. The phase-space volume 
can be expressed as $(2\pi)^3 \ud^3 \textbf{J}$, and the phase-space coordinates, $\textbf{x}$, are periodic functions of 
$\boldsymbol{\theta}$ in the manner of $\textbf{x}(\boldsymbol{\theta}+2\pi k, \textbf{J})=\textbf{x}(\boldsymbol{\theta},\textbf{J})$. 

More recently, a series of methods have been proposed for fast evaluations of actions in different potentials. The methods include 
the cylindrical adiabatic approximations \citep{2010MNRAS.401.2318B} and refinements \citep{2011MNRAS.413.1889B,2012MNRAS.419.1546S}, 
the st{\"a}ckel fudge \citep{2012MNRAS.426.1324B,2015MNRAS.447.2479S}, and the local st{\"a}ckel fitting \citep{2012MNRAS.426..128S}. 
More details about these and other methods are available in a review paper of \cite{2016MNRAS.457.2107S}. Softwares which can be used 
to calculate and test actions, to compute phase-space positions and velocities given actions and to infer the underlying potential 
given observed positions and velocities of tracers have been released as well \citep[e.g.][]{2015ApJS..216...29B,2015ascl.soft12020S,
2016MNRAS.456.1982B,2016ApJ...830...97T,2018ascl.soft05008V}. 

While evaluating actions is becoming more feasible, accurate and efficient, certain functional forms have to be proposed for the distribution 
of actions, in order to enable further applications to studies of Galactic disks and stellar halos. In 2010, Binney \cite{2010MNRAS.401.2318B} 
discussed analytical distribution functions for the Galactic disk, and the slightly refined form discussed in a follow-up paper by 
Binney ad McMillan \cite{2011MNRAS.413.1889B} is

\begin{equation}
 f(J_r,L_z,J_z)=f_{\sigma_r}(J_r,L_z)\times \frac{\nu_z}{2\pi\sigma_z^2}e^{-\nu_zJ_z/\sigma_z^2}.
\end{equation}

$f_{\sigma_r}(J_r,L_z)$ models the motion parallel to the disk

\begin{equation}
 f_{\sigma_r}(J_r,L_z)=\frac{\Omega \Sigma}{\pi\sigma_r^2 \kappa}|_{R_c} [1+\mathrm{tanh}(L_z/L_o)]e^{-\kappa J_r/\sigma_r^2}.
\end{equation}

$\Omega(L_z)$, which depends on $L_z$, is the circular frequency for the angular momentum $L_z$. $\kappa(L_z)$, which also depends on $L_z$, 
is the radial epicycle frequency, and $\nu(L_z)$ is its vertical counterpart. $\Sigma(L_z)$, which can be parametrized as $\Sigma_0e^{-(R-R_c)/R_d}$, 
is the radial surface density profile of the disk, with $R_c(L_z)$ the radius of the circular orbit with angular momentum $L_z$. $R_d$ is the 
scale length of the disk at the solar radius. The factor $\mathrm{tanh}(L_z/L_0)$ is a chosen odd function for rotation, where $L_0$ is a 
constant that determines the steepness of the rotation curve in the central region of solid-body rotation. On large radius where $r\times 
v_c$ is much larger than $L_0$, $1+\mathrm{tanh}(L_z/L_0)$ simply eliminates the contribution from counter-rotating stars. $\sigma_z(L_z)$ 
and $\sigma_r(L_z)$ are the vertical and radial velocity dispersions, which can be parametrized as 

\begin{equation}
 \sigma_r(L_z)=\sigma_{r0}e^{q(R_0-R_c)/R_d}
\end{equation}
and

\begin{equation}
 \sigma_z(L_z)=\sigma_{z0}e^{q(R_0-R_c)/R_d}. 
\end{equation}


This was obtained by assuming that the velocity dispersions decline exponentially in radius with a scale length roughly twice that of the 
surface density, so $q\sim 0.5$. Binney \cite{2010MNRAS.401.2318B} also discussed the integrated distribution function for stars with 
different ages in the thin disk, while for the thick disk, a single population can be assumed. The distribution function of the Galactic disk 
had been extended to have an analytical dependence on the metallicity of stars \citep{2015MNRAS.449.3479S}. Moreover, perturbed distribution 
functions by spiral arms were discussed by other studies as well \citep{2016MNRAS.457.2569M,2017ApJ...839...61T}.

The above distribution function was applied to 16,269 G-type dwarf stars from SEGUE by Bovy and Rix in 2013 \cite{2013ApJ...779..115B} to 
successfully infer quantities such as the mass of the disk, the total local surface density and the shape of the radial profile of dark 
matter halo within 12~kpc from the Galactic center. Using $\sim$200,000 giant stars from the RAVE survey, Piffl et al. in 2014 
\cite{2014MNRAS.445.3133P} constrained the vertical density profile within $\sim$1.5~kpc to the Galactic plane. The analytical distribution 
function was also used to fit the kinematics of RAVE stars and predict the vertical profile, which showed very good agreement with the 
observed profiles. Their results suggest that the chosen functional form of the distribution function is capable of approximating the truth. 

To constrain the mass distribution of our MW out to large distances, analytical distribution functions in terms of actions for the stellar 
halo with double power-law tracer density profile are required. Both Posti et al. in 2015 \cite{2015MNRAS.447.3060P} and Williams and Evans in 
2015 \cite{2015MNRAS.448.1360W} had discussed the action distribution function for double power-law tracer density profiles. The discussions were 
based on choosing a certain functional form for the action distribution, which can be reduced to the expected behavior of the distribution 
at the small and large scale regimes. In the following, we briefly introduce how the function is chosen based on the deductions made by 
\cite{2015MNRAS.447.3060P}. 

A family of models for double power law density profiles is

\begin{equation}
 \rho_\ast(r)=\frac{\rho_0}{(r/r_b)^\alpha(1+r/r_b)^{\beta-\alpha}},
\end{equation}

where $\alpha$ and $\beta$ are the two power-law indexes, and $r_b$ is the breaking radius \citep{2008gady.book.....B}. 

When $r\ll r_b$, the enclosed mass is $M(r)\propto r^{3-\alpha}$. Hence $\frac{\ud \Phi}{\ud r}\propto r^{1-\alpha}$ or 
$\Delta \Phi(r)\propto r^{2-\alpha}$. On the other hand, it can be proved that for a power-law potential $\Phi(r)\propto r^a$ ($a=2-\alpha$), 
once the position is scaled by a factor of $x'=\xi x$, the energy and action are scaled in ways of $E'=\xi^a E$ and 
$\textbf{J}'=\xi^{1+a/2}\textbf{J}$ respectively. Hence one can figure out that the Hamiltonian should be of the form 
$H(\textbf{J})=[h(\textbf{J})]^{a/(1+a/2)}$, where $h(\textbf{J})$ is some homogeneous function of $h(\xi \textbf{J})=\xi h(\textbf{J})$. 
Referring to the Poisson equation, one can have $\rho_\ast\propto|\Phi|^{1-2/a}$ for a power-law potential. From the Eddington 
formula (Eqn.~\ref{eq:isoform}), it is not difficult to derive that $f(E)\propto E^{-(4+a)/2a}$. Considering $H(\textbf{J})=[h(\textbf{J})]^{a/(1+a/2)}$ 
and set back $a=2-\alpha$, one can have the behavior of the double power-law distribution function on very small 
scales as $f(\textbf{J})=[h(\textbf{J})]^{-(6-\alpha)/(4-\alpha)}$. 

When $r$ approaches to infinity, the potential is Keplerian, i.e., $\Phi(r)\propto r^{-1}$ and $\rho_\ast\propto|\Phi|^\beta$. 
The Hamiltonian takes the form of $H(\textbf{J})=[g(\textbf{J})]^{-2}$ (see \cite{2008gady.book.....B} for more details), 
where $g(\textbf{J})$ is some homogeneous function. Referring to the Eddington's formula (Eqn.~\ref{eq:isoform}) with 
$\rho_\ast\propto|\Phi|^\beta$, one can have $f(E)\propto E^{\beta-3/2}$. Substituting $H(\textbf{J})=[g(\textbf{J})]^{-2}$ 
into $f(E)\propto E^{\beta-3/2}$, the behavior of the double-power law distribution function at infinity is 
$f(\textbf{J})=[g(\textbf{J})]^{-2\beta+3}$. 

\cite{2015MNRAS.447.3060P} proposed the functional form to connect the two limiting behaviors above as 

\begin{equation}
 f(\textbf{J})=\frac{M_0}{J_0^{3}}\frac{[1+J_0/h(\textbf{J})]^{(6-\alpha)/(4-\alpha)}}{[1+g[\textbf{J}]/J_0]^{2\beta-3}}.
\end{equation}

Details about the choices of homogeneous functions of $h(\textbf{J})$ and $g(\textbf{J})$ can be found in \cite{2014MNRAS.442.1405W}, 
\cite{2015MNRAS.447.3060P} and \cite{2015MNRAS.448.1360W}. Das and Binney in 2016 \cite{2016MNRAS.460.1725D} have extended such 
distribution function to include metallicity dependence. Moreover, Binney and Wong in 2017 \cite{2017MNRAS.467.2446B} have adopted 
the above disk+halo distribution functions in action space to model globular clusters in our MW. 

Using the action distribution function based on double power-law density profiles developed by \citep{2015MNRAS.448.1360W} 
and adopting a much simpler power-law potential, Williams and Evans in 2015 \cite{2015MNRAS.454..698W} constrained the enclosed 
mass within 50~kpc of our MW to be $\sim4.5\times10^{11}\msun$, based on about 4,000 BHB stars from SDSS. Velocity anisotropy 
was constrained to be $\beta\sim 0.4$ at $\sim$15~kpc and $\beta\sim 0.7$ at $\sim$60~kpc. 

Following the approach of Binney and Wong in 2017 \cite{2017MNRAS.467.2446B} and using the distribution functions for Galactic 
disk and the outer stellar halo described above, Posti et al. in 2019 \cite{2019A&A...621A..56P} constrained the mass of our MW  
through the recently estimated proper motions of 75 globular clusters from \textit{Gaia} DR2 and 16 other globular clusters from 
\textit{HST} \citep{2018ApJ...862...52S}. 52 globular clusters without proper motions from \cite{1996AJ....112.1487H} were also 
used in their analysis. In addition to the adopted mass-concentration relation to fix the concentration parameter, 
Posti et al. \cite{2019A&A...621A..56P} also fixed the parameters for the Galactic disk and bulge to observational constraints 
made by Piffl et al. in 2014 \cite{2014MNRAS.445.3133P}. Part of the parameters in their modeling of radial and vertical velocity 
dispersions were fixed as well. The double power-law distribution function of the halo was simplified by fixing it to have a 
constant density core in phase space, and the halo was allowed to be prolated. Their results are consistent with a constant 
and slight radially biased halo of $\beta\sim 0.20\pm 0.07$. The masses for the Galaxy and dark matter within 20~kpc were 
constrained to be $1.91_{-0.15}^{+0.17}\times 10^{11}\msun$ and $1.37_{-0.11}^{+0.12}\times 10^{11}\msun$, respectively. 
The virial mass was extrapolated to be $M_{200}=1.1\pm0.3 \times 10^{12}\msun$.

Based on the kinematics of member stars from \textit{Gaia} DR2, Vasiliev in 2019 \cite{2019MNRAS.484.2832V} derived proper motions for 
150 globular clusters in our MW. Similar to the previous studies of Binney and Wong in 2017 \cite{2017MNRAS.467.2446B} and Posti et al. 
in 2019 \cite{2019A&A...621A..56P}, Vasiliev \cite{2019MNRAS.484.2832V} adopted the action distribution function to constrain the mass 
distribution, and found a spherical halo is preferred, with the mass enclosed inside 50~kpc and 100~kpc being $5.4_{-0.8}^{+1.1}\times 
10^{11}\msun$ and $8.5_{-2.0}^{+3.3}\times 10^{11}\msun$, respectively. The virial mass was extrapolated as $M_{200}=1.0_{-0.5}^{+1.5}
\times10^{12}\msun$. In the end, we note that both \cite{2019A&A...621A..56P} and \cite{2019MNRAS.484.2832V} have modeled measurement 
uncertainties, and have approximated the selection function of globular clusters to be complete. 

Very recently, Callingham et al. in 2020 \cite{2020arXiv200107742C} investigated how the contraction of dark matter halos caused by the 
accumulation of baryons in the central regions can affect the action distributions. They have developed an iterative algorithm 
to contract dark matter halos, and have used this algorithm to predict the density and velocity distribution of the MW's contracted 
dark matter halo.

\subsection{Template-based distribution functions}
\label{sec:DFsim}
So far we have introduced the method of fitting a given model distribution function to the observed positions and velocities of dynamical 
tracers such as halo stars, satellite galaxies and globular clusters. However, this approach strongly depends on whether the adopted functional 
form of the model distribution is realistic or not \citep{2015MNRAS.453..377W}. In this subsection, we further introduce distribution 
functions that are generalized from simulation templates. Because the templates as well as their universality are extracted empirically from 
simulations, these methods fall in between dynamics based distribution function method and satellite phenomenology that we will introduce in 
section~\ref{sec:SatPheno}.  

The phase-space distribution of tracers can be easily extracted from simulations. However, to make these distribution functions useful for 
dynamical inference, proper parametrization of these distribution functions are needed. It is well-known that the halo density profile can 
be well described by the universal NFW function parametrized by a scale density, $\rho_s$, and a scale radius, $r_s$, or equivalently a mass 
and a concentration parameter. Given this universality, it is natural to expect that the full phase-space distribution may also be universal,
once the phase-space coordinates are scaled by appropriate combinations of NFW parameters. That is~\cite{2019ApJ...886...69L}
\begin{equation}
    f(\bm{r},\bm{v})=\frac{N_{\rm tot}}{r^3_s v^3_s} \tilde{f}(\tilde{\bm{r}}, \tilde{\bm{v}}),
\end{equation} where the probability density in $(\tilde{\bm{r}}, \tilde{\bm{v}})$ space, $\tilde{f}(\tilde{\bm{r}}, \tilde{\bm{v}})\equiv d^6P/d^3\tilde{x}d^3\tilde{v}$, 
is approximately the same for any halo. Here we have defined $\tilde{r}={r}/r_s$ and $\tilde{v}={v}/v_s$, with $v_s=\sqrt{G\rho_s r_s^2}$. $\tilde{f}(\tilde{\bm{x}}, 
\tilde{\bm{v}})$ can be extracted numerically as a template distribution function. Once this is done, the full distribution function is known for any halo parameters, and 
can be fit against data to obtain best-fit halo parameters. Note the universality of this distribution function over a wide range of halo parameters remain to be explicitly 
tested. However, it is believed that the universality should at least hold locally over a small range of halo parameters.

In 2008, Wojtak et al. \cite{2008MNRAS.388..815W} first studied such a rescaled distribution function of dark matter particles in simulated cluster halos. Instead of 
working in $(\bm{r}, \bm{v})$ space, they study the distribution function as a function of the rescaled energy, $\tilde{E}=E/v_s^2$, and rescaled angular momentum, 
$\tilde{L}=L/r_s v_s$, because Jeans theorem implies the distribution function can be expressed as functions of these integrals of motion. 

In 2017, Li et al. \cite{2017ApJ...850..116L} first exploited the idea of using a template distribution function for satellite galaxies to infer the MW halo mass. 
In this pioneering work, the template is also built in energy and angular momentum space. However, instead of using a full distribution function, they built their 
template as well as the likelihood function using the distribution of energy and angular momentum parameters, $p(E,L)=\ud^2 P/\ud E \ud L$. As $E$ and $L$ are not 
direct observables, the likelihood function in $(E,L)$ space leads to a biased halo mass estimator. In addition, their distribution function is parametrized with a 
single halo mass parameter instead of two NFW parameters, so that the halo concentration cannot be constrained. A halo-to-halo scatter is also found beside the 
overall bias mentioned before, which is attributed to the variation in the distribution function due to different halo formation histories.  

Callingham et al. in 2019 \cite{2019MNRAS.484.5453C} adapted the Li et al. \cite{2017ApJ...850..116L} method and applied it to classical satellites in our 
MW. MW-like galaxies selected from the cosmological hydrodynamical \textsc{Eagle} simulation \citep{2015MNRAS.446..521S,2015MNRAS.450.1937C} are used to build a 
template distribution function and to calibrate the estimator bias. An independent set of halos from the \textsc{auriga} simulations are used to further test the 
method and the bias calibration. Applied to the observed classical satellites, the virial mass of our MW was found to be $M_{200}=1.17_{-0.15}^{+0.21}\times 10^{12}\msun$. 
Combined with independent measurements from other studies, which provided the enclosed mass within a smaller radius, the halo concentration was estimated as $c_{200}=10.9_{-2.0}^{+2.6}$. 

Very recently, Li et al. in 2019 \citep{2019ApJ...886...69L} improved their method published in 2017 \citep{2017ApJ...850..116L}. Starting from a distribution of 
orbital parameters, the complete probability distribution function in $(r,E,L)$ space can be derived as
\begin{equation}
    p(r, E, L)=p(r|E,L)p(E,L).
\end{equation} The second part of this distribution can be obtained from the template distribution function, $p(E,L)=p(\tilde{E},\tilde{L})\ud \tilde{E}\ud\tilde{L}/\ud 
E \ud L$. For a steady-state distribution function, the distribution along each orbit, $p(r|E,L)=\ud P(r|E,L)/\ud r$ is given by Equation~\eqref{eqn:oPDF} (see section~\ref{sec:DForbit}). 
The distribution function in $(\bm{r},\bm{v})$ space is related to $p(r,E,L)$ by a coordinate transformation, 
\begin{equation}
    p({\bm{r}},{\bm{v}})=\frac{|v_r|}{8\pi^2L}p(r,E,L),\label{eqn:DFconvert}
\end{equation}
where $p(\bm{r},\bm{v})=f(\bm{r},\bm{v})/N_{\rm tot}$ is the normalized (or probability) distribution function. 

Once the model distribution function given by Equation~\eqref{eqn:DFconvert} is obtained, a likelihood estimator can be constructed given the observed $(\bm{r},\bm{v})$ 
of each tracer. This is a proper likelihood to use compared to those used in \cite{2017ApJ...850..116L} and \cite{2019MNRAS.484.5453C}, and thus free from the 
systematic bias due to improper likelihood function. In a follow-up paper, Li et al. \cite{li2019constraining} applied this new estimator to a sample of 28 satellite galaxies
between 40 and 300 kpc, with proper motions taken from \textsc{Gaia} DR2. Using a template distribution extracted from the \textsc{Eagle} simulation, the Milky Way 
halo mass was best constrained to be $M_{200}=1.23_{-0.18}^{+0.21}\times 10^{12}\msun$, and the concentration was constrained as $c_{200}=9.4_{-2.1}^{ +2.8}$.
 Combined with the rotation curve measured by halo stars, tighter constraints were given as $ M_{200}=1.26_{-0.15}^{+0.17} \times 10^{12}\msun$ and
  $c_{200}=10.4_{-1.9}^{+2.3}$. Using multiple tracer populations is thus very helpful to better infer the halo concentration. Dependencies on the adopted templates 
  are also discussed in \cite{li2019constraining}.

\subsection{Free-form distribution functions}
\label{sec:DForbit}


In this subsection, we briefly introduce a set of methods with more general assumptions about the distributions. These methods
generally do not assume a fixed functional form of the distribution function, but rather allow for a very flexible
distribution function to be constrained by the data itself. 

The starting point of these methods is the steady-state assumption. If a system is in a steady-state, then phase-space continuity (i.e., the collisionless Boltzmann equation) 
implies that the distribution of particles along each orbit is determined by the travel time distribution on the orbit~\citep{2016MNRAS.456.1003H}, i.e., 
\begin{equation}
    \ud P(x|\mathrm{orbit}) \propto \ud t(x).\label{eqn:oPDF}
\end{equation} 
Han et al. in 2016 \cite{2016MNRAS.456.1003H} also explicitly showed that the above equation is equivalent to the Jeans theorem. With this conditional distribution along each orbit, the construction 
of a full distribution function still needs to specify the distribution of orbits, which can be constrained by the data itself while fitting for the underlying potential. 

The most classical method of this family is perhaps the Schwarzschild method~\citep{1979ApJ...232..236S,2018MNRAS.473.3000Z}, which works by parameterizing the distribution of orbits 
with histograms in orbit space. The number of orbits at each grid point in the orbital parameter space (or the orbit library), is left as a free parameter. For each orbit, the 
distribution along the orbit can be computed by Equation~\eqref{eqn:oPDF} once a potential model is assumed. These combined then predict the phase-space distribution, which can be 
compared against the observed distributions to solve for the distribution of orbits as well as the best-fit potential. As this method numerically builds a distribution function that 
is binned in orbit space, it can work for any potential and for incomplete phase-space data. On the other hand, as the number of parameters (including the gridded orbit counts) is 
large and the orbits need to be integrated numerically, this method is usually computationally expensive.

In order to build smoothly varying histograms of orbits, Bovy et al. (2010)~\cite{2010ApJ...711.1157B} proposed to model the histogram with a Gaussian process with some hyper 
parameters that are further marginalized during the inference. They applied their method to infer the potential of the Solar system using planets as tracers. Magorrian in 
2014~\cite{2014MNRAS.437.2230M} proposed to model the distribution of orbits with an arbitrary number of Gaussians in action space, and then marginalizing over the proposed 
prior distribution of the parameters of the Gaussians. Because some assumptions on the form of the distribution and on the priors are still needed, these methods are still 
not fully assumption free. They exist as a trade-off among model flexibility, model smoothness and computational efficiency. 

In fact, if the full phase-space coordinates of particles are available, the distribution of orbits can be specified by the data itself instead of being proposed with a library. 
This is because once a model potential is assumed, the observed phase-space coordinates of each particle then fully specify its orbital parameters. This is essentially the key
difference between the orbital Probability Density Function (oPDF) method proposed in Han et al. in 2016 \cite{2016MNRAS.456.1003H} and the Schwarzschild method. The use of the
data-inferred distribution of orbits in place of a gridded library significantly simplifies the inference of the potential, at the cost of losing the flexibility to handle missing
dimensions in the data. This method has been used to study the dynamical state of simulated MW halos in \cite{2016MNRAS.456.1017H,2017MNRAS.470.2351W,Han2019IAU}. Their analyses 
have revealed a stochastic scatter in the best-fit mass and concentration parameters, which can be as large as a factor of 2 when halo stars are used as tracers. These have been 
interpreted as being caused by phase-correlations in the tracer particles, which violate the steady-state assumption of the model. We give more discussions on such biases in section~\ref{sec:sims}. 
Note that this stochastic bias undermines the precision of any steady-state method. As \cite{2018MNRAS.476.5669W} explicitly demonstrated, the SJE, which is a completely different 
steady-state method, exhibits a very similar bias when tested on the same set of simulated halos. While the stochastic bias using stars is large, the bias using dark matter 
particles is much smaller, $\sigma_M\sim 20\%$. It is also shown that satellite galaxies have a dynamical state close to dark matter particles, and are thus expected to be better 
dynamical tracers than halo stars~\citep{Han2019IAU}.  

When the spatial coordinate of particles are specified by the action angles, Equation~\eqref{eqn:oPDF} translates to a uniform distribution in angle, as the angles evolve uniformly 
in time. For each assumed potential, one can convert the spatial coordinates of each particle to action angles. The true potential can be found as one that reproduces a uniform 
angle distribution. In practice, this is achieved by minimizing some distances between the converted and the expected distributions. Beloborodov and Levin in 2004 \cite{2004ApJ...613..224B} 
first proposed two such minimum distance estimators. However, as shown in \cite{2016MNRAS.456.1003H}, these estimators are usually less efficient than likelihood estimators such 
as the oPDF and suffer from strong degeneracies between the halo mass and concentration parameters.

In the end, we note that none of the above methods in this sub-section have been applied to real data of the MW.

\section{Modeling the stripping and evolution of tidal debries: stellar streams}
\label{sec:stream}

As mentioned in Sec.~\ref{sec:intro}, stellar streams are formed by stripped stars from satellite galaxies or from globular clusters
through tidal forces. These tidally formed stellar streams (tidal streams or debries) contain a wealth of information on structure 
formation, galaxy evolution, dynamics of progenitor satellites and the underlying potential. 

Early studies of tidal debris in our MW and nearby galaxies used photometric plates \citep[e.g.][]{1995AJ....109.2553G,
1997A&A...320..776L,2000A&A...359..907L,2003AJ....126..815L}. However, tidal streams in the MW can extend tens of degrees 
over the sky, and thus surveys covering large areas are crucial for detecting them. Nowadays, deep and large sky surveys 
have enabled detections of tidal streams in both our MW and nearby galaxies. For example, the Sloan Digital Sky Survey 
\citep[][SDSS]{2000AJ....120.1579Y} has enabled a numerous number of detections of tidal streams in our MW, which 
are either associated with known globular clusters, satellite galaxies or without obvious associations \citep[e.g.][]{
2001ApJ...548L.165O,2003AJ....126.2385O,2006ApJ...639L..17G,2006ApJ...641L..37G,2006ApJ...643L..17G,2006ApJ...645L..37G,
2006ApJ...637L..29B,2007ApJ...658..337B,2009ApJ...693.1118G,2011MNRAS.416..393B,2012ApJ...760L...6B,
2013ApJ...769L..23G,2014ApJ...790L..10G}. 

More and more imaging surveys have added to the growing list of detected tidal structures in our MW, including the study of Koposov 
et al. in 2014 \cite{2014MNRAS.442L..85K} using the VST ATLAS survey \citep[][]{2013Msngr.154...38S}, the findings by Martin 
et al. in 2014 \cite{2014ApJ...787...19M} based on the Pan-Andromeda Archaelogical Survey \citep[][PAndAS]{2009Natur.461...66M}, 
Grillmair et al. in 2013 \cite{2013ApJ...769L..23G} using data from the 2MASS Point Source Catalog \citep{2006AJ....131.1163S} 
and the Wide Field Infrared Survey Explorer (WISE) All-Sky Release \citep{2010AJ....140.1868W}, Bernard et al. in 2014
\cite{2014MNRAS.443L..84B} based on the Pan-STARRS1 \citep[PS1;][]{2010SPIE.7733E..0EK} 3$\pi$ survey, Shipp et al. in 2018
\cite{2018ApJ...862..114S} with the Dark Energy Survey \citep[DES;][]{2011AJ....141..185R}, and some other studies 
\citep[e.g.][]{2015MNRAS.454.3613S,2017ApJ...840L..25M}. Streams and tidal features are also commonly detected in nearby 
galaxies thanks to deep photometry \citep[e.g.][]{2009AJ....138.1417T, 2010AJ....140..962M,2019arXiv191108497K}.


Almost all of the above detections were made from photometric data, and only few combined radial velocities from 
spectrocopic data and proper motions. With available velocity information, in fact some tidal streams were either 
solely or partly detected in velocity space \citep{1999Natur.402...53H,2007AJ....134.1579K,2009ApJ...700L..61N,
2011ApJ...728..102W,2012ApJ...753..145C,2013ApJ...776...26S,2015ApJ...803...56S,2018MNRAS.475.1537M}. There were also 
efforts of looking for debris and substructures in action space \citep[e.g.][]{2018ApJ...856L..26M,2018MNRAS.478.5449M}
or through machine learning approaches \citep[e.g.][]{2018ApJ...863...26Y,2019arXiv191007538Y}.

With on-going and up-coming spectroscopic surveys (see those mentioned in the introduction), increased proper motion 
data from \textit{Gaia}, and even deeper imaging surveys in the future such as the Large Synoptic Survey Telescope 
\citep[LSST;][]{2008arXiv0805.2366I}, we expect growing observations of tidal debris and increasing kinematical data 
of resolved stars with 3-dimensional velocities associated to tidal streams. Because tidal streams can extend over 
large distances, their dynamics are sensitive to both the depth and shape of the Galactic gravitational potential 
\citep[e.g.][]{2004ApJ...610L..97H,2005ApJ...619..800J,2012MNRAS.424L..16L,2004ApJ...601..242M,2016ApJ...833...31B}, 
and it was proposed that dark matter substructures can induce localized fluctuations and gaps along such long streams, 
which can be used to detect dark matter subhalos and dark streams \citep[e.g.][]{2009ApJ...705L.223C,2011ApJ...731...58Y,
2016MNRAS.463..102E,2016MNRAS.457.3817S,2016PhRvL.116l1301B,2017MNRAS.466..628B,2018arXiv180406854B}. Despite the richness 
of data and the valuable dynamical information, most of the studies on tidal streams were theoretical \citep[e.g.][]{
2009MNRAS.400..548E,2009MNRAS.399L.160E,2011MNRAS.417..198V,2016PhRvL.116l1301B}, or qualitative and empirical \citep{
2006ApJ...642L.137B,2006ApJ...651..167F,2012MNRAS.424L..16L}, or based on numerical simulations without contaminations 
and errors \citep[e.g.][]{2014ApJ...794....4P,2014ApJ...795...94B,2014MNRAS.439.2678D,2017ApJ...836..234S,2018ApJ...867..101B,
2019MNRAS.487.2718M,2019MNRAS.483.1427L}. 

A number of methods have been developed to model observed tidal streams. These include orbit fitting 
\citep[e.g.][]{2010ApJ...712..260K,2010ApJ...711...32N,2013ApJ...773L...4V} to the tidal stream and the remnant of the 
progenitor, if the progenitor still survives and the association can be identified, N-body simulations \citep{
2005ApJ...619..807L,2010ApJ...714..229L,2017ApJ...836...92D}, approaches of particle releasing/spraying \citep{
2012MNRAS.420.2700K,2014ApJ...795...94B,2014MNRAS.445.3788G,2015ApJ...803...80K}, semi-analytic approaches \citep{
2017ApJ...836...92D,2017ApJ...847...42D} and action angle distribution of tidal debries \citep[e.g.][]{1999Natur.402...53H,
2013MNRAS.433.1826S,2014MNRAS.443..423S,2014ApJ...795...95B}. Relatively fewer studies had specifically constrained the mass 
of our MW, among which only two measurements are selected into Fig.~\ref{fig:massplot} that provided virial mass estimates 
with statistical errors. 

Early in 1995, Lin et al. \cite{1995ApJ...439..652L} made orbit modeling of the observed distances and motions of the 
Magellanic clouds and segments along the Magellanic stream\footnote{Later studies based on proper motions of Magellanic 
Clouds have reported that the stream might be formed through local interactions between large and small Magellanic Clouds. 
See Sec.~\ref{sec:MC} for details.}, and through their modeling the mass within 100~kpc of our MW was estimated to be 
$5.5\pm 1.0 \times 10^{11}\msun$. 

Koposov et al. in 2010 \cite{2010ApJ...712..260K} made orbit fitting to the 6-dimensional phase-space map of the thin 
but extended (60 degrees) GD-1 stream, and placed strong constraints on the local circular velocity (221$\pm$18~km/s) at 
the solar orbital radius. More recently, with new data from \textit{Gaia}, SEGUE and LAMOST for the GD-1 stream, Malhan and 
Ibata in 2019 \cite{2019MNRAS.486.2995M} constrained the local circular velocity to be $244\pm 4 $~km/s, and the mass within 
20~kpc to the Galactic center was estimated as $2.5\pm 0.2 \times 10^{11}\msun$. 
 
Through orbit fitting to BHB stars in the so-called ``Orphan stream'' discovered by Grillmair \cite{2006ApJ...645L..37G} and Belokurov 
et al. \cite{2006ApJ...642L.137B} in 2006, which spans about 60 degree over the sky, the total mass within 60~kpc of our MW 
was estimated to be $\sim$2.7$\times 10^{11}\msun$ by Newberg et al. in 2010 \cite{2010ApJ...711...32N}. The mass out to 240~kpc 
was extrapolated to be $\sim6.9\times 10^{11}\msun$ assuming a log potential. More recently, Hendel et al. in 2018 \cite{2018MNRAS.479..570H} 
conducted orbit fitting to RR Lyrae stars in the Orphan stream, and an upper limit on the MW mass enclosed within 60~kpc was 
constrained to be $5.6_{-1.1}^{+1.2}\times 10^{11}\msun$.

It is often assumed that the orbit of the progenitor is traced by the stream and the motions of stripped stars are all aligned 
with the stream track. The assumptions might not be strictly valid \citep[e.g.][]{2009MNRAS.400..548E,2019MNRAS.485.4726K}.
In addition, it was pointed out by Lux et al. \cite{2013MNRAS.436.2386L} and by Sanders and Binney \cite{2013MNRAS.433.1813S} in 2013 
that a single orbit fitting to the observed dynamics of a tidal stream may lead to significant biases. Thus, realistic modelings of 
not only the orbit of the progenitor, but how stars along the stream are stripped and evolved are necessary.

N-body simulations are powerful tools to model and understand the formation histories of tidal streams and their progenitors. 
Comparing M giant stars along the Sagittarius stream and with N-body simulations and test particle orbits, Law et al. 
in 2005 \cite{2005ApJ...619..807L} constrained the total mass within 50~kpc of our MW to be in the range of $\sim$3.8 to 
5.6$\times 10^{11}\msun$. However, with N-body simulations, it is very expensive to properly explore the parameter 
space and obtain a best-fit model potential with a robust confidence region. As a result, the number of studies relying 
on N-body simulations to explore the parameter space is very limited at the current stage. 

Gibbons et al. in 2014 \cite{2014MNRAS.445.3788G}, Bowden et al. in 2015 \cite{2015MNRAS.449.1391B} 
and K{\"u}pper et al. in 2015 \cite{2015ApJ...803...80K} subsequently proposed less expensive approaches of generating tidal 
streams, which involves steadily releasing particles through the two Lagrangian points of the progenitor and evolving the released 
particles within given potential models. The approach is less expensive compared with standard N-body simulations. The one proposed 
by K{\"u}pper et al. in 2015 \cite{2015ApJ...803...80K} was called the \textsc{streakline}. The initial velocities of particles can 
be modeled through the velocity of the progenitor, modulated to match the instantaneous angular velocity of the object center with 
respect to the galactic center, plus some scatters \citep{2012MNRAS.420.2700K,2014ApJ...795...94B}. Particles released from the two 
Lagrangian points formed the leading and trailing arms of the stream. In particular, K{\"u}pper et al. \cite{2015ApJ...803...80K} 
chose to ignore the scatter and fit the coldest model stream to observed density peaks along the Palomar 5 stream, with its 
trailing stream extending 23.2 degrees and leading arm cut off by the survey edge. The virial mass was found to be $M_{200}=
1.69\pm0.42\times 10^{12}\msun$. Within the apocenter of Palomar 5 ($\sim$19~kpc), the enclosed mass of the Galaxy (disk+bulge)
was estimated to be $2.14_{-0.35}^{+0.38}\times 10^{11} \msun$. The circular velocity at the solar radius was constrained as 
$253\pm16$~km/s.

The method proposed by Gibbons et al. in 2014 \cite{2014MNRAS.445.3788G} also relied on releasing 
particles through the Lagrangian points and they applied their method to constrain the mass of our MW as well. They in addition
considered the progenitor's gravity, which was shown to be very crucial in order to bring consistency with direct N-body simulations. 
Applying the method to the famous Sagittarius stream, the total mass within 100~kpc was constrained to be 4.1$\pm$0.7$\times 10^{11}
\msun$. The mass was extrapolated to 200~kpc as 0.56$\pm$0.12$\times 10^{12}\msun$, i.e., a ``light'' MW. 

Recently, it was reported that the stellar motions in the Orphan stream are misaligned with the stream track\citep{2019MNRAS.485.4726K}. 
Based on the Gibbons et al. 2014 method, Erkal et al. in 2019 \cite{2019MNRAS.487.2685E} found the motion-track misalignment can be well 
explained by the LMC perturbation to the MW potential. Jointly fitting an LMC mass and a MW potential, the MW mass within 50~kpc was 
estimated to be $3.8_{-0.11}^{+0.14}\times 10^{11}\msun$. 

Dierickx et al. in 2017 \cite{2017ApJ...847...42D} considered tidal stripping and dynamical friction in their modeling 
of the progenitor, following a semi-analytical approach. Their approach was based on orbit fitting to the satellite remnant and 
the Sagittarius stream, but instead of integrating the current position and velocity of the progenitor back in time, 
Dierickx et al. \cite{2017ApJ...847...42D} modeled the progenitor forward in time and tried a series of initial velocities and 
orbit angles for different masses of the MW and progenitor masses of the stream. It was found that massive MW halos had difficulties 
to reproduce the velocity and distance of the progenitor simultaneously, resulting in an upper limit to the virial mass of our MW as $10^{12}\msun$.


\section{The timing argument and local Hubble flow: the motion of MW and M31}
\label{sec:timing}

In this section we will introduce how to constrain the mass of our MW and the Local Group by modeling the relative motion of MW and M31, 
and the motion of distant satellite galaxies in our Local Group. Measurements in this section fall in the category of ``timing \& LG dyn.''
in Fig.~\ref{fig:massplot}.

\subsection{Timing argument}
M31 is the massive companion of our MW Galaxy. Currently MW and M31 are approaching each other. Our Universe is expanding, 
but gravitational forces can reverse the expansion locally. In our Local Group (hereafter LG), MW and M31 are the two 
dominating galaxies, and their distances to the nearest external bright galaxy are much larger than the separation between 
themselves. The fact that they are approaching each other can thus be used as constraints of the mass associated with them. 

Early in 1995, Kahn and Woltjer \cite{1959ApJ...130..705K} pointed out that galaxies were at zero separation at the Big Bang, 
and thus they must have passed through apocenters at least once in order to be approaching each other today. This requires that 
the apocentric distance of the orbit must be larger than the current separation between them, and half of the ortibal period 
should be smaller than the age of the Universe. These requirements help to provide a lower limit on the total mass of our MW 
and M31. 

There are evidences that the tangential velocity of M31 with respect to our MW is negligible. Ignoring the tangential velocity 
and cosmic expansion and further assuming point mass, the equation of energy conservation along the orbit can be written as

\begin{equation}
 \frac{1}{2}(\frac{\ud r}{\ud t})^2-\frac{GM}{r}=-\frac{1}{2}\frac{GM}{a},
\end{equation}
where $M$ is the total mass of MW and M31, $r$ is the separation, $\frac{\ud r}{\ud t}$ is the relative velocity, $a=GM/(-2E)$ 
and $2a$ is the maximum value of $r$ on the orbit, $E$ is the orbital energy.

The solution to the separation and velocity can be obtained by introducing an angle-like quantity of $\eta$, which is referred 
as the eccentric anomaly 

\begin{align}
\begin{array}{l}
{\rm (i)} :  r=a(1-\cos 2\eta),\\
{\rm (ii)} : \frac{\ud r}{\ud t}=\sqrt{\frac{GM}{a}}\frac{\sin 2\eta}{1-\cos 2\eta}.
\end{array}
\label{eqn:orbits}
\end{align}

$\eta$ is related to time, $t$, through the following equation and can be solved numerically

\begin{equation}
 \eta-\sin 2\eta=(\frac{GM}{a^3})^{1/2}2 t.
\end{equation}

Equation~\ref{eqn:orbits} can be used to constrain the total mass, $M$, for our LG. Given the observed separation and 
velocities of MW and M31, plus the age of our Universe, the solution to the above equations with a single apocentric passage 
gives the lowest limit of the total mass. Such a lower limit of the LG mass was measured to be $5\times10^{12}\msun$ by 
Kahn and Woltjer in 1995 \cite{1959ApJ...130..705K}. 

More recently, using proper motion data from the multi-epoch \textit{HST}/ACS photometry, van der Marel et al. in 2012 
\cite{2012ApJ...753....8V} concluded that the tangential component of the M31 velocity with respect to MW is statistically 
consistent with being negligible, and the M31 orbit towards MW is radial. They revised the LG timing mass to be 
$4.93\pm 1.63 \times10^{12}\msun$.

Similar approaches can be adopted to constrain our MW mass, by considering the system formed by our MW and distant dwarf 
satellite galaxies. The estimated total mass mainly reflects the mass of our MW, as satellites are sub-dominant. Based on 
the MW-Leo I system and assuming radial orbits, Zaritsky et al. in 1989 \cite{1989ApJ...345..759Z} derived a lower mass limit 
of $1.3\times10^{12}\msun$ for our MW. More recently, Li and White in 2008 \cite{2008MNRAS.384.1459L} calibrated the MW 
timing mass, using halo pairs in analogy to MW and Leo I in the cosmological Millennium simulation \citep{2005Natur.435..629S}. 
Basically, the timing approach were applied to halo pairs in the simulation, and the timing mass can be compared with the 
true virial mass in the simulation to quantify the bias. The virial mass of our MW was calibrated to be $M_{200}=2.43\times10^{12}\msun$
with a 95\% lower confidence limit of $0.8\times10^{12}\msun$, which was at the massive end compared with other contemporary 
measurements.

Applying the timing approach to the MW-Leo I system required the boundedness of Leo I. Besides, none-zero tangential velocity 
of Leo I would increase the timing mass. Using two epochs of HSC WCS/WFC observations spanning a five-year time baseline, 
Sohn et al. in 2013 \cite{2013ApJ...768..139S} measured the proper motion of Leo I. The tangential velocity with respect to 
the Galactic center was estimated to be $\sim101.0\pm34.4$~km/s, which is inconsistent with radial orbits. With the observed 
radial and tangential velocities of Leo I, \cite{2013ApJ...768..139S} concluded the boundedness of Leo I to our MW. Solving 
the complete and non-radial timing equation, the orbit of Leo I inferred a virial mass of our MW as $M_{200}=2.65_{-1.36}^{+1.58}\msun$\footnote{
We have converted the original virial mass provided in the paper to $M_{200}$ by dividing a factor of 1.19, which is the 
value provided in their paper, based on the NFW halo profile with concentration of 9.5. Not only observational errors, but 
also the cosmic scatter are included in the errors. \cite{2013ApJ...768..139S} calculated the cosmic scatter based on a 
similar catalog used by the earlier study of Li and White \cite{2008MNRAS.384.1459L}. Note the virial mass was calibrated by 
\cite{2013ApJ...768..139S} against subhalos in numerical simulations having similar tangential velocities to Leo I as well. }

Instead of modeling a single object, Zaritsky et al. in 2019 \cite{2020ApJ...888..114Z} applied the 
timing argument to a sample of 32 stars with Galactocentric distances larger than 60~kpc. The timing mass was calibrated and 
compared with the suite of \textsc{auriga} simulations to obtain a statistical estimate of our MW virial mass in the range of 
$0.91\times 10^{12}\msun<M_{200}<2.13\times 10^{12}\msun$ (90\% confidence).

Benisty et al. in 2019 considered different numbers of past encounters between MW and M31 and tried different gravity 
models \citep{2019arXiv190403153B} in their estimates of the LG mass under the timing framework, though past encounters do not seem to 
be supported by recent {\it Gaia} data \citep[e.g.][]{2019NatAs...3..932G}. 

Very recently, Zhai et al. in 2020 \cite{2020ApJ...890...27Z} looked for MW-31 like systems in numerical simulations, and they 
found that higher tangential velocities correspond to higher total mass and also affect the individual mass distribution of MW and M31 
analogs. The typical host halo mass of MW is $1.5_{-0.7}^{+1.4}\times 10^{12}\msun$ for radial orbits between MW and M31, and 
$2.5_{-1.4}^{+2.2}\times 10^{12}\msun$ for low-ellipticity orbits.

\subsection{The local Hubble flow}
\label{sec:LHF}
The timing approach can also be applied to model the relative motion of nearby galaxies in the local volume towards the LG. 
Assuming the companion galaxies are massless, it can help to constrain the LG mass. Under the timing framework, Pe{\~n}arrubia
et al. in 2014 \cite{2014MNRAS.443.2204P} specifically modeled the dynamics of galaxies in the local volume of an expanding 
universe (Eqn.~\ref{eqn:expand}), using published distances and velocities of nearby galaxies within 3~Mpc where the gravitational 
force reverses the expansion 

\begin{equation}
 \frac{\mathrm{d}^2r}{\mathrm{d} t^2}=-\frac{GM}{r^2}+H_0^2\Omega_\Lambda r.
 \label{eqn:expand}
\end{equation}

M is the total mass of LG. $r$ is defined with respect to the LG barycenter. 

Basically, given the LG barycenter (or mass ratio between MW and M31) and the circular velocity at the solar radius, which were treated as 
model parameters, the distance ($r(t_0)$) and radial velocity ($V(t_0)$) of an observed galaxy to the LG barycenter at current epoch $t_0$ 
can be calculated. Then after choosing a small initial radius of the galaxy, its initial velocity can be solved through Eqn.~\ref{eqn:expand}, 
by requiring that the integrated distance at time $t_0$ agrees with the distance, $r(t_0)$, calculated in the previous step. Note $H_0$, 
$\Omega_\Lambda$ and $M$ are model parameters in this step. In the end, the integrated radial velocity of the galaxy at $t_0$ can be compared 
with $V(t_0)$ through the likelihood function, in order to find the set of best-fit parameters for the circular velocity at the solar radius, 
the LG barycenter, the LG mass, the cosmological constant and the Hubble constant. \cite{2014MNRAS.443.2204P} derived the LG mass to be $2.3\pm 
0.7 \times10^{12}\msun$ and a mass ratio between MW and M31 as $0.54_{-0.17}^{+0.23}$. Hence the virial mass of our MW was estimated as 
$0.8_{-0.3}^{+0.4}\times10^{12}\msun$. In their analysis, both the LG quadrupole and the time variance of the potential were considered 
and discussed, which had negligible effects to the results.


In two follow-up papers, Pe{\~n}arrubia et al. in 2016 \cite{2016MNRAS.456L..54P} and Pe{\~n}arrubia and Fattahi in 2017,
\cite{2017MNRAS.468.1300P}), the effects of the LMC and the fraction of mass outside the virial radius that perturbs the local Hubble flow 
were taken into considerations. The LMC can change the barycenter velocity of nearby galaxies and lead to an updated virial 
mass of our MW as $1.04_{-0.23}^{+0.26}\times 10^{12}\msun$. Moreover, using a set of hydrodynamical simulations of MW-like halos and 
galaxies from the APOSTLE project \citep[A Project of Simulations of The Local Environment;][]{2016MNRAS.457..844F,2016MNRAS.457.1931S}, 
it was found that a relatively large fraction of the mass perturbing the local Hubble flow and driving the relative trajectory of the 
main galaxies is not contained within the halo virial radius. Adopting the outer halo profiles in N-body simulations to calibrate the 
virial mass, it was reported that the mass given by \cite{2016MNRAS.456L..54P} should be divided by a factor of 1.2, to give the actual 
mass within the virial radius, which led to the constraint of $M_{200}=\sim0.87\times10^{12}\msun$.

\subsection{Momentum of MW and M31}
If the LG is sufficiently isolated from nearby galaxy groups and matter distributions, plus the assumption that the LG mass is dominated 
by MW and M31, the total momentum of MW and M31 should be close to zero in the rest frame of the LG. If the velocity vector of MW and M31 
with respect to the LG barycenter can be measured, the mass ratio between MW and M31 can then be further constrained \citep{2014MNRAS.443.1688D}. 
Compared with the timing argument, the orbit does not have to be assumed as radial Keplerian orbits. The velocity of MW and M31 with 
respect to the barycenter of the LG can be decomposed into two components, the Heliocentric velocity of MW (or M31) and the solar motion 
with respect to the LG barycenter. 

The solar motion, as have been described in Sec.~\ref{sec:localobs}, is a combination of the velocity of the LSR and solar peculiar 
velocity with respect to the LSR, and can be measured through the apparent motion of Sgr A* or through modeling the distances and 
velocities of maser sources. The Heliocentric motion of MW is simply the solar motion added with a negative sign. The Heliocentric 
velocity of M31 has be measured through spectroscopic observations for the line-of-sight component \citep{2012AJ....144....4M}, and 
its proper motion has been measured through high-precision astrometry of \textit{HST} \citep{2012ApJ...753....7S}. 

The solar motion with respect to the LG barycenter can be constrained through observations of distant satellite galaxies with a 
Bayesian approach. Basically, the radial velocities of distant satellites are the observed Heliocentric velocities plus the solar 
motion with respect to the LG barycenter. These radial velocities can be assumed to follow a Gaussian distribution such that 
galaxies within the LG move randomly with uncorrelated motions. This can be used to construct the likelihood and constrain the 
solar motion with respect to the LG barycenter. 

Based on the ideas above, Diaz et al. in 2014 \cite{2014MNRAS.443.1688D} estimated the mass ratio between M31 and MW as $10^{0.36\pm 0.29}$. 
Combined with the virial theorem, the total mass of LG was estimated as $2.5\pm0.4\times10^{12}\msun$, and hence the mass of MW was 
constrained as $0.8\pm0.5\times10^{12}\msun$.

\section{Satellite phenomenology: matching observed satellites to simulations}
\label{sec:SatPheno}

The population of satellite galaxies in our MW offer various approaches to measure the virial mass of our MW. We have already 
introduced the example of measuring the MW timing mass based on the MW-Leo I pair (see Sec.~\ref{sec:timing} for details) 
with calibrations against numerical simulations. MW satellite galaxies have also been used as dynamical tracers together 
with globular clusters and halo stars in the SJE modeling and distribution functions (see Sec.~\ref{sec:jeans} and 
Sec.~\ref{sec:DF}), though whether satellites are in dynamical equilibrium awaits further checks. In this section, we 
introduce the efforts which constrain the virial mass of our MW by comparing observed bright dwarf spheroidal satellite 
galaxies in our MW and subhalos in numerical simulations. Some of the studies attempted to select subhalos in simulations 
that are analogous to observed MW satellites, and sample the simulated systems under the Bayesian framework to obtain the 
most likely virial mass for our MW \citep{2011MNRAS.414.1560B, 2013ApJ...770...96G, 2017MNRAS.468.3428P,2018ApJ...857...78P}. 
Some studies simply looked at the fraction of MW satellite-like systems as a function of the virial mass of the host halo in 
simulations \citep{2014MNRAS.445.2049C,2014MNRAS.437..959B}. There are also attempts which relied on empirical relations derived 
from simulations that link observed satellite properties to the virial mass of the host halo \citep{2007MNRAS.379.1464S}. 
The group of measurements introduced in this section fall in the category of ``Satellite Phenomen'' in Fig.~\ref{fig:massplot}.

Such comparisons and calibrations stem on the fact that satellite galaxies can be directly linked to subhalos in N-body 
or hydrodynamical simulations. This is an advantage compared with stars and globular clusters, as modern hydrodynamical 
simulations do not have enough power to resolve individual stars, whereas particle painting/tagging approaches have to 
rely on semi-analytical modeling of stellar evolution and phase-space sampling \citep[e.g.][]{2015MNRAS.446.2274L}. 

To properly link observed satellites to simulated subhalos, available proper motions are crucial. With high precision 
astrometric instruments and imaging data taken at different epochs, accurate proper motions for about ten classical MW
satellite galaxies and the Magellanic Clouds had already been measured \citep[e.g.][]{2002AJ....124.3198P,2003AJ....126.2346P,
2004AJ....128..687D,2005ApJ...618L..25D,2005ApJ...631L..49D,2005AJ....130...95P,2006AJ....131.1445P,2006ApJ...652.1213K,
2006ApJ...638..772K,2007AJ....133..818P,2008AJ....135.1024P,2008ASSP....5..199P,2013ApJ...768..139S,2016AJ....152..166P}.
The classical satellites are bright spheroidal dwarfs, including Leo I and the Sagittarius dwarf introduced in Sec.~\ref{sec:HVS},
Sec.~\ref{sec:stream} and Sec.~\ref{sec:timing}. These satellites have already been used as dynamical tracers in many 
previous studies based on the SJE, distribution functions or tidal streams. Recently, the proper motions of 9 classical 
dwarf spheroids, the ultra faint satellite galaxy, Bootes I, and the Magellanic Clouds in our MW were either refined or 
further measured based on \textit{Gaia} DR2 \citep[e.g.][]{2018A&A...616A..12G}.

\subsection{Magellanic Clouds}
\label{sec:MC}

Among the satellite galaxies of our MW, the Large and Small Magellanic Clouds (LMC and SMC) are of great interests. 
They are very likely accreted by our MW as a group given the similarity in their orbits \citep{2006ApJ...652.1213K}, 
which is in good consistency with simulation results \citep[e.g.][]{2008ApJ...686L..61D}. 

For galaxies with LMC stellar mass, the typical host halo mass is $\sim2\times 10^{11}\msun$ before being accreted by a 
more massive host halo and becoming satellites. The host halo mass of SMC is approximately a factor of 2 to 3 smaller.
Interestingly, LMC stellar mass galaxies with an SMC mass satellite are very rare and are typically ${\sim}50\%$ 
more massive than LMC sized objects, which suggest that the LMC have been as massive 
as $\sim3\times 10^{11}\msun$ at infall \cite{2016MNRAS.456L..54P,2018MNRAS.479..284S,2019MNRAS.483.2185C}. Hence 
the MW's two brightest satellites are massive objects which contribute a considerable fraction of the 
total MW mass. Such massive satellites are rare in halos smaller than $M_{200}\sim10^{12}\msun$ in numerical simulations, 
but are more common if the host halo is more massive than $M_{200}\sim2.0\times 10^{12}\msun$ (see discussions in Boylan-Kolchin 
et al. in 2010 \citep{2010MNRAS.406..896B}). The probability for such massive satellites to be within MW-like host galaxies 
as predicted by numerical simulations \citep{2011ApJ...743..117B,2011MNRAS.414.1560B,2017MNRAS.468.3428P,2018MNRAS.479..284S} 
is in very good agreement with other extra-Galactic MW-like galaxies in SDSS \citep{2011ApJ...738..102T,2011ApJ...733...62L}.

Before the three-dimensional velocities of MCs were actually measured, it was conventionally believed that MCs have accomplished 
multiple passages orbiting the MW \citep[e.g.][]{1980PASJ...32..581M,1994MNRAS.266..567G}. The argument was motivated by the 
existence of the long and coherent Magellanic stream, which is a young stream of HI gas spanning 150$^\circ$ along the sky and 
was believed to be formed by tidal forces. However, with the measured proper motion, Kallivayalil et al. in 2006 \cite{2006ApJ...638..772K} 
updated the total velocity of the LMC to be $\sim$380~km/s, larger than the commonly assumed velocity in old studies. In addition, it 
was found that the observed three-dimensional velocities of the LMC were not aligned with the Magellanic stream. 
Thus the Magellanic stream might have formed through local interactions between the LMC 
and SMC, rather than formed by the MW's tidal or ram-pressure stripping (see the paper by Besla et al. in 2010 \cite{2010ApJ...721L..97B}). 
Follow-up papers based on the measured high velocity of the massive LMC argued that the LMC was very likely accreted late and 
on its first passage near the orbit pericenter \citep{2011MNRAS.414.1560B,2017MNRAS.464.3825P,2007ApJ...668..949B,2011ApJ...743...40B,
2013ApJ...770...96G}. The time it spends close to the orbit pericenter is short due to its high speed, which might explain why LMC 
analogues in numerical simulations matched in Galactocentric distances and velocities are rare. 

Looking for subhalos which have similar masses and velocities as that of the LMC in the cosmological MillenniumII simulation 
\citep[MRII;][]{2009MNRAS.398.1150B}, Boylan-Kolchin et al. in 2011 \cite{2011MNRAS.414.1560B} claimed that the virial mass 
of our MW is unlikely smaller than $M_{200}\sim1.25\times10^{12}\msun$. A similar conclusion was reached by Patel et al. 
in 2017 \cite{2017MNRAS.464.3825P} based on the high resolution and dark matter only run of the cosmological Illustris 
simulation \citep{2014MNRAS.445..175G,2014MNRAS.444.1518V,2014Natur.509..177V,2015A&C....13...12N}. The orbital energies 
of LMC analogues in the Illustris simulation favor a MW halo mass of $1.5\times10^{12}\msun$.

Tighter constraints on the mass of our MW have been subsequently made by Busha et al. in 2011 
\cite{2011ApJ...743...40B}, Gonz{\'a}lez et al. in 2013 \cite{2013ApJ...770...96G} and Patel et al. in 2017 
\cite{2017MNRAS.468.3428P}, by comparing the observed positions, velocities and masses of MCs with MC-analogues 
in numerical simulations under the Bayesian framework. Based on dark matter halos in the cosmological Bolshoi 
simulation \citep{2011ApJ...740..102K}, Busha et al.~in 2011 \cite{2011ApJ...743...40B} statistically sampled 
subhalos in a large population of host dark matter halos. The observed Galactocentric distances, total velocities 
and the circular velocities of both the LMC and SMC were used to construct the likelihood that a halo of a given 
mass can host two satellites with these properties, and the prior was represented by the sample of halos in 
the simulation. The posterior probability distribution function was calculated through importance sampling. 
The virial mass of our MW was constrained to be $M_{200}=1.0_{-0.4}^{+0.7}\times 10^{12}\msun$.


Gonz{\'a}lez et al. in 2013 \cite{2013ApJ...770...96G} further investigated the effect of the LG environment on estimating 
the virial mass of our MW. It was found that satellites in host halos of LG-like environments tend to have slightly larger 
velocities, but it does not significantly affect the likelihood. Gonz{\'a}lez et al. \cite{2013ApJ...770...96G} derived the 
virial mass of our MW to be $M_{200}=1.15_{-0.34}^{+0.48}\times 10^{12}\msun$, which is in good agreement with the earlier 
measurement by Busha et al. \cite{2011ApJ...743...40B}. 

Based on the dark matter only run of the Illustris simulation, Patel et al. in 2017 \cite{2017MNRAS.468.3428P} have selected 
satellite galaxies in host halos with different virial masses, using the observed Galactocentric distance, total velocity, 
circular velocity and the specific angular momentum of the LMC (SMC was not used in their analysis), and have employed a similar 
Bayesian analysis as Busha et al. \cite{2011ApJ...743...40B}. Patel et al. have found that the specific angular momentum of 
satellites is well conserved, and the virial mass of our MW is $M_{200}=0.83^{+0.77}_{-0.55}\times 10^{12} \msun$. 
In a later study, Patel et al. in 2018 \citep{2018ApJ...857...78P} extended the method to all MW satellites with available 
proper motions. They found that the scatter among mass estimates based on individual satellite can be reduced by using the 
specific angular momentums instead of a satellite's velocity. Joint constraints based on all classical satellites suggested 
a virial mass of $0.68_{-0.26}^{+0.23}\times10^{12}\msun$. If one were to exclude the Sagittarius dwarf satellite, the 
measured virial mass would be $0.78_{-0.28}^{+0.29}\times10^{12}\msun$.

\subsection{$V_\mathrm{max}$ distributions, orbital ellipticities and velocity dispersions of classical satellite galaxies}

In addition to MCs, there are about 10 classical dwarf spheroidal satellites in our MW. Early attempts of using these classical 
satellites to constrain the MW virial mass involved using the velocity dispersion of the population. Sales et al. in 2007 
\cite{2007MNRAS.379.1464S} analyzed subhalos and satellite galaxies in a suite of N-body and hydrodynamical simulations. 
They found that the spatial and kinematic distributions of satellites trace well that of dark matter, and that the velocity 
dispersion of the satellites is closely related to the virial velocity of the host halo, $\sigma_\mathrm{sat}/V_\mathrm{vir}
\sim 0.9\pm 0.2$. Applying the relation to the velocity dispersion of MW classical satellites, the virial velocity of the 
MW was constrained to be $109\pm 22$~km/s. This corresponds to a very low MW virial mass of $M_{200}=0.58_{-0.20}^{+0.24}\msun$.

As mentioned above, MCs are massive, whose maximum circular velocities, $V_\mathrm{max}$, are greater than 60~km/s. On the 
other hand, most classical satellites in our MW have $V_\mathrm{max}$ smaller than 30~km/s, except for the Sagittarius 
dwarf. So basically, at most three satellite galaxies of our MW (the LMC, SMC and potentially the Sagittarius dwarf) have 
$V_\mathrm{max}\geq30$~km/s, whereas all the other satellites have $V_\mathrm{max}<30$~km/s. There is an apparent lack of 
objects with $V_\mathrm{max}$ between 30 and 60~km/s, which leads to the so-called ``too big to fail'' problem that our 
MW does not have enough massive satellite galaxies (only three) to match the number of massive subhalos in numerical 
simulations. The problem can be resolved if the virial mass of our MW becomes smaller than 1$\times10^{12}\msun$ 
\cite{2012MNRAS.424.2715W,2014MNRAS.445.1820C}, but as have been introduced in the previous section, the existence of 
LMC and SMC in $\sim10^{12}\msun$ halos is rare. 

Using the cosmological MillenniumII simulation, Cautun et al.~in 2014 \cite{2014MNRAS.445.2049C} found that the virial mass of our 
MW should satisfy $M_{200}\leq 1.4\times10^{12}\msun$ to meet the condition of only three satellites with $V_\mathrm{max}\geq30$~km/s, 
whereas the condition of hosting LMC and SMC-like subhalos requires $M_{200}> 1\times10^{12}\msun$. Combining the two requirements, 
the most plausible virial mass for a halo to host a MW-like population of subhalos was estimated to be $M_{200}=0.78_{-0.33}^{+0.57}
\times10^{12}\msun$. The confidence region is given by the fraction of halos in the simulation which have at most three 
subhalos with $V_\mathrm{max}\geq 30$~km/s, and at least two subhalos with $V_\mathrm{max}\geq 60$~km/s.

Based on MW-like halos in the Aquarius simulation \citep{2008MNRAS.391.1685S}, Barber et al. in 2014 \cite{2014MNRAS.437..959B} 
investigated the orbital ellipticity distribution of subhalos. They have found that the orbital ellipticity distribution of subhalos 
which can plausibly host luminous satellites show little halo-to-halo variations in cosmological simulations. Given a set of fiducial MW 
virial masses, Barber et al. \cite{2014MNRAS.437..959B} inferred the orbit ellipticity of nine galactic classical satellites, 
which were then compared with the simulation-based orbital ellipticity distribution. The virial mass of our MW was constrained 
to be $M_{200}=1.10_{-0.29}^{+0.45}\times 10^{12} \msun$, in order to bring consistency between observed and simulation based 
orbit ellipticity distributions.

\section{Other methods}
\label{sec:other}

\subsection{Total mass estimated from baryonic mass fraction}
A dynamics-free lower mass limit has been estimated for our MW by Zaritsky and Courtois in 2017 \cite{2017MNRAS.465.3724Z}, based on the 
total baryonic matter and the cosmological baryon fraction. To estimate the total baryonic matter, Zaritsky and Courtois \cite{2017MNRAS.465.3724Z} 
used a sample of MW-like disk galaxies which have measured stellar mass and cold disk gas mass, while the mass confined in hot and cold 
halo gas was taken from other studies \citep{2012ApJ...756L...8G,2014ApJ...792....8W,2015ApJ...800...14M,2016ApJ...828L..12N}. The total 
baryonic mass was converted to the projected total mass assuming the baryon fraction in MW-like galaxies is the same as inferred by the 
cosmological baryon fraction. Based on the mass distribution of their 151 MW-like galaxies and the measurement uncertainties of cold and 
hot gas mass, they estimated a 10\% lower percentile of $7.7\times10^{11}\msun$ and a median of $1.2\times10^{12}\msun$.


\section{Summary and discussion}
\label{sec:disc}

The last two decades have seen a multitude of determinations of the virial mass of our Galaxy using a diversity of methods and tracer 
populations. This review is an attempt to summarize and classify the various approaches used in literature, and to highlight potential 
ways in which future progress can be made. The numerous studies and methods used are best outlined by Fig.~\ref{fig:massplot}, which 
encapsulates the previous determinations of the MW virial mass. It shows that there are no less than 47 individual MW mass measurements 
using seven broad classes of methods (see Table~\ref{tbl:summary} for a short summary of these classes).

\begin{figure}[H]
    \centering
    \vskip -.3cm
    \includegraphics[width=0.49\textwidth]{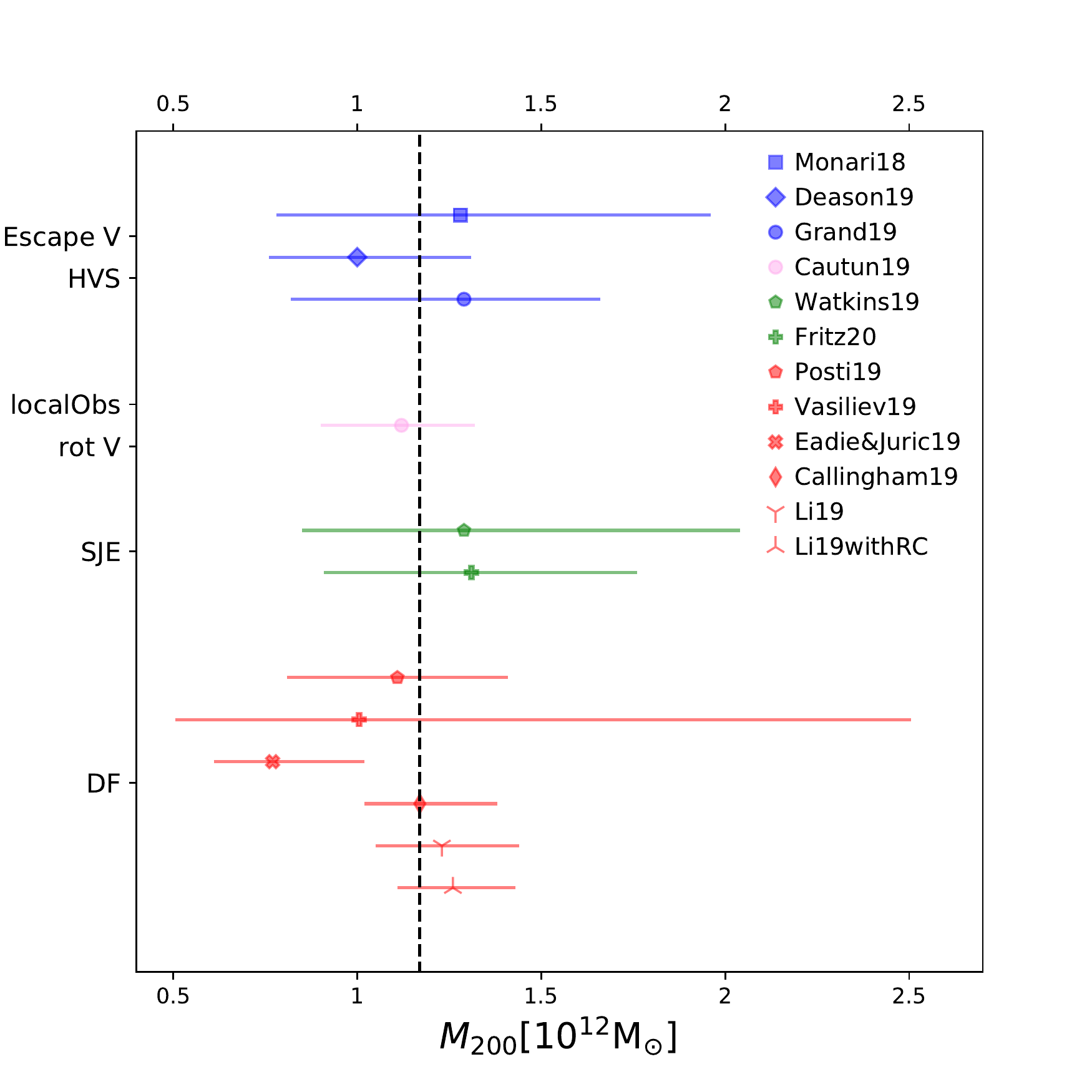}
    \vskip -.3cm
    \caption{ Similar to Fig.~\ref{fig:massplot}, but it shows only recent MW mass measurements that made use of the \textit{Gaia} DR2 data. 
    Such measurements typically use a large number of tracers with full 6D phase-space data and are less affected by systematic 
    uncertainties (see main text). For reference, the vertical dashed line     marks the median value among all 11 measurements. Note that 
    this plot shows a narrower $M_{200}$ range than Fig.~\ref{fig:massplot} (it roughly corresponds to the range between the two vertical 
    dotted lines in Fig.~\ref{fig:massplot}). The errorbars correspond to $68\%$ confidence intervals.
    }
    \label{fig:massplot_gaiadr2}
\end{figure}

Our compilation of total mass measurements highlights that the virial mass of our Galaxy is still uncertain to within at least a factor 
of two, and that probably its value lies in the interval $[0.5,2.0]\times10^{12}~\msun$. The large uncertainty interval is a byproduct 
of two effects. Firstly, some of the first studies in the subject had access to a limited and potentially biased number of observations, 
typically with large measurement errors, and thus those results have large uncertainties associated to them. Secondly and more worryingly, 
some measurements have potentially systematic biases that are not taken into account in the quoted error bars. This explains why at least 
a couple of results (e.g. see some of the distribution function results in Fig.~\ref{fig:massplot}) have very small error bars, however 
those measurements do not overlap within reasonable confidence limits.

We also find systematic differences in the inferred MW virial mass between classes of methods. For example, the determinations based on 
high velocity stars and also on the assumption that the MW satellites are bound (Sec.~\ref{sec:HVS}) typically argue for a heavy MW, with 
a virial mass higher than $1.25\times10^{12}~\msun$. Similarly, the timing argument (Sec.~\ref{sec:timing}), which provides a lower 
limit to the virial mass of our MW, often prefers a heavy MW as well. A massive MW is also helpful to explain how come our Galaxy 
simultaneously hosts two very bright satellite galaxies, the LMC and the SMC. In contrast, other classes of methods, such as the dynamics 
of MW satellites and those of the LG typically suggest MW virial masses below $1\times10^{12}~\msun$, which, for example, are 
needed to alleviate the too big to fail problem. The other methods, such as the those based on the spherical Jeans equation or on modeling 
the distribution function, prefer a mass of about ${\sim}1.0\times10^{12}~\msun$, with a roughly equal number of measurements above and 
below this value.

As we just discussed, many mass measurements, especially early ones, are affected by either small sample size or incomplete phase-space data 
(e.g. proper motions) that can lead to large statistical and, more importantly, systematic uncertainties. The recent \textit{Gaia} DR2 data, combined 
with other complementary observations, offer an exquisite data set with millions of stars and tens of satellite galaxies and globular clusters with 
full position and velocity information. In Fig.~\ref{fig:massplot_gaiadr2}, we present the subset of MW total mass measurements that use the 
\textit{Gaia} DR2 data. Interestingly, these recent measurements show good agreement with each other and even between different classes of methods,
with most results overlapping within their 1-$\sigma$ uncertainties. The good agreement between methods and the small uncertainties are at least partly 
due to the availability of \textit{Gaia} DR2 proper motions. This means that there is no need to make many of the assumptions employed by earlier studies, 
which had to cope with incomplete data, and thus these recent estimates are not affected by the systematics arising from these assumptions. Furthermore, 
many of the methods shown in Fig.~\ref{fig:massplot_gaiadr2} have been validated and calibrated against numerical simulations (for details see Section 
\ref{sec:sims}) and thus have more realistic and better understood uncertainties. We note that many of the studies shown in the figure used overlapping 
data sets. For example, i) the Deason et al. and Grand et al. studies are based on the same stellar halo high velocity data, while the latter used an 
improved method calibrated with numerical simulations; ii) the Watkins et al., Posti \& Helmi, Vasiliev, and Eadie \& Juri{\'c} studies are based on 
the dynamics of halo globular clusters; and iii) the Callingham et al. and Li et al. studies used respectively the classical and all Galactic satellites 
and both methods are based on empirical phase-space distribution functions determined from the same hydrodynamical 
simulation. Nonetheless, the good agreement between the various estimates shown in Fig.~\ref{fig:massplot_gaiadr2} is indicating that we are converging 
towards a higher precision determination of the MW total mass.

\subsection{A summary of different methods}

We provide a summary of the various classes of methods in Table~\ref{tbl:summary},
including the tracers they have used and their modeling uncertainties.

Beyond the limitations listed in the table, all methods can be potentially affected by sample selection effects and hidden observational 
systematics. For example, both Vasiliev \cite{2019MNRAS.484.2832V} and  Posti et al. \cite{2019A&A...621A..56P} in 2019 have assumed their 
sample of globular clusters to be complete, which might not be exactly true. In addition, since many of the previous measurements rely 
critically on the radial velocity of tracer objects from one or multiple spectroscopic surveys, systematic differences among different 
surveys may cause discrepancies between studies using different data or result in systematic biases in the same study combining different 
data. For example, comparisons of the radial velocities measured from LAMOST, RAVE, SEGUE and APOGEE have revealed a systematic offset of 
$\sim$5~km/s between LAMOST and the other three surveys \citep[e.g.][]{2015RAA....15.1095L,2015MNRAS.448..822X,2017MNRAS.472.3979S}. The 
cause for the offset is still unknown, but it is shown to be independent of stellar
properties and signal-to-noise ratios, and have been corrected for in the LAMOST data.

Besides, all methods can potentially suffer from uncertainties arising from a sparse 
number of tracers at large distances, and hence the measured total or virial mass of the MW largely relies on model extrapolations 
to large Galactocentric distances where there are not enough tracers to provide tight constraints. Furthermore, regardless of the method 
itself and the type of tracers used, parametrized potential models for both baryonic and dark matter have to be adopted for most of the 
measurements. Improper potential models can lead to ill-constrained mass, especially for the extrapolated virial mass, which is largely
model dependent. For studies trying to fit multiple components (disc, bulge and halo components for example), degeneracies exist among 
these different components. The readers can find such examples in Fig.~6 of Kafle et al. in 2014 \citep{2014ApJ...794...59K}. A 
slightly overestimated stellar disk component brings a slightly underestimated enclosed dark matter mass, but a much more 
underestimated total mass, especially when no tracer objects in the outer halo are available.

In the next subsection, we summarize the role of modern numerical simulations. Since 
the mass distribution is known in simulations, testing the methods using cosmological simulations provides the most straight-forward 
way to validate model assumptions and to characterize systematic uncertainties. In addition, mock observations constructed 
from simulation data are also helpful for assessing sample selection effects. 

\begin{table*}[t]
\footnotesize
\caption{
    A summary of the classes of mass determination methods discussed in this review. 
    }
\tabcolsep 18pt 
\tabcolsep 10pt
\begin{tabular*}{\textwidth}{ p{0.2\textwidth} p{0.25\textwidth} p{0.5\textwidth} }
\toprule
  Method & Tracer Population & Uncertainties and Difficulties  \\\hline
  Escape velocity & high velocity halo stars, & Prior for the power law index, $k$, of the high velocity tail distribution.  \\
  (Sec.~\ref{sec:HVS})         &                            & Interlopers such as disk stars and substructures.  \\
                  & exotic hypervelocity stars, & Uncertain mechanisms of formation for hypervelocity stars.  \\
                  & high velocity satellites & Whether satellites are bound or not. \\\hline
  Rotation velocity & terminal velocities of ISM within $R_0$,  & Modeling the deviation from axis-symmetric assumption. \\
   of the inner MW     & circular velocities of masers/disk stars, & Need to combine with other methods and more distant tracers to \\
  (Sec.~\ref{sec:localobs})&  solar neighborhood stars   &  \ \ infer the mass out to large Galactocentric distances. \\\hline
  Spherical Jeans Equation & halo stars, satellites, globular clusters & Violation of the steady state and spherical assumptions.\\ 
  (Sec.~\ref{sec:jeans})&                                          & Unknown velocity anisotropy due to unavailable proper motions.\\
                        &                                          & Different tracer populations may have different velocity anisotropies.\\\hline
  Distribution function & halo stars, satellites, globular clusters & Violation of the steady state assumption due to phase correlations. \\ 
  (Sec.~\ref{sec:DF})   &                                            & Violation of the spherical symmetry if spherical assumption is made. \\ 
     &                                           & Validity of the functional form. \\\hline
  modeling tidal debries & stellar streams, survived progenitors & Streams do not follow exactly the orbits of progenitors.\\
  (Sec.~\ref{sec:stream})&                                       & Contamination by stars not belonging to the stream. \\
                         &                                       & Stripped stars might be re-accreted by the progenitor. \\
                         &                                       & Single orbit fitting can introduce significant biases. \\
                         &                                       & Tidal stripping and dynamical friction of the progenitor  \\
                         &                                       & \ \ is challenging to directly integrate backwards in time.  \\
                                                  &  & N-body simulations are expensive to explore the parameter space. \\\hline
  Timing argument & MW versus M31, & Non-zero tangential motions. \\
  and LG dynamics &  distant satellites of LG & Mass contributed by massive satellites such as MCs. \\
  (Sec.~\ref{sec:timing})&         & Mass outside the virial radius contributing to the local reversal \\
                &        & \ \ of cosmic expansion. \\\hline
  Empirical distributions and &luminous satellite galaxies& Simulation-based empirical relations and calibrations are largely\\
  relations based on      &                    & \ \ model dependent.\\
  simulated subhalos                   &                    & Simulated halos and subhalos do no fairly represent the properties \\
  (Sec.~\ref{sec:SatPheno})                   &                    & \ \ and merger histories of host and subhalos of our MW.\\
\bottomrule
\end{tabular*}
\label{tbl:summary}
\end{table*}

\subsection{The role of numerical simulations}
\label{sec:sims}
Various methods rely on numerical simulations to infer the mass of our MW. Part of the studies used simulations to either 
directly infer plausible ranges for nuisance parameters or indirectly circumvent unconstrained parameters in their modeling. 
For example, as have been mentioned in Sec.~\ref{sec:HVS}, Smith et al. in 2007 \cite{2007MNRAS.379..755S}, Piffl et al. in 
2014 \cite{2014A&A...562A..91P} and Deason et al. in 2019 \citep{2019MNRAS.485.3514D} used cosmological simulations to infer 
the range of the power-law index in their high velocity tail distribution of halo stars. In addition, Xue et al. in 
2008 \cite{2008ApJ...684.1143X} adopted the distributions of radial versus circular velocities of star particles in two simulated 
halos from hydrodynamical simulations. The Galactic circular velocities were determined by matching the observed distributions 
of radial versus circular velocities to those in simulations. Their approach helped to circumvent the problem of unknown proper 
motions or velocity anisotropies of tracers. 

Some studies entirely depended on empirical relations \citep{2007MNRAS.379.1464S} and probability distributions \citep{2011ApJ...743...40B,
2013ApJ...770...96G,2014MNRAS.445.2049C,2014MNRAS.437..959B,2017MNRAS.468.3428P,2019MNRAS.484.5453C} drawn from MW-like systems 
in simulations, which were then used to infer the virial mass of our MW. These efforts have been introduced in detail in Sec.~\ref{sec:DFsim} 
and Sec.~\ref{sec:SatPheno}. 


Other studies have relied on numerical simulations to validate their dynamical modeling, and some have calibrated 
their inferred mass by comparing between recovered and true masses in simulations. In the following, we 
summarize some of these attempts. 

The early attempt of validating and calibrating the recovered MW mass through cosmological simulations can be traced 
back to the study of Li and White in 2008 \cite{2008MNRAS.384.1459L}, who have applied the timing argument approach (see details 
in Sec.~\ref{sec:timing}) to halo pairs from the cosmological Millennium simulation. They calibrated the difference 
between MW timing mass and the true virial mass, using halo pairs similar to MW and Leo I. The usage of MW and Leo I 
pair requires the boundedness of Leo I to our MW, which was validated by Boylan-Kolchin et al. in 2013 \cite{2013ApJ...768..140B} 
based on subhalos in cosmological simulations. 


Pe{\~n}arrubia et a. in 2016 \cite{2016MNRAS.456L..54P} constrained the virial mass of our MW through the dynamics of 
local Hubble flow under the framework of the timing argument (see details in Sec.~\ref{sec:LHF}). Their measurement 
in 2016 has considered the effect of the LMC, and was calibrated by Pe{\~n}arrubia and Fattahi in 2017 \cite{2017MNRAS.468.1300P}, 
by estimating the mass outside the virial radius of MW-like host halos using cosmological hydrodynamical simulations \citep{2016MNRAS.456L..54P}.

In terms of the distribution function method (see Sec.~\ref{sec:DF}). Wang et al. in 2015 \cite{2015MNRAS.453..377W} have extended the 
approach to the NFW potentials, and have applied it to five MW-like galaxies from the Aquarius simulation, with stars generated from the 
particle tagging technique of \cite{2010MNRAS.406..744C}. The best-fit virial masses were biased from their true values to varying levels, 
and it was concluded that the cause for the bias varied from halo to halo. 
Besides, although it seems reasonable to assume that the binding energy and angular momentum terms can be decoupled from each other, the 
$\beta$ parameter in Eqn.~\ref{eq:anisoform} only stands for the true averaged velocity anisotropies for dark matter particles. For stars, 
their true velocity anisotropies are in fact larger than half of the best-fit power-law index in Eqn.~\ref{eq:anisoform}. This reflects 
the inability of the adopted form of the distribution function to correctly match the actual distribution of stars. 
In fact, by applying the oPDF method (see Sec.~\ref{sec:DForbit} for details), which is a free-form distribution function method, to the 
Aquarius halos, it was shown that the major systematic biases found in \cite{2015MNRAS.453..377W} can be removed for all the halos. This 
demonstrates the biases found in \cite{2015MNRAS.453..377W} can be mostly attributed to the failure of the assumed function form in 
matching the actual distribution. Thus it is critical to avoid introducing incorrect or strong model assumptions in the construction of 
a distribution function.

\begin{figure}[H]
\includegraphics[width=0.5\textwidth]{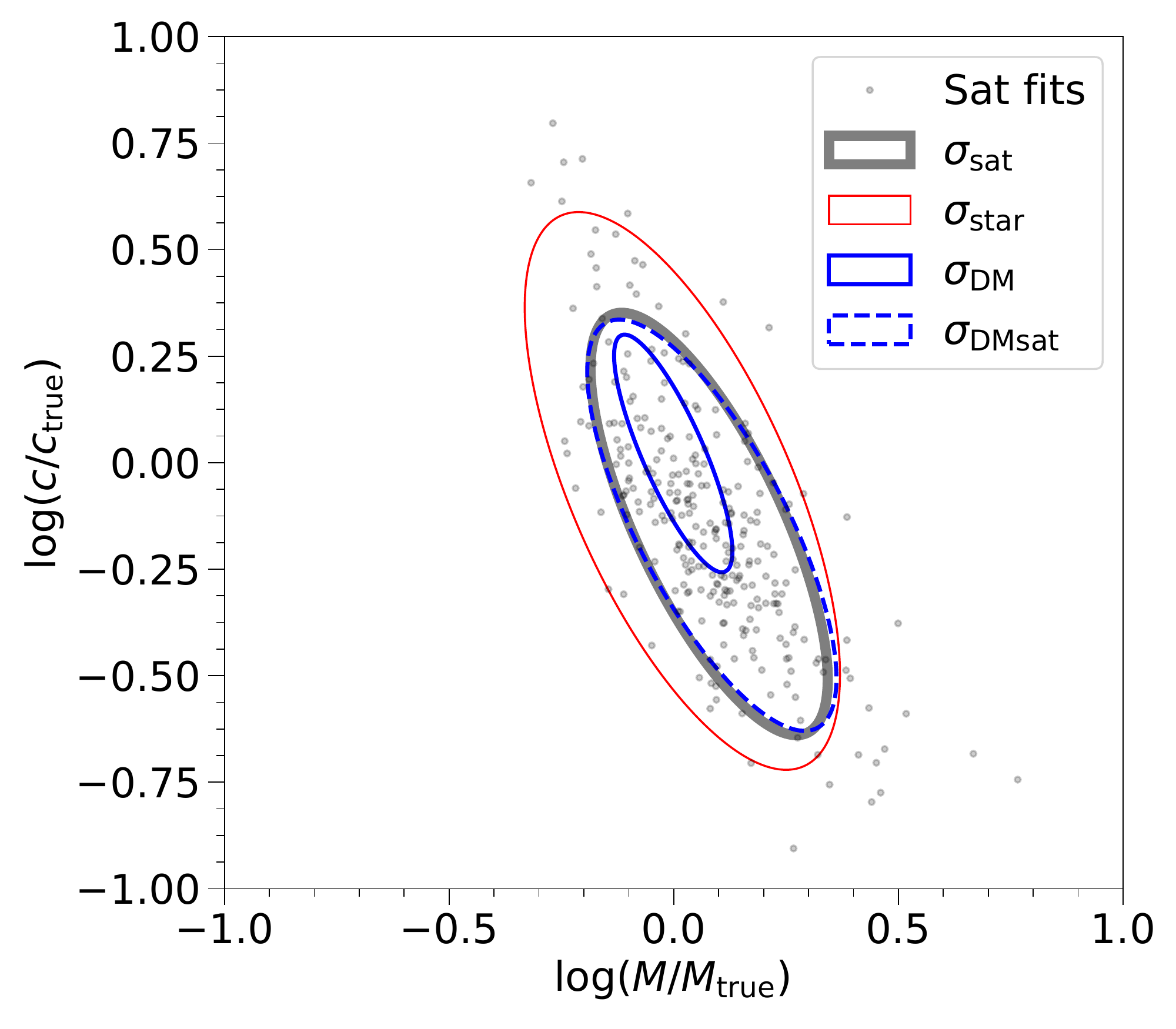}
\caption{Biases of steady-state modeling using different tracers. A sample of MW-sized halos are selected from the Millennium-II 
simulation, with satellite galaxies extracted from the semi-analytic galaxy formation model~\cite{Guo11}. The grey points show the 
fits to individual halos using satellites as tracers. The best-fit parameters are normalized by the true parameter values. The thick 
grey ellipse ($\sigma_{\rm sat}$) shows the total $1-\sigma$ scatter of the data points. The red ($\sigma_{\rm star}$) and blue 
($\sigma_{\rm DM}$) solid ellipses show the total scatters in the fits using stars and dark matter particles as tracers 
respectively~\cite{2017MNRAS.470.2351W}. For stars and dark matter particles, the number of tracers is $\sim 10^5$ for each halo, 
so that the total scatters are dominated by systematic biases. For satellites, the number of tracers is $\sim 100$. For a fair 
comparison between satellites and dark matter, the blue dashed ellipse ($\sigma_{\rm DMsat}$) shows the total scatter associated 
with dark matter particles down-sampled to the have the same number as that of satellites. Figure reproduced from \cite{Han2019IAU}.}
\label{fig:tracers_comp}
\end{figure}

Wang et al. in 2017 \cite{2017MNRAS.470.2351W} have further applied the oPDF method to $\sim1000$ MW sized dark matter halos from the 
cosmological Millennium II simulation and 24 MW/M31-like galaxies from the APOSTLE hydrodynamical simulations. On average, the best-fit 
halo properties were unbiased, while significant individual biases exist for most halos (see e.g., Fig.~\ref{fig:tracers_comp}). Such 
individual biases can be as large as a factor of 2 to 3 when star particles were used as dynamical tracers. They found that these biases 
can be mostly attributed to correlated phase-space structures that violate the steady-state assumption. In the presence of phase correlations, 
the number of independent tracer particles is smaller than the actual number of tracers. This leads to stochastic biases in the parameter 
estimates, the distribution of which are determined by the effective number of phase independent particles. As the oPDF method only makes 
use of the steady-state property (Equation~\eqref{eqn:oPDF}) in modeling the dynamics, these results suggest that there is only limited 
information that can be extracted from the data under a steady-state assumption. Such a limiting precision of steady-state modeling was 
further confirmed by Wang et al. (2018) \cite{2018MNRAS.476.5669W}, who found very similar amount of stochastic biases using the spherical 
Jeans equation (SJE, see section~\ref{sec:jeans}). Even though the methodologies of the oPDF and the Jeans equation are very different, they 
can be both derived from the collisionless Boltzmann equation under the steady-state assumption.  

This steady-state information limit, though affecting all steady-state methods, is still different for different tracers. As found by Han et al. 
2019~\cite{Han2019IAU}, the dynamical state of satellite galaxies is found to be close to that of dark matter particles. As a result, satellite 
galaxies exhibit a smaller stochastic bias compared to halo stars, as shown in Fig.~\ref{fig:tracers_comp}.  

In addition to picking a better steady-state tracer, another way of getting over the steady-state information limit is to use 
additional information beyond the steady-state assumption. Such information can be provided, for example, by the halo-dependent 
distribution of orbits that can be extracted semi-empirically from simulations. This is exactly what is done in the template-based 
distribution function method such as \cite{2019ApJ...886...69L}, \cite{li2019constraining} and \cite{2019MNRAS.484.5453C}. As tested 
in \cite{2019ApJ...886...69L} using simulated halos, their method is able to achieve a smaller systematic uncertainty than the oPDF 
method if a correct template is used.

Besides the dynamical state of the tracers, Wang et al. \cite{2017MNRAS.470.2351W} have tested the validity of modeling the 
underlying potential as an NFW profile, and have found that biases arise when using tracers within 20~kpc from the galaxy center,
because the inner profiles deviate from the NFW model due to baryonic physics and the existence of galaxy disks. After excluding 
these innermost tracers, the NFW model returns on average unbiased mass estimates, and the scatter is very similar to that based 
on true potential templates directly extracted from the simulations. Deviations from spherical symmetry and the existence of a 
companion halo are also found to contribute to their systematic biases~\citep{2017MNRAS.470.2351W,2018MNRAS.476.5669W}. For 
typical applications of the SJE and derived mass estimators, additional systematic biases arise when $\beta$ is treated as a free 
but constant parameter~\citep{2018MNRAS.476.5669W,2018MNRAS.475.4434K} and when the radial density profile of tracer objects 
is not properly measured or modeled \citep{2020arXiv200102651F}.

In two very recent papers \citep{2020arXiv200102651F,2020arXiv200111030E}, the Watkins et al. \cite{2010MNRAS.406..264W} mass estimator was 
validated and calibrated against simulations. Fritz et al. in 2020 \cite{2020arXiv200102651F} used dark matter only simulations to test and 
calibrate their measurements based on 45 satellite galaxies. After applying the mass estimator to simulations, significant biases were reported, 
which were mainly due to the deviation of satellite density profiles from a single power law. \cite{2020arXiv200102651F} also discussed 
systematic uncertainties arising from the LMC and LMC satellites. In another study, Erkal et al. in 2020 \cite{2020arXiv200111030E} 
pointed out that since the infall of the massive LMC can induce a substantial reflex motion in the MW \citep[e.g.][]{2015ApJ...802..128G,
2019MNRAS.487.2685E,2020MNRAS.tmpL..25P}, this can make our MW, in particular the outer stellar halo, be out of equilibrium. Under this 
picture, \cite{2020arXiv200111030E} investigated how the non-equilibrium affects the performance of the Watkins et al. mass estimator and 
how this reflex motion affects the mass estimated by using Leo I \citep{2013ApJ...768..140B}. They found if the mean reflex motion is not 
accounted for, the mass estimator can have systematic biases which are always positive and can be as large as 50\%. In addition, the LMC 
can significantly increase the speed of Leo I and cause overestimation of the MW mass. 

A series of studies by Eadie et al. \citep{2015ApJ...806...54E,2016ApJ...829..108E,2017ApJ...835..167E,2019ApJ...875..159E} 
between 2015 and 2019 constrained the mass of our MW by modeling the phase-space distribution of globular clusters. In 
their paper published in 2018 \citep{2018ApJ...865...72E}, old and metal poor star particles in hydrodynamical simulations 
were used as globular cluster analogs to test and validate their hierarchical Bayesian-based dynamical method. Eadie et al. have 
found that the virial mass of the host halo in the simulations can sometimes be well recovered, but sometimes not. The main cause 
behind the incorrect estimates is due to the model itself: it has difficulties to simultaneously predict the inner and 
outer regions of the true mass profile. This limitation is probably due to the single power-law potential 
model \citep{2012MNRAS.424L..44D}. Using only the outer-most tracers, where the underlying potential can be better 
approximated as a single power law, results can be improved to provide more accurate mass determinations.

Compared with satellite galaxies, halo stars and globular clusters, the modeling of spatially extended tidal streams is more 
complicated. A single orbit fitting can introduce significant biases, and integrations done backward in time makes it hard to 
directly incorporate tidal stripping and dynamical friction for the progenitor. Moreover, the tidal stream does not 
strictly follow the orbit of the progenitor. As we have mentioned in Sec.~\ref{sec:stream}, N-body simulations are powerful 
approaches of modeling tidal streams and the remnant of their progenitors than single orbit fitting, but N-body simulations 
are very expensive to explore the parameter space \citep{2005ApJ...619..807L,2010ApJ...714..229L,2017ApJ...836...92D}. Other 
alternative and less expensive approaches such as particle releasing/spraying methods \citep{2012MNRAS.420.2700K,2014ApJ...795...94B,
2014MNRAS.445.3788G,2015ApJ...803...80K} and semi-analytic approaches \citep{2017ApJ...836...92D,2017ApJ...847...42D} have 
been invoked as well. 

Most of the earlier attempts of fitting the high velocity tail distribution of halo stars relied on numerical simulations to 
infer the plausible range of the nuisance parameter, and only until very recently, Grand et al. in 2019 \cite{2019MNRAS.487L..72G} 
applied the method to simulated star particles in the suite of \textsc{auriga} simulations. They reported that substructures 
can affect the modeling of high velocity tail distributions. The median of recovered halo virial masses fell below the true 
values by $\sim$20\%, and the scatter can be as large as a factor of 2. 

The several examples just discussed highlight the essential role played by numerical simulations for validating and calibrating 
many mass estimation methods. However, these simulation-based validations and calibrations are not free of uncertainties. It is 
probably straight-forward and more robust to link satellite galaxies to simulated subhalos, which is less dependent of the baryonic 
physics adopted in the modeling. Dark matter particles are less dependent of baryonic physics as well, except for the very 
inner regions of dark matter halos, but it is not safe to directly extend validations based on dark matter particles to halo 
stars, as they have very different spatial and velocity distributions. Nowadays, star particles can be either directly simulated in 
hydrodynamical simulations or painted through semi-analytic approaches based on dark matter only simulations. Nonetheless, these 
simulations might still have limitations due to how the underlying baryonic physics is simulated or modeled, as our current 
understanding towards galaxy formation and evolution is still uncertain. Fortunately, our understanding of the infall and stripping 
of star particles under gravitational forces is more robust, and largely insensitive to baryonic physics. Compared with stars, 
there could be even more uncertainties behind the modeling of globular clusters, because the current understanding of the formation 
and evolution of globular clusters and their connections to dwarf galaxies is still uncertain \citep[e.g.][]{2015ApJ...802...30Z,
2018ApJ...858...37Z}. Also, very importantly, as has been pointed out by \cite{2019MNRAS.484.5453C} and \cite{2019MNRAS.485.3514D}, 
to test or design methods that take into account the specifics of the MW system, we would need a large sample of simulated systems 
whose mass growth is as close as possible to that of our own Galaxy. Despite these uncertain aspects, numerical simulations have 
been very helpful in aiding the dynamical modeling of various tracers and hence they are powerful tools for constraining the mass 
of our MW.

\subsection{Future prospects}

The near future is expected to bring a wealth of new observational data that will allow for a much more precise and hopefully 
accurate MW total mass determination. The abundance of new data will lead to new measurements of the MW mass distribution, such as the 
radial mass profile as well as the shape and orientation of the dark matter halo. To fully exploit the upcoming data, it is also necessary 
to improve the theoretical modeling of the MW system, such as going beyond spherical or axis-symmetric approximations, and accounting 
for non-equilibrium effects in the tracer distribution. In the following, we will discuss some of the upcoming Galactic observational
campaigns as well as theoretical improvements that will be essential to correctly interpret and model the data.

The first and second data releases of the \textit{Gaia} mission \cite{2001A&A...369..339P,2016A&A...595A...2G,2018A&A...616A...1G} 
have already revolutionized our understanding of the dynamics of the bulge, disk and halo stars, of the growth and chemical evolution, 
and of the orbits of stellar streams, globular clusters and satellite galaxies in our MW. And as we have demonstrated in 
Fig.~\ref{fig:massplot_gaiadr2}, measurements of the MW total mass using {\it Gaia} DR2 data are converging towards a higher precision.
This has been possible due to the precise \textit{Gaia} measurements of proper motions for billions of stars. 

Future \textit{Gaia} data releases\footnote{For the most up-to-date schedule see \url{https://www.cosmos.esa.int/web/gaia/release} .} 
will improve further the data quality and the number of measured stars, which are essential for accurate modeling of the MW gravitational 
potential. However, the exquisite observational data set provided by the \textit{Gaia} mission will mostly consist of nearby stars 
(e.g. most of the \textit{Gaia} DR2 stars with useful proper motion and parallax measurements are found within $\sim$3~kpc from the Sun). To 
accurately determine the MW mass, we need measurements at large Galactocentric radii, but at such large radii \textit{Gaia} will only provide 
proper motion data for a small subset of intrinsically bright stars. This will be complemented by the deep and multi-epoch LSST imaging survey 
\citep{2008arXiv0805.2366I} that will provide proper motion measurements and photometric distances for individual stars $\sim$4 magnitudes 
deeper than \textit{Gaia}. This can potentially enable 6D phases space measurements for individual stars out to several tenths to hundreds of 
kpc and will be essential for determining the dynamics of the outer MW halo.

Among the billions of stars released by \textit{Gaia} so far, only $\sim$7.2 million stars have available radial velocities (these are 
mostly stars brighter than $G=13$).  Encouragingly, many ground based follow-up spectroscopic surveys have been proposed to provide 
additional radial velocity measurements of faint stars, and especially of stars at large Galactocentric distances. To mention a few 
examples, these future surveys include the 4-metre Multi-Object Spectroscopic Telescope \citep[4MOST;][]{2016SPIE.9908E..1OD} in the 
south, the Subaru Prime Focus Spectroscopy \citep[PFS;][]{2014PASJ...66R...1T}, LAMOST-2, The Milky Way Mapper (MWM) program of SDSS-V 
\citep{2017arXiv171103234K}, the Dark Energy Spectroscopic Instrument \citep[DESI;][]{2016arXiv161100036D}, the WEAVE 
survey \citep{2014EAS....67..227B}, the MaunaKea Spectroscopic Explorer (MSE) in the north and the K2 RR Lyrae 
Survey along the Ecliptic \citep{2017EPJWC.16004004S}. 

Future surveys, such as LSST, are predicted to increase the number of satellite galaxies by a factor of two to ten, depending on 
model assumptions \citep{2018MNRAS.479.2853N,2019ARA&A..57..375S}. Most such objects are predicted to be found at large distances and 
thus represent a potentially powerful window into the dynamics of the outer halo. Gravitationally bound objects, such as dwarf galaxies 
and globular clusters, are especially useful since one can average over the individual measurements of their member stars to obtain 
much more precise positions and velocities than possible for single stars. LSST will also help to identify more low-surface-brightness 
objects and stellar streams, which are largely missed by current photometric surveys.

The topic of this review, the total mass of the MW, is the first step in characterizing the mass distribution surrounding our Galaxy. 
Future data sets will allow us to characterize the MW halo in much more details, such as in more precisely determining the concentration, 
shape and orientation of the dark matter halo. These aspects are also important for having an accurate determination of the MW total mass since, 
as the statistical uncertainties decrease, the errors will be dominated by systematic effects. For example, not accounting for the non-spherical
shape of the dark matter halo can lead to biases in the inferred halo mass \cite[e.g.][]{2018MNRAS.476.5669W}. Future models will also have to 
take into account that the shape and orientation of the halo changes with radius. In particular, previous works have argued that the outer MW 
halo is misaligned with the MW disk \cite{2010ApJ...714..229L}, which, given stability considerations, implies that the Galactic dark matter
halo has a dramatic change in orientation from being aligned with the disk at distances ${\lesssim}20~\rm{kpc}$ to being perpendicular to the
stellar disk at large distances \cite{2013ApJ...773L...4V,2019arXiv190402719S}.

The MW halo beyond several tens of kpc is likely to be out of equilibrium since the typical timescale at those distances is ${\sim}1~\rm{Gyr}$ 
and longer. Furthermore, recent observational measurements as well as theoretical considerations suggest that the MW's brightest satellite, the 
LMC, is rather massive, having had a total mass at infall of about ${\sim}2.5\times10^{11}~\msun$ \cite{2016MNRAS.456L..54P,2018MNRAS.479..284S,
2019MNRAS.483.2185C,2019A&A...623A.129F,2019MNRAS.487.2685E,2019arXiv190709484E}. This corresponds to ${\sim}20\%$ of the total MW mass, and 
thus the LMC is a major perturber of the Galactic potential. Previous studies have already highlighted the impact of a massive LMC on the orbit 
of the Sagittarius, Tucana III and Orphan streams \cite{2015ApJ...802..128G,2018MNRAS.481.3148E,2019MNRAS.487.2685E}, and have predicted significant 
LMC-induced disturbances in the density and velocity distribution of the MW stellar halo \cite{2019ApJ...884...51G}. In particular, Erkal et al. in 
2020 \cite{2020arXiv200111030E} pointed out that if the LMC is ignored, the MW mass can be overestimated by as much as 50\%. In addition, other 
massive satellite galaxies can also induce non-equilibrium features in the inner region of the MW, such as the Sagittarius perturbations of the 
stellar disk \cite[e.g.][]{2018MNRAS.481..286L}. To optimally use the wealth of upcoming observational data, future models will need to account 
for departures from equilibrium, which could be done by a combination of modeling individual perturbers, such as the LMC and Sagittarius dwarfs, 
and by statistically accounting for smaller non-equilibrium structures, such as diffuse stellar streams and shells, whose causes are harder to 
pinpoint. 

New developments will also be required in terms of cosmological simulations that better describe the MW. The current state-of-the-art simulations 
of MW-mass systems \cite[e.g.][]{2016MNRAS.457.1931S,2017MNRAS.467..179G}, while very useful, are rather limited since there are few (several tens) 
such systems, their resolution is insufficient to optimally compare with observations \cite[e.g.][]{2018MNRAS.481.1726G}, and the simulated systems 
do not necessarily reproduce the MW formation history. To better understand our Galaxy, future simulations should, beside having more realistic galaxy 
formation physics and higher resolution, be selected to reproduce key aspects in the evolution of the MW, such as a Gaia-Enceladus-Sausage 
\cite{2018Natur.563...85H,2018MNRAS.478..611B} type of merger at high redshift followed by a long period of uneventful satellite galaxy accretions.

{\bf Acknowledgements:}
WW is grateful for useful suggestions made by Yipeng Jing, Carlos Frenk, Alis Deason and Zhao-yu Li. This work was supported by NSFC grant 11973032, 
11890691, National Key Basic Research and Development Program of China (No.2018YFA0404504) and JSPS Grant-in-Aid for Scientific Research JP17K14271.

\InterestConflict{The authors declare that they have no conflict of interest.}


\bibliographystyle{abbrv}
\bibliography{master}
 
%
%

\begin{appendix}

\section{MW mass measurements}
\label{app:measurements}

\clearpage

\begin{table*}
\footnotesize
\begin{threeparttable}
\caption{MW mass measurements}\label{tab1}
\doublerulesep 0.1pt \tabcolsep 13pt 

\begin{tabular}{c|ccc}
\toprule
    \rc Method &  Reference  & $M_{200} [10^{12}\msun]$   & \makecell{$M(<r_\mathrm{tracer}~[\mathrm{kpc}]) [10^{11}\msun]$ \\ 
					or $v_\mathrm{circ}(r_\mathrm{tracer}~[\mathrm{kpc}]) [\mathrm{km/s}]$ \\ 
					or $v_\mathrm{esc}(r_\mathrm{tracer}~[\mathrm{kpc}]) [\mathrm{km/s}]$ }
\\    \hline
\multirow{6}{*}{Escape Vel of HVS} 
&  \tnote{1)}{\ \ \ }Smith 07 &  $1.42^{+0.49}_{-0.36}$ & $v_\mathrm{esc}(R_0)=544_{-46}^{+64}$\\
&  \tnote{1)}{\ \ \ }Piffl 14 & $1.60^{+0.29}_{-0.25}$ & $v_\mathrm{esc}(R_0)=533_{-41}^{+54}$ \\
&  {\ \ \ }Williams 17 &                                     & 
    \makecell{$v_\mathrm{esc}(R_0)=521_{-30}^{+46}$ \\
              $v_\mathrm{esc}(50)=379_{-28}^{+34}$ \\
              $M(50)=2.98_{-0.52}^{+0.69}$ }\\
&  Monari 18 & $1.28^{+0.68}_{-0.50}$ & $v_\mathrm{esc}(R_0)=580\pm63$ \\
&  Deason 19 & $1.00^{+0.31}_{-0.24}$ & $v_\mathrm{esc}(R_0)=528_{-25}^{+24}$ \\
&  Grand 19 & $1.29^{+0.37}_{-0.47}$ & \\

\hline
   Leo I &  Boylan-Kolchin 13 & $1.34^{+0.41}_{-0.31}$ & \\

\hline
  \multirow{11}{*}{LocalObs rot V} &  Dehnen \& Binney 98 &  & $M(100)\sim 3.41-6.95 $\\
  & Klypin 02 & $\sim 0.86 $ & \\
  & McMillan 11 & $1.26\pm0.24$ & $v_\mathrm{circ}(R_0)=239\pm5$\\
  & \tnote{2)}{\ \ \ }Nesti \& Saucci\  13 BUR & $1.11^{+1.45}_{-0.55}$ & \makecell{$M(50)=4.5_{-1.8}^{+3.2}$\\
                                              $M(100)=6.7_{-3.0}^{+6.1}$}\\
  & Nesti \& Saucci 13 NFW & $1.53^{+2.10}_{-0.70}$ & \makecell{$M(50)=4.8_{-1.4}^{+1.8}$ \\
                          $M(100)=8.1_{-2.9}^{+5.5}$} \\
  & Irrgang 13 model II &                        & \makecell{$M(50)=4.6\pm0.3$ \\
                     $M(100)=7.9_{-0.8}^{+0.6}$ \\
                     $M(200)=12._{-2.}^{+1.}$} \\
  & Irrgang 13 model III &                        & \makecell{$M(50)=8.1_{-1.5}^{+1.3}$ \\
                     $M(100)=16.7\pm4.6$ \\
                     $M(200)=30._{-11.}^{+12.}$} \\
  & Bovy 15 & $\sim$0.7 & \\
  & Bajkova \& Bobylev 16 &  & $M(200)=7.5\pm1.9$ \\
  & McMillan 17 & 1.3$\pm$0.3 & $v_\mathrm{circ}(R_0)=232.8\pm3.0$\\
  & Cautun 19 & 1.12$_{-0.22}^{+0.20}$ &  \\
  & Karukes 19 & 0.89$_{-0.08}^{+0.10}$ &  \\

\hline
  \multirow{14}{*}{SJE} &  Battaglia 05 & $0.7^{+1.2}_{-0.2}$ & $M(120)=5.4_{-1.4}^{+2.0}$\\
  & Xue 08 & $0.84^{+0.3}_{-0.2}$ & $M(60)=4.0\pm 0.7$\\
  & \tnote{3)}{\ \ \ }Gnedin 10 & 1.3$\pm$0.3 & $M(80)=6.9_{-1.2}^{+3.0}$\\
  & \tnote{4)}{\ \ \ }Watkins 10 & 1.17$\pm$0.3 & \makecell{$M(100)=6.9\pm 1.9$\\
          $M(200)=10.0\pm 2.3$\\
          $M(300)=14.1\pm 3.1$}\\
  & Kafle 12 & $0.77^{+0.40}_{-0.30}$ & $M(25)\sim 2.1$\\
  & Deason 12 &  & $M(150)\sim 5-10$ \\
  & \tnote{5)}{\ \ \ }Kafle 14 &  $0.71^{+0.31}_{-0.16}$ & \\
  & Bhattacharjee 14 &   & $M(200)\ge 6.8 \pm 4.1$ \\
  & Huang 16 & $0.85^{+0.07}_{-0.08}$ & $v_\mathrm{circ}(98.97)=147.72\pm 23.55$ \\
  & Ablimit \& Zhao 17 &  &\makecell{$M(50)=3.75\pm1.33$\\
                         $v_\mathrm{circ}(50)=180.00\pm 31.92$ }\\
  & Zhai 18 & $1.11^{+0.24}_{-0.20}$ & \\
  & Sohn 18 & $1.71^{+0.97}_{-0.79}$ & $M(39.5)=6.1_{-1.2}^{+1.8}$\\
  & Watkins 19 & $1.29^{+0.75}_{-0.44}$ & \makecell{$M(21.1)=2.1_{-0.3}^{+0.4}$\\
              $M(39.5)=4.2_{-0.6}^{+0.7}$}\\
  & Fritz 20 & $1.31^{+0.45}_{-0.40}$ & \makecell{$M(64)=5.8_{-1.4}^{+1.5}$\\
              $M(273)=14.3_{-3.2}^{+3.5}$}\\

\bottomrule
\end{tabular}

\begin{tablenotes}
\item[1)] The error of the escape velocity is 90\% confidence region.
\item[2)] The original 2-$\sigma$ errors are shrinked by 10\%.
\item[3)] The virial radius was adopted to be 300~kpc.
\item[4)] The virial radius was adopted to be 300~kpc.

If considering systematic uncertainties
due to velocity anisotropy, $\beta$, the error becomes $M_{200}=1.17_{-0.20}^{+1.30}\msun$.
\item[5)] Velocity dispersion measured on K giants out to 160~kpc.
\end{tablenotes}
\end{threeparttable}
\label{tbl:measurements}
\end{table*}


\begin{table*}
\footnotesize
\begin{threeparttable}
\doublerulesep 0.1pt \tabcolsep 13pt 

\begin{tabular}{c|ccc}
\toprule
  \rc  Method & Reference & $M_{200} [10^{12}\msun]$ &\makecell{ $M(<r_\mathrm{tracer}~\mathrm{kpc}) [10^{11}\msun]$ \\or $v_\mathrm{circ}(r_\mathrm{tracer}~[\mathrm{kpc}]) [\mathrm{km/s}]$ }\\
  \hline
  \multirow{18}{*}{DF}                                     
  & \tnote{6)}{\ \ \ }Kochanek  96 w Leo I&   & $M(50)=5.1_{-1.1}^{+1.3}$ \\
  & \tnote{6)}{\ \ \ }Kochanek  96 no Leo I&   & $M(50)=3.9_{-0.7}^{+1.6}$ \\
  & \tnote{7)}{\ \ \ }Wilkinson \& Evans\  99 & $1.9^{+3.6}_{-1.7}$ & $M(50)=5.4_{-3.6}^{+0.2}$ \\
  & \tnote{7)}{\ \ \ \ }Sakamoto 03 w Leo I & $2.5^{+0.5}_{-1.0}$ & $M(50)=5.5_{-0.2}^{+0.0}$ \\
  & Sakamoto 03 no Leo I & $1.8^{+0.4}_{-0.7}$ & $M(50)=5.4_{-0.4}^{+0.1}$\\
  & Deason 12b & $0.94^{+0.22}_{-0.20}$ & $M(50)=4.2\pm0.4$\\
  & \tnote{8)}{\ \ \ \ }Eadie 15 & $1.55^{+0.18}_{-0.13}$ & $M(260)=13.7_{-1.0}^{+1.4}$\\
  & \tnote{8)}{\ \ \ \ }Eadie 15 no Pal 3 & $1.36^{+0.15}_{-0.10}$ & \\
  & Williams \& Evans 15 &  & $M(50)\sim4.5$ \\
  & \tnote{8)}{\ \ \ \ }Eadie 16 & $0.68^{+0.07}_{-0.08}$ & $M(125)=5.22_{-0.43}^{+0.41}$\\
  & \tnote{8)}{\ \ \ \ }Eadie 16 no $<$10~kpc GC & $0.90^{+0.18}_{-0.33}$ & \\
  & \tnote{8)}{\ \ \ \ }Eadie 17 & $0.86_{-0.19}^{+0.23}$ & $M(125)=6.3\pm1.1$\\
  & Posti 19 & $1.11\pm 0.30$ & \makecell{$M(20,\mathrm{total})=1.91_{-0.15}^{+0.17}$\\
            $M(20,\mathrm{DM})=1.37_{-0.11}^{+0.12}$}\\
  & \tnote{8)}{\ \ \ \ }Eadie \& Juri{\'c} 19 & $0.77_{-0.16}^{+0.25}$ & \makecell{$M(25)=2.6_{-0.6}^{+1.0}$\\
            $M(50)=3.7_{-0.8}^{+1.4}$\\
            $M(100)=5.3_{-1.2}^{+2.1}$}\\
  & \tnote{9)}{\ \ \ \ }Vasiliev 19 & $1.0^{+1.5}_{-0.5}$ & \makecell{$M(50)=5.4_{-0.8}^{+1.1}$ \\
               $M(100)=8.5_{-2.0}^{+3.3}$ }\\
  & Callingham 19 & $1.17_{-0.15}^{+0.21}$ & \\
  & Li 19 & $1.23_{-0.18}^{+0.21}$ & \\
  & Li 19 with rotation curve & $1.26_{-0.15}^{+0.17}$ & \\

\hline
  \multirow{9}{*}{   stream     }     
  & Lin 95 &      & $M(100)\sim5.5\pm1.0$\\
  & Law 05 &      & $M(50)\sim3.8-5.6$\\
  & Koposov 10 &      & $v_\mathrm{circ}(R_0)=221\pm 18$\\
  & Newberg 10 &             & $M(60)\sim 2.7$\\
  & \tnote{10)}{\ \ \ \ }Gibbons 14 & 0.56$\pm$0.12 & \makecell{$M(50)=2.9\pm 0.5$\\
             $M(100)=4.1\pm 0.7$}\\
  & K{\"u}pper 15 & 1.69$\pm$0.42 & \makecell{$M(19,\mathrm{Galaxy})=2.14_{-0.35}^{+0.38}$\\
             $v_\mathrm{circ}(R_0)=253\pm 16$}\\
  & Malhan \& Ibata 19 &      & \makecell{$M(20)=2.5\pm0.2$\\
             $v_\mathrm{circ}(R_0)=244\pm 4$}\\
  & Hendel 18 &      & $M(60)<5.6_{-1.1}^{+1.2}$\\
  & Erkal 19 &      & $M(50)=3.8_{-0.11}^{+0.14}$\\

\hline
  \multirow{7}{*}{timing \& LG dyn. }
  & Li \& White 08 & $2.43^{+0.49}_{-0.53}$ & \\
  & Sohn 13 & $2.65^{+1.58}_{-1.36}$ & \\
  & Diaz 14 & $0.8\pm 0.5$ & \\
  & Pe{\~n}arrubia 14 & $0.80^{+0.40}_{-0.30}$ & $v_\mathrm{circ}(R_0)=245\pm23$\\
  & Pe{\~n}arrubia 16 & $0.87^{+0.22}_{-0.19}$ & \\
  & \tnote{6)}{\ \ \ \ }Zaritsky 19 & $0.91-2.13$ & \\
  & Zhai 20 & $1.5_{-0.7}^{+1.4}-2.5_{-1.4}^{+2.2}$ &  \\



\bottomrule
\end{tabular}

\begin{tablenotes}
\item[6)] 90\% confidence region.
\item[7)] Error includes systematics.

The virial mass is the flat rotation model parameter.
\item[8)] 95\% credible regions are quoted for Eadie 15 and 17.

50\% credible regions are quoted for Eadie 16 and for the mass enclosed within 25, 50 and 100~kpc by Eadie \& Juri{\'c} 19.

Error of $M_{200}$ by Eadie \& Juri{\'c} 19 is the 68\% credible region reading from their
figures.

Eadie 15 estimate is the Hernquist mass.

Eadie 19 estimate has inner cut of 20~kpc.
\item[9)] Error of $M_{200}$ includes uncertainties in the outer slope of the halo density profile.
\item[10)] Maximum distance extends to $\sim$260~kpc for Leo I.
\end{tablenotes}
\end{threeparttable}
\label{tbl:measurements2}
\end{table*}


\begin{table*}
\footnotesize
\begin{threeparttable}
\doublerulesep 0.1pt \tabcolsep 13pt 

\begin{tabular}{c|ccc}
\toprule
  \rc Method & Reference & $M_{200} [10^{12}\msun]$ & $M(<r_\mathrm{tracer}~\mathrm{kpc}) [10^{11}\msun]$ \\
\hline
  \multirow{8}{*}{satellite phenomenology} 
  & Busha 11 & $1.0^{+0.7}_{-0.4}$ & \\
  & Gonz{\'a}lez 13 & $1.15^{+0.48}_{-0.34}$ & \\
  & Patel 17 & $0.83^{+0.77}_{-0.55}$ & \\
  & Sales 07 & $0.58^{+0.24}_{-0.20}$ & \\
  & Barber 14 & $1.10^{+0.45}_{-0.29}$ & \\
  & Cautun 14 & $0.78^{+0.57}_{-0.33}$ & \\
  & Patel 18 & $0.68^{+0.23}_{-0.26}$ & \\
  & Patel 18 no Sagittarius & $0.78^{+0.29}_{-0.28}$ & \\
\hline
  Dynamics-free &  Zaritsky \& Courtois 17 & $\sim1.2$ & \\

\bottomrule
\end{tabular}

\end{threeparttable}
\label{tbl:measurements3}
\end{table*}

\end{appendix}

\end{multicols}
\end{document}